\documentclass[preprint,12pt]{elsarticle}

\usepackage{amssymb}
\usepackage{amsmath}

\usepackage{microtype}
\usepackage{booktabs}
\usepackage{array}
\usepackage{tabularx}
\usepackage{ragged2e}
\usepackage{adjustbox}
\usepackage{placeins}
\usepackage{xurl}
\usepackage{capt-of}

\usepackage{hyperref}
\hypersetup{breaklinks=true}

\newcolumntype{Y}{>{\RaggedRight\arraybackslash}X}
\newcolumntype{P}[1]{>{\RaggedRight\arraybackslash}p{#1}}

\begin{document}

\begin{frontmatter}

\title{Making Agent-Mediated Contributions Governable: A Project-Level Governance Manifest for Open-Source AI Collaboration}

\author[sxufe-ba,sxufe-exp]{Jinjin Gao}
\ead{gaojinjin@sxufe.edu.cn}

\author[nuc-cs]{Luyang Li}
\ead{liluyang@nuc.edu.cn}

\author[sxufe-ba]{Shufen Guo}
\ead{guosf@sxufe.edu.cn}

\author[warwick-cs]{Ligang He}
\ead{ligang.he@warwick.ac.uk}

\author[sxufe-ba,sxufe-info]{Xiaoning Sun\corref{cor1}}
\ead{arborxs@163.com}

\affiliation[sxufe-ba]{organization={School of Business Administration, Shanxi University of Finance and Economics},%Department and Organization
            %addressline={}, 
            city={Taiyuan},
            postcode={030006}, 
            state={Shanxi},
            country={China}}

\affiliation[nuc-cs]{organization={School of Computer Science and Technology, North University of China},%Department and Organization
            %addressline={}, 
            city={Taiyuan},
            postcode={030051}, 
            state={Shanxi},
            country={China}}

\affiliation[sxufe-info]{organization={School of Information, Shanxi University of Finance and Economics},%Department and Organization
            %addressline={}, 
            city={Taiyuan},
            postcode={030006}, 
            state={Shanxi},
            country={China}}

\affiliation[warwick-cs]{organization={Department of Computer Science, University of Warwick},%Department and Organization
            %addressline={}, 
            city={Coventry},
            postcode={CV4 7AL}, 
            state={West Midlands},
            country={United Kingdom}}

\affiliation[sxufe-exp]{organization={Experiment and Practical Training Center, Shanxi University of Finance and Economics},%Department and Organization
            %addressline={}, 
            city={Taiyuan},
            postcode={030006}, 
            state={Shanxi},
            country={China}}

\cortext[cor1]{Corresponding author}

\begin{abstract}
Generative AI and coding agents are intensifying a central governance tension in open-source software (OSS): they scale contribution generation faster than maintainers can assess risk, evidence, and accountability. Existing responses strengthen agent-readability and traceability: instruction files help agents work in repositories, while disclosure, behavioral traces, and provenance records make AI participation visible. Governability addresses a further need: project rules must organize contribution-specific risk, evidence, accountability, and review-gate states. We theorize the required organizational arrangement as project-side governability infrastructure, which allocates risk rules, evidence obligations, accountability states, review gates, and decision rights. A diagnostic audit of 50 GitHub repositories finds widespread general OSS governance artifacts, observable agent-readability, and uneven, fragmented AI-governance cues. No audited repository provides a project-wide arrangement that institutionalizes shared rules, preparation obligations, verification rights, and maintainer decision authority across AI-mediated contribution workflows. We develop the Agent Governance Manifest (AGM) to instantiate this governance logic as a repository-hosted boundary resource and bidirectional contract linking contributor-side evidence preparation with maintainer-side verification. In a controlled reviewer-side evaluation with 15 participants and 75 task-level outputs, AGM-supported materials improved exact risk-label recovery (37/38 vs.\ 15/37) and perceived review support (6.14 vs.\ 3.27 on a 1--7 scale). In a contributor-side feasibility check, 15 participants completed 45 tasks; all final packages represented the core governance state correctly, and 41 passed strict structural validation. The study develops a three-layer framework of agent-readability, traceability, and governability, theorizes agent-mediated contributions as governable boundary objects, and advances compliance-enabling digital innovation governance while preserving maintainer decision authority.
\end{abstract}


\end{frontmatter}

\section{Introduction}
\label{sec:introduction}

Open-source software (OSS) has become a central infrastructure for digital innovation. Modern software ecosystems depend on open repositories for frameworks, cloud infrastructure, security components, machine-learning libraries, data-processing systems, and AI application platforms. OSS projects therefore operate as distributed innovation systems~\citep{brunswicker2025openCollaboration,kowalski2026ecosystemSmartification} in which contributors, maintainers, firms, foundations, and users coordinate around shared digital artifacts and open digital knowledge infrastructures~\citep{mager2026openDigitalKnowledgeInfrastructures}. Classic OSS research shows that this openness is sustained by governance arrangements that protect shared resources~\citep{omahony2003guardingCommons}, establish legitimate authority~\citep{omahonyFerraro2007emergenceGovernance}, and design participation structures~\citep{westOmahony2008participationArchitecture}.

Generative AI and coding agents are now entering this innovation infrastructure. LLMs (Large Language Models) increasingly support software-engineering tasks~\citep{fan2023llmSoftwareEngineering,hou2024llmSoftwareEngineeringReview}, and contemporary agentic systems can navigate repositories, inspect files, generate patches, write tests, revise documentation, summarize issues, run commands, and prepare contribution materials~\citep{wang2025openhands,li2025aiTeammatesSE30}. AI now participates in the processes through which OSS contributions are generated, supported with evidence, and submitted. This creates a generation--verification asymmetry: AI can reduce the cost of producing contribution artifacts, but it does not automatically reduce the cost of determining whether those artifacts are correct, safe, maintainable, adequately supported, and accountable.

Recent work documents several parts of this shift. Datasets such as AIDev~\citep{li2026aiddev} observe agentic software development across repository workflows. Agent-readable instruction files, including \texttt{AGENTS.md}~\citep{gloaguen2026agentsMd}, \texttt{CLAUDE.md}, Copilot instructions, and tool-specific repository guidance~\citep{mofidi2026contextEngineeringAIagents,shepard2026repositoryGuidance}, help agents understand how to build, test, and modify projects. Trace and provenance proposals such as Agent Trace~\citep{cursor2026agentTrace} seek to record how agentic contributions are produced. OSS governance research has also examined responses such as prohibition, disclosure, accountability, verification, and tooling rules \citep{yang2026beyondBanningAI}. This literature establishes two governance functions: agent-readability supports agent participation in repository work, while traceability records how AI-mediated contributions are produced. These developments also expose a distinct governability question: how project-level rules allocate risk, evidence obligations, accountability signals, and review gates across contribution workflows.

\begin{figure}[!htbp]
    \centering
    \includegraphics[width=0.82\linewidth]{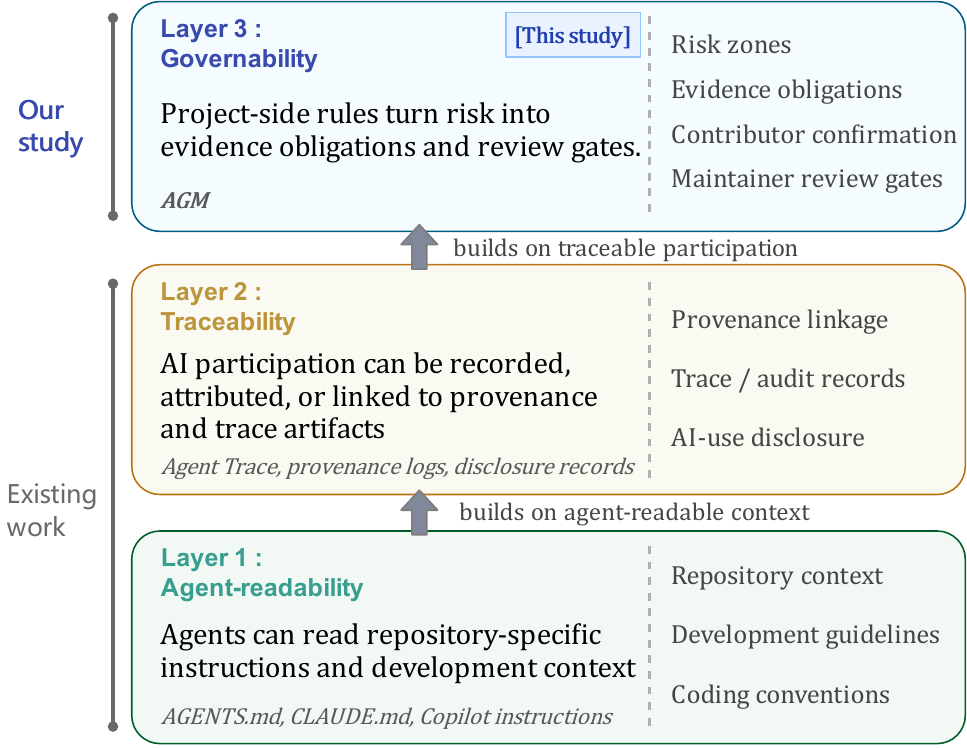}
    \caption{Three-layer framework for AI-mediated OSS governance. Agent-readability enables repository-specific context to be accessed by agents, traceability makes AI participation recordable, and governability carries project-side risk rules into evidence obligations, contributor confirmation, and maintainer review gates.}
    \label{fig:three_layer_governance}
\end{figure}

We conceptualize this gap through an agent-readability--traceability--governability framework for AI-mediated OSS governance, shown in Figure~\ref{fig:three_layer_governance}. Agent-readability makes repositories interpretable to coding agents. Traceability makes AI participation recordable. Governability determines whether project-level rules make AI-mediated contributions reviewable, evidence-oriented, and actionable at review time. Throughout the paper, \textit{AI-mediated OSS governance} denotes the broader governance setting, while \textit{agent-mediated contribution} denotes contribution work substantially prepared, supported with evidence, or submitted with the assistance of coding agents.

Widespread OSS governance artifacts and visible agent-readable instructions provide an important foundation, while review-facing evidence, confirmation, and gate structures remain sparse and unevenly coordinated. For projects that allow AI-mediated contributions under specified conditions, the practical question is whether each contribution is risk-classified, evidence-supported, human-accountable, and ready for review before limited maintainer review capacity is committed to final acceptance decisions. In organizational information-processing terms~\citep{tushmanNadler1978informationProcessing,almarzouq2015informationProcessingOSS}, AI-mediated contribution creates uncertainty when maintainers lack review-relevant information and equivocality when the same contribution can support competing interpretations of risk, responsibility, or readiness~\citep{daftWeick1984interpretationSystems}. The management challenge is to shift structured evidence preparation toward contributors and their tools while preserving maintainer authority over verification and judgment.

To address this challenge, the paper proposes the \textit{Agent Governance Manifest} (AGM), a repository-hosted project governance manifest for AI-mediated OSS collaboration. AGM carries project-specific governance expectations through risk zones, evidence standards, accountability boundaries, agent-use expectations, and maintainer-facing review gates. During contribution review, those expectations are enacted through contribution-specific evidence packages, contributor-confirmation declarations, missing-evidence reports, and gate states. The resulting materials make AI-mediated contributions reviewable under project governance while acceptance decisions remain with maintainers.

This design adopts a service-oriented, compliance-enabling view of OSS AI governance for voluntary communities, where transaction costs and community legitimacy shape participation. AGM provides a clearer, reusable path for responsible contribution by expressing project expectations as agent-readable rules, human-readable explanations, contribution-side evidence obligations, and maintainer-facing review signals. It gives governability a repository-hosted form alongside agent-readability and traceability, carrying project-defined rules across contributor-side preparation and maintainer-side verification.

The study follows a diagnostic--design--evaluation sequence. It first audits 50 public GitHub repositories and then develops AGM as a project-level governance manifest. The artifact is evaluated through two structured assessments: a controlled within-participant reviewer-side evaluation with 15 participants and 75 task-level outputs, and a contributor-side feasibility check in which a separate cohort of 15 participants completed 45 AGM-supported contribution tasks. The audit links formal governance documents with 23,237 pull requests, 19,884 issues, repository metadata, file-level change records, and observable AI-related contribution signals. The evaluation asks whether AGM improves recovery of risk, evidence, accountability, and gate states, and whether contributor-side agents can prepare evidence packages for human confirmation while maintainer review remains a separate authority.

The paper addresses four research questions:

\begin{itemize}
    \item \textbf{RQ1:} To what extent do existing OSS governance artifacts provide the project-level rules and review-stage mechanisms needed to classify, verify, and gate agent-mediated contributions?

    \item \textbf{RQ2:} How can a project-level governance manifest carry project-defined AI contribution rules into contribution-specific evidence obligations and maintainer-side governance gates?

    \item \textbf{RQ3:} Does AGM improve reviewer-side recovery of risk, evidence, accountability, and gate states when project-level governance rules are enacted during contribution review?

    \item \textbf{RQ4:} Can contributor-side agents prepare AGM-based evidence packages in a manageable way while preserving the boundary between contributor confirmation and maintainer review?
\end{itemize}

The paper's central theoretical contribution is the concept of project-side governability infrastructure: the organizational arrangement that makes agent-mediated contributions reviewable under project governance. The argument distinguishes agent-readability, traceability, and governability, documents a project-wide governability gap while separating fragmented cues from coordinated arrangements, develops AGM as a review-facing, repository-hosted governance resource, and evaluates the preparation--verification mechanism on both sides of the contribution workflow. By examining contributor-side preparation and reviewer-side recovery together, the evaluation treats governability as a relationship across the workflow rather than a property of a single tool or artifact. Together, these contributions position AI-mediated OSS governance as compliance-enabling infrastructure for responsible contribution and maintainer judgment.

\section{Literature Review and Theoretical Background}
\label{sec:literature_review}

Studying agent-mediated contributions at the project level separates the repository as a technical host from the project as a governance actor. Repositories host code, issues, pull requests, policies, and automation; projects define rules, roles, responsibilities, community expectations, and contribution legitimacy. This perspective centers how OSS projects organize contribution readiness, accountability, and review obligations.

\subsection{Open-Source Projects as Distributed Governance Systems}
\label{subsec:oss_projects_governance_systems}

OSS projects are distributed socio-technical systems in which contribution rules, role structures, review practices, release procedures, community norms, and technical infrastructures jointly shape how innovation is produced and accepted. Governance emerges through coordinated arrangements that establish legitimate authority~\citep{omahonyFerraro2007emergenceGovernance}, configure multidimensional authority structures~\citep{markus2007ossGovernance}, and make distributed work actionable through coordination processes~\citep{shaikhHenfridsson2017coordinationProcesses}. Sponsored projects additionally design participation architectures that balance external accessibility with project control~\citep{westOmahony2008participationArchitecture}.

Pull-based development~\citep{gousios2014pullBasedModel,alami2022pullRequestGovernance} is a central coordination mechanism: contributors propose changes, maintainers evaluate them, and projects use review workflows to filter, revise, accept, or reject contributions. Over time, this workflow becomes embedded in contribution guidelines, issue and pull-request templates, project-health practices~\citep{linaker2022ossProjectHealth}, and documented role structures or decision rights~\citep{oliveira2026governancePracticeRoles,noori2026founderLeadershipGovernance}. Contribution review therefore serves as both technical quality control and a governance process through which projects allocate responsibility, preserve continuity, and manage limited maintainer review capacity.

Because OSS relies on voluntary and heterogeneous participation, governance must coordinate work without assuming an employment hierarchy. Research on OSS participation shows that governance arrangements shape whether contributors remain involved and move toward more sustained forms of contribution~\citep{shah2006motivationGovernance}. Sustainability and collective-governance research likewise connects project health to institutional arrangements for participation, stewardship, and role transitions~\citep{yin2022ossSustainabilityInstitutionalNetworks,noori2025collectiveGovernanceBaseline}. From an organizational information-processing perspective~\citep{tushmanNadler1978informationProcessing,almarzouq2015informationProcessingOSS}, projects must match their information-processing capacity to uncertainty and interdependence while reducing equivocality when actors interpret the same contribution differently~\citep{daftWeick1984interpretationSystems}. AI-mediated contribution intensifies these demands by increasing the need to establish evidence, risk, responsibility, and readiness before acceptance.

\subsection{AI-Assisted and Agentic Contribution in OSS Workflows}
\label{subsec:agent_mediated_contributions}

LLMs now support code generation, explanation, debugging, testing, review, and documentation~\citep{fan2023llmSoftwareEngineering,hou2024llmSoftwareEngineeringReview}. More recent systems move toward agent-mediated development~\citep{wang2025openhands,li2025aiTeammatesSE30,cruz2025redefiningProgrammer}, where agents interpret repository context, plan changes, modify files, run
commands, and generate patches or pull-request-like outputs. Empirical work on GitHub Copilot~\citep{peng2023githubCopilotProductivity}, Cursor~\citep{he2025cursorSpeedCostQuality}, and collaborative OSS development~\citep{song2026generativeAICollaborativeOSS} shows that GenAI tools can accelerate code production and reshape developer workflows, while
introducing trade-offs among speed, cost, quality, and coordination.

AI-assisted and agentic contribution is now observable in public repository workflows. Large datasets such as AIDev~\citep{li2026aiddev} collect agentic pull requests across repositories and tools, while complementary studies examine agentic coding PRs~\citep{watanabe2026agenticCodingPullRequests}, task-level agent performance~\citep{rahman2026taskLevelEvaluationAgents},
lightly reviewed LLM-agentic changes~\citep{branco2026lgtm}, and maintenance consequences~\citep{sawada2026agentGeneratedCodeMaintenance}. This literature also documents concerns about security-sensitive contributions~\citep{siddiq2026securityAgeAITeammates}, maintenance-heavy agent-generated code~\citep{sawada2026agentGeneratedCodeMaintenance}, and
low-quality AI-generated contribution streams~\citep{baltes2026endlessStreamAISlop,sen2026githubAISlop,iyer2026aiCodingAgentsPrPipeline}. Together, these findings establish the empirical conditions for examining how projects organize review when contribution production expands faster than maintainer verification capacity.

\subsection{GenAI Governance Responses in Open-Source Communities}
\label{subsec:genai_governance_responses}

OSS communities have begun to respond through prohibition, disclosure requirements, human-accountability rules, tool-use guidelines, and evidence-oriented expectations. Recent work maps GenAI governance in OSS communities beyond simple bans, showing that projects are experimenting with different responses to AI-assisted issues, pull requests, reviews, and security reports \citep{yang2026beyondBanningAI}. Policies on AI-use disclosure and human-in-the-loop expectations~\citep{hora2026aiPolicyDisclosureHumanLoop,fedora2025aiAssistedContributionsPolicy} and machine-contributor governance~\citep{manita2026machineContributorGovernance} likewise emphasize responsibility for AI-assisted contributions.

These responses express governance intent, while operational enactment requires project rules that specify required evidence, implicated risks, contributor-confirmation states, and blocking conditions. Disclosure improves visibility; reviewability additionally depends on these contribution- and review-stage requirements. AI attribution~\citep{kraishan2025aiAttributionParadox} can also become socially strategic, as developers may emphasize, hide, or ambiguously frame GenAI involvement depending on perceived norms and criteria~\citep{tufano2026developersGenerativeAI}. Work on executable governance and machine-readable policies suggests a pathway from textual principles to operational rules \citep{datla2025executableGovernanceAI}. For OSS AI governance, the resulting question is whether project policies take a form that contributors and agents can read, that specifies evidence obligations, and that can be enacted consistently during review.

\subsection{Agent-Readable and Traceable Infrastructures}
\label{subsec:agent_readable_traceable_infrastructures}

A parallel stream makes repositories more legible to AI agents and agent activity more traceable to humans. Agent instruction files such as \texttt{AGENTS.md}~\citep{gloaguen2026agentsMd} and related context-engineering practices~\citep{mofidi2026contextEngineeringAIagents,shepard2026repositoryGuidance} provide repository-specific guidance about builds, tests, navigation, coding conventions, and workflows. Traceability and provenance infrastructures, including Agent Trace~\citep{cursor2026agentTrace} and related proposals~\citep{vispute2026reasoningProvenanceAgents,cai2026ghostInTheAgent}, seek to record what agents did, which files or lines were affected, and how AI-mediated work can be attributed or audited.

Repository guidance supports agent navigation, while provenance records support auditability. Governability adds a distinct review-stage function: specifying the evidence a contribution must provide, the project risks it triggers, the human responsibility it carries, and the status of its governance gates. This functional distinction is consistent with boundary-resource views of shared technical artifacts~\citep{mayer2025genAIBoundaryResource} and private-ordering accounts of operationalized governance expectations~\citep{su2026platformGovernancePrivateOrdering}. The analytical focus consequently shifts from agent behavior to the project rules that organize contribution review.

\subsection{From Governance Strategies to Project-Level Governability}
\label{subsec:project_level_governability_gap}

Taken together, the literature establishes the conditions of the governance problem while leaving its project-level organizational mechanism undertheorized. OSS governance operates through configurations of authority, participation structures, and coordination processes~\citep{omahonyFerraro2007emergenceGovernance,markus2007ossGovernance,shaikhHenfridsson2017coordinationProcesses}; coding agents make agent-mediated contribution observable at scale~\citep{li2026aiddev}; communities encode policies and accountability expectations for AI-mediated work~\citep{yang2026beyondBanningAI}; and agent-readable and traceable infrastructures guide agents and record their actions~\citep{gloaguen2026agentsMd,cursor2026agentTrace}. The unresolved organizational question is how project-level governance expectations acquire contribution-level force during preparation and review. Addressing it requires a functional distinction among agent-readability, traceability, and governability; an empirical test of whether repositories host review-facing arrangements; and an account of how compliance-enabling infrastructure can support responsible preparation while preserving maintainer authority.

Three management-theory conversations clarify this organizational-design problem. As coding agents expand contribution-production capacity, maintainers face greater uncertainty and equivocality about risk, evidence, responsibility, and review readiness. Information-processing research explains the need to redistribute information preparation and interpretation across the workflow~\citep{tushmanNadler1978informationProcessing,daftWeick1984interpretationSystems}; boundary-resource research explains how shared artifacts mediate participation and control across heterogeneous actors~\citep{ghazawnehHenfridsson2013boundaryResources,mayer2025genAIBoundaryResource}; and OSS governance research situates those artifacts within project-defined authority and coordination~\citep{westOmahony2008participationArchitecture,shaikhHenfridsson2017coordinationProcesses}. This synthesis motivates project-side governability infrastructure as the arrangement that coordinates project rules across contribution preparation and review through repository-hosted governance resources.

\section{Problem Analysis and Conceptual Framework}
\label{sec:conceptual_framework}

The conceptual framework develops a mechanism-centered account of generation--verification asymmetry and derives the project-side design requirements that motivate AGM\@. It traces how project-defined governance enters contribution-specific review through repository-hosted resources, distinguishes agent-readability, traceability, and governability as different organizational functions, and specifies the evidence and authority relationships that connect contribution preparation with maintainer verification.

\subsection{Generation--Verification Asymmetry as an Information-Processing Problem}
\label{subsec:generation_verification_asymmetry}

AI-mediated contribution refers to OSS contribution activity in which generative AI systems or coding agents participate in producing, modifying, explaining, testing, documenting, or reviewing project artifacts. The contribution may still be submitted by a human account and may be extensively edited by a human contributor. Its governance significance lies in the changed cost structure of contribution production.

Generation--verification asymmetry arises when the marginal cost of producing contribution artifacts falls faster than the marginal cost of assessing their correctness, safety, maintainability, evidentiary support, and accountability. Verification therefore remains a distinct organizational burden even when generation becomes faster. Maintainers must still determine whether a change is sufficiently tested, consistent with project norms, linked to visible human responsibility, and ready for acceptance.

This asymmetry is a socio-technical information-processing problem~\citep{tushmanNadler1978informationProcessing,almarzouq2015informationProcessingOSS}. Uncertainty arises when maintainers lack review-relevant information about production, testing, responsibility, or risk. Equivocality arises when the available artifacts support competing interpretations of those states or of review readiness~\citep{daftWeick1984interpretationSystems}. OSS pull-request governance~\citep{gousios2014pullBasedModel,alami2022pullRequestGovernance} operates under limited maintainer review capacity, distributed trust, project norms, and evidence-based judgment, as also reflected in OSS health and collective-governance research~\citep{linaker2022ossProjectHealth,noori2025collectiveGovernanceBaseline}. Project-side governability infrastructure responds to these information-processing demands by externalizing project-defined governance states before review, shifting maintainer work from open-ended reconstruction toward targeted verification while preserving maintainer decision rights.

\subsection{Project-to-Contribution Governance Enactment}
\label{subsec:project_repository_contribution_layers}

To explain how project-level governance acquires contribution-specific force during review, the framework distinguishes four related concepts at the project, repository, contribution, and cross-actor collaboration levels. Together, they form a project-to-contribution governance enactment chain.

At the project level, \textit{project-side governability infrastructure} is the organizational arrangement that defines and allocates risk classifications, evidence obligations, accountability states, review gates, and decision rights. At the repository level, a \textit{repository-hosted governance resource} carries these rules in a shared, inspectable form accessible to contributors, agents, maintainers, and workflow tools. AGM occupies this artifact level as the repository-hosted resource developed in the study.

At the contribution level, \textit{contribution governability} is the change-level state produced when project rules are enacted for a specific contribution, making its governance state assessable during review. At the cross-actor collaboration level, the evidence-bearing contribution becomes a \textit{governable boundary object} when that state is structured for interpretation across contributors, agents, maintainers, and supporting platforms. AGM therefore functions as a shared boundary resource that supports this contribution-level enactment.

This enactment chain anchors contribution-specific review states in project-defined authority and a shared repository-hosted resource. It builds on OSS research that locates authority in coordination processes~\citep{shaikhHenfridsson2017coordinationProcesses} and on boundary-resource and boundary-object theory, in which shared artifacts mediate external participation and project control~\citep{ghazawnehHenfridsson2013boundaryResources,handler2017ossBoundaryObject,mayer2025genAIBoundaryResource}. It extends those views by explaining how a repository-hosted resource stabilizes project-defined governance meanings across actors and gives those meanings contribution-specific force during review.

\subsection{Readability, Traceability, and Governability}
\label{subsec:readability_traceability_governability}

At the functional level, the framework distinguishes three layers of AI-mediated OSS governance: agent-readability, traceability, and governability, as shown in Figure~\ref{fig:three_layer_governance}. Their division of labor provides the conceptual basis for the design requirements developed below.

\textit{Agent-readability} concerns the ability of coding agents to understand how work should be performed in a repository. Agent-readable artifacts specify commands, conventions, project structure, tests, and workflows, reducing ambiguity on the contributor side of agent-mediated production.

\textit{Traceability} concerns the observable history of AI-mediated work. Provenance mechanisms record how generated artifacts relate to tools, models, conversations, files, line ranges, revisions, or hashes and thereby support ex post audit. Review readiness remains a separate governance judgment.

\textit{Governability} organizes a contribution for review under project-level rules. At review time, it is expressed through four dimensions: risk classifiability, evidence inspectability, accountability visibility, and gate-state recoverability. The three layers therefore address different organizational needs: agent-readability supports contribution production, traceability preserves production history for audit, and governability supplies the normative structure for preparation and review. Agent-readable guidance and traceability records provide contextual inputs to that structure. The distinction is consequential because a repository can be highly readable to agents and richly traceable while still leaving reviewers to reconstruct project-specific obligations. Governability supplies the missing link between information and authority by specifying what evidence is owed, who may verify it, and which gate states follow.

\subsection{Distributed Trust and Evidence-Based Verification}
\label{subsec:identity_trust_evidence_verification}

OSS governance has historically relied on a combination of identity-based, reputation-based, and process-based trust. Maintainers learn to trust contributors through prior interactions, code quality, responsiveness, domain expertise, and adherence to project norms. Projects also use tests, CI workflows, templates, reviews, security policies, project-health practices~\citep{linaker2022ossProjectHealth}, and project-health metric or dashboard infrastructures~\citep{chaoss2026terminology,linuxFoundation2025lfxInsights,ivchenko2026communityHealthMetrics} to reduce reliance on individual trust alone.

AI-mediated contribution changes the relative importance of these trust mechanisms. Contributor identity remains important, but it becomes less informative about the production process of a specific contribution. A contribution submitted by a known contributor may be heavily AI-assisted; a contribution submitted by a new contributor may be generated by an agent but manually reviewed; a contribution may contain a mix of human-written and AI-generated material. In these situations, review depends on both contributor identity and the evidence accompanying the specific change, especially because AI attribution and GenAI disclosure or self-reporting practices~\citep{tufano2026developersGenerativeAI,hora2026aiPolicyDisclosureHumanLoop} may be incomplete, strategic, or difficult to interpret.

Contributor reputation remains important, but it cannot fully stand in for evidence about a specific change. Maintainers need contribution-level evidence that allows them to assess scope, validation, provenance, known limitations, and human responsibility without reconstructing the contributor's entire generative process. The governance shift is to attach inspectable claims to the contribution itself, so review can proceed on an evidentiary basis even when attribution is incomplete or ambiguous.

Evidence obligations should vary with verification risk. Routine changes may require a lightweight account, whereas security-sensitive or high-consequence changes require stronger validation and explicit human confirmation. Proportionality avoids both under-governing consequential changes and overburdening low-risk participation. Structured evidence makes these differentiated obligations inspectable at review time and provides a common basis for judging whether a change is ready for technical review. Placing evidence preparation upstream also redistributes work across the workflow: contributors and their agents prepare the facts that make a change assessable, while maintainers retain the task of challenging those facts and exercising judgment.

\subsection{Bidirectional Governance Contracts}
\label{subsec:bidirectional_governance_contracts}

Project-level AI contribution governance connects contributor-side preparation with maintainer-side verification through a bidirectional contract. OSS projects exercise authority through community-legitimated and configurational arrangements rather than a conventional employment hierarchy~\citep{omahonyFerraro2007emergenceGovernance,shaikhHenfridsson2017coordinationProcesses}. Informed by this governance setting and executable-governance thinking~\citep{datla2025executableGovernanceAI}, the contract is a shared, machine- and human-readable specification that reciprocally allocates preparation obligations, verification rights, review-support functions, and final decision rights when a contribution is submitted.

On the contributor side, the contract defines risk-sensitive evidence obligations whose strength rises with verification risk. Higher-risk changes may also require a contributor-confirmation declaration: a machine-readable accountability state indicating that a human contributor has inspected and accepted responsibility for the submitted evidence and change scope.

On the maintainer side, the same contract defines review gates. Maintainers, possibly assisted by maintainer-side review-support agents, can inspect the evidence package, identify missing or placeholder material, interpret risk and contributor-confirmation states, and decide whether the contribution can proceed to technical review or final acceptance consideration. The contract therefore moves part of the review-preparation work upstream while preserving final decision authority for maintainers.

This bidirectional structure gives project rules reciprocal force across preparation and review. Contributor-side actors bear preparation obligations, while maintainers retain verification and decision authority. Disclosure policies, machine-contributor rules, submission templates, agent instructions, and provenance logs remain relevant inputs~\citep{yang2026beyondBanningAI,fedora2025aiAssistedContributionsPolicy,manita2026machineContributorGovernance}, while AGM provides the canonical governance resource through which a project specifies how evidence, confirmation, and routing requirements apply before a contribution becomes review-ready. By clarifying what contributors and their tools should prepare and making unresolved obligations visible, the contract enables compliance while preserving project authority in a voluntary participation setting~\citep{shah2006motivationGovernance,westOmahony2008participationArchitecture}.

\subsection{Design Requirements for Executable AI Governance}
\label{subsec:design_requirements_executable_governance}

Building on the preceding analysis of generation--verification asymmetry, project-to-contribution governance enactment, and the functional distinction among agent-readability, traceability, and governability, we derive five design requirements for executable AI governance in OSS projects, summarized in Table~\ref{tab:design_requirements_executable_governance}. These requirements synthesize OSS governance research and information-processing theory~\citep{almarzouq2015informationProcessingOSS}, GenAI policy responses~\citep{yang2026beyondBanningAI}, agent-readable repository guidance~\citep{gloaguen2026agentsMd}, provenance infrastructure~\citep{cursor2026agentTrace}, and executable-governance thinking~\citep{datla2025executableGovernanceAI}.

\begin{table}[!ht]
\centering
\caption{Design requirements and AGM implications for project-level executable AI governance}
\label{tab:design_requirements_executable_governance}
\footnotesize
\setlength{\tabcolsep}{3pt}
\renewcommand{\arraystretch}{1.05}
\begin{tabularx}{\textwidth}{P{0.24\linewidth} Y Y}
\hline
\textbf{Requirement} & \textbf{Governance rationale} & \textbf{Implication for AGM} \\
\hline
Human- and machine-readability & Governance rules must be interpretable by contributors, maintainers, coding agents, and maintainer-side review-support agents without relying only on ambiguous prose or opaque schemas. & Use structured manifest fields, controlled values, and human-readable explanations. \\
Risk sensitivity & Verification burden differs across project areas, change types, downstream risks, and maintainer review costs. & Map files, modules, or project areas to risk zones and graduated evidence obligations. \\
Structured evidence & AI-use disclosure and provenance can indicate participation or origin, but they do not by themselves establish review readiness. & Require contribution-specific evidence packages, test summaries, provenance summaries, linked context, and missing-evidence reports. \\
Bidirectionality & Governance must connect contributor-side preparation with maintainer-side verification through the same project-defined rule set. & Use the manifest as a shared governance contract for generating contributor obligations and maintainer-facing review packets. \\
Human decision authority & Automation should organize evidence, surface missing materials, and highlight gate states without replacing maintainer judgment. & Separate review-support outputs, contributor-confirmation declarations, governance-gate states, and final maintainer decision rights. \\
\hline
\end{tabularx}
\end{table}

These requirements define governability through both informational and relational components. Human- and machine-readability, risk sensitivity, and structured evidence provide the informational basis for review-ready contributions. Bidirectionality and human decision authority define the relationship between contributor-side preparation and maintainer-side verification: the same project rules structure evidence production and review support, while final judgment remains with maintainers.

These requirements position formalization as enabling infrastructure. Enabling formalization treats procedures as organizational technologies that help participants understand, perform, and repair their work~\citep{adler1996twoTypesBureaucracy}; empirical research further shows that experience-based development, experimentation, and transparency can make formal systems more usable by the people whose work they organize~\citep{woutersWilderom2008enablingFormalization}. Applied to voluntary OSS collaboration~\citep{shah2006motivationGovernance}, AGM makes project expectations, evidence gaps, repair points, and review states visible, giving willing contributors a clearer and lower-friction path to responsible contribution while helping maintainers diagnose unresolved obligations. This design aligns formal rules with the practical conditions of voluntary participation while retaining proportional controls and final maintainer authority.

\section{Design of the Agent Governance Manifest}
\label{sec:agent_governance_manifest}

The \textit{Agent Governance Manifest} (AGM) instantiates the design requirements from Section~\ref{sec:conceptual_framework} as a repository-hosted governance resource for project-side governability infrastructure. It gives the project-level arrangement a stable, repository-visible carrier and makes one governance specification available across contribution preparation and maintainer review. The resource stabilizes the governance relationships defined in the conceptual framework while preserving human maintainer authority.

The reference prototype represents this specification in both a human-readable Markdown document and a structured YAML file. It is accompanied by a community-draft specification. Other representation formats can serve the same design so long as the governance rules remain consistently readable by human contributors, coding agents, maintainer-side review-support agents, and maintainers. AGM can also incorporate external provenance signals~\citep{cursor2026agentTrace,vispute2026reasoningProvenanceAgents,cai2026ghostInTheAgent}, such as file- or line-level AI attribution records, as contextual inputs to contribution-level governance states. The public reference prototype and community-draft specification are summarized in Appendix~\ref{app:artifact_availability}; the main text focuses on the design logic, operational workflow, core mechanisms, and boundaries of the artifact.

\subsection{Design Logic and Governance Principles}
\label{subsec:agm_design_logic}

Five governance principles carry the design requirements in Table~\ref{tab:design_requirements_executable_governance} into artifact-level commitments. Several principles may support one requirement, and each principle may advance more than one requirement.

The cross-cutting design logic is a service-oriented approach to repository-hosted governance resources. OSS participation architectures must make contribution accessible without dissolving project control~\citep{westOmahony2008participationArchitecture}. AGM addresses this tension by making project expectations explicit, discoverable, followable, and verifiable: it separates lightweight discovery pointers from canonical governance rules, links risk zones to proportional evidence obligations, and presents review packets as diagnostic aids for maintainers. This design supports explicit, reviewable contribution preparation while preserving maintainer authority and proportional participation costs.

\textit{Evidence-based governance} requires claims about review readiness to be supported by inspectable contribution-specific evidence rather than disclosure alone. The central question is whether claims can be examined and challenged; AI-use disclosure alone provides only contextual visibility. AGM organizes evidence around review-relevant claims and visible human responsibility. These independently checkable artifacts extend human-in-the-loop policy expectations~\citep{hora2026aiPolicyDisclosureHumanLoop,fedora2025aiAssistedContributionsPolicy} into contribution-level preparation and reduce the attribution ambiguity highlighted in OSS GenAI use research~\citep{kraishan2025aiAttributionParadox}.

\textit{Risk-zoned governance} links the evidence burden to verification risk. Uniform requirements would either under-govern consequential changes or overburden routine participation; AGM therefore allows projects to calibrate evidence and confirmation requirements to the areas a contribution affects. Calibration is itself a governance decision because it determines where project scrutiny intensifies and where participation remains lightweight. This principle responds to security concerns around agentic PRs~\citep{siddiq2026securityAgeAITeammates} and to broader evidence-gating strategies in AI governance~\citep{yang2026beyondBanningAI,manita2026machineContributorGovernance}.

The \textit{bidirectional governance contract} gives artifact-level form to the allocation developed in Section~\ref{subsec:bidirectional_governance_contracts}. The same specification must be useful on both sides of the workflow: it guides evidence preparation and provides maintainers with a basis for verification, drawing on executable-governance thinking~\citep{datla2025executableGovernanceAI}. By making that allocation legible across human and machine actors, AGM serves as a shared boundary resource~\citep{handler2017ossBoundaryObject,mayer2025genAIBoundaryResource} that aligns contributor-side preparation with maintainer-side verification.

\textit{Evidence-centered disclosure and process privacy} establish a selective transparency boundary. AGM requires independently checkable review evidence while retaining private reasoning within the contributor-side process. Selective transparency keeps formalization focused on what review requires and makes review independence compatible with contributor autonomy. This boundary addresses the social and strategic ambiguity of disclosure and attribution~\citep{kraishan2025aiAttributionParadox,tufano2026developersGenerativeAI}.

\textit{Review independence and human responsibility} position automated systems as organizers of evidence rather than decision makers. Review-support agents can identify risk, missing evidence, and unresolved gates under human oversight; final authority over merge, revision, rejection, or escalation remains with human maintainers. This division of labor sharpens and better directs maintainer judgment while responding to evidence on lightly reviewed agentic PRs~\citep{branco2026lgtm}, maintenance consequences of agent-generated code~\citep{sawada2026agentGeneratedCodeMaintenance}, and task-level limits in agent evaluation~\citep{rahman2026taskLevelEvaluationAgents}.

Together, these principles define how a service-oriented governance resource should behave when project rules enter contribution and review workflows. AGM makes the conditions for review readiness explicit and consistently inspectable, giving participants a clearer path to responsible contribution under project authority.

\subsection{Operational Workflow of the Manifest}
\label{subsec:agm_operational_workflow}

\begin{figure}[t]
\centering
\includegraphics[width=1.0\linewidth]{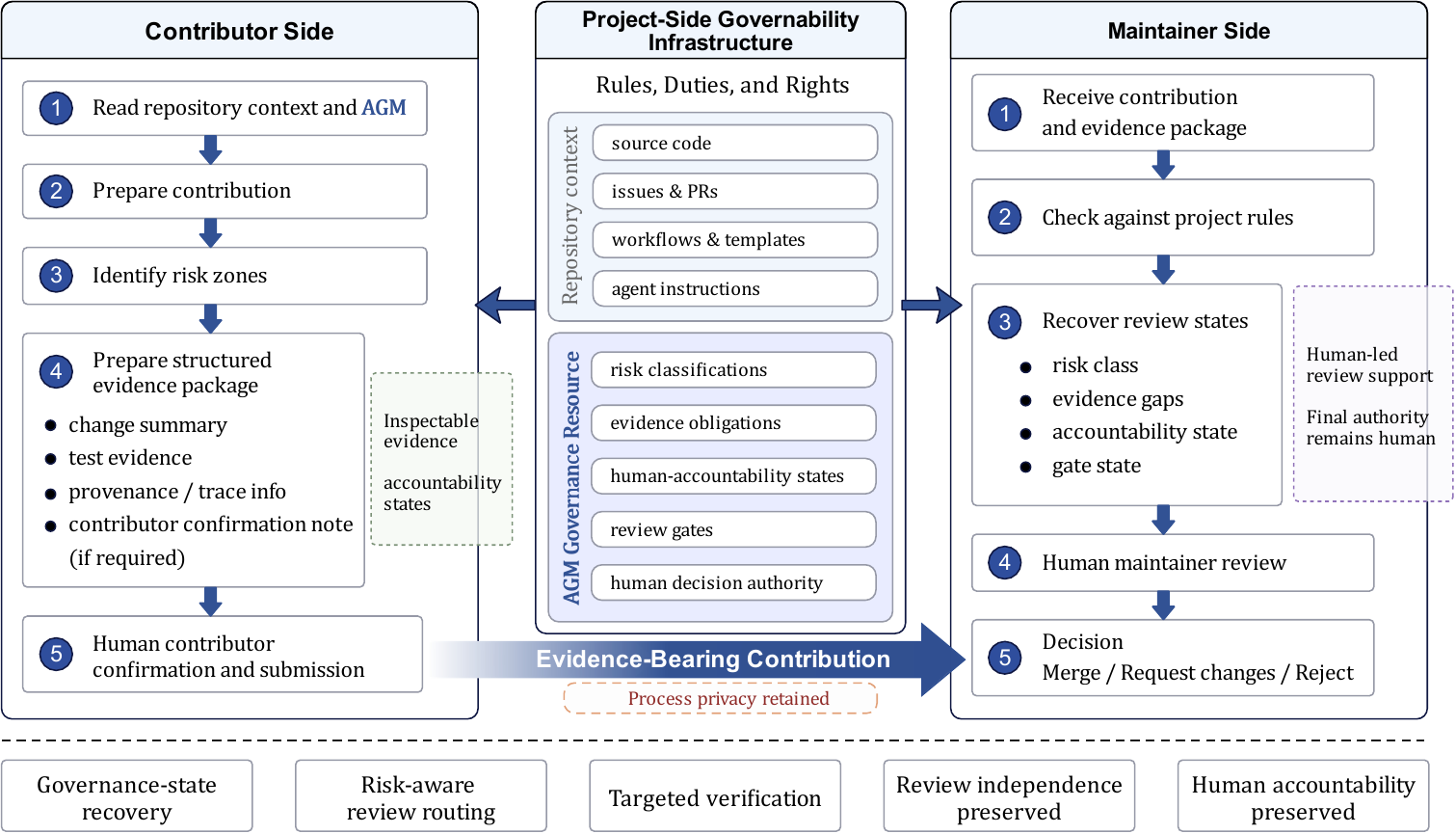}
\caption{AGM workflow across contribution preparation and maintainer review. Project-side governability infrastructure defines shared rules, duties, and rights, which the repository-hosted AGM governance resource carries into the workflow. The resulting evidence-bearing contribution supports governance-state recovery during review, while human accountability and final maintainer decision authority are preserved.}
\label{fig:agm_workflow}
\end{figure}

Figure~\ref{fig:agm_workflow} shows how AGM links contributor-side preparation with maintainer-side verification. The workflow carries project-defined rules, duties, and rights from project-side governability infrastructure into contribution-specific evidence and review states.

The workflow begins from a common project specification carried by AGM\@. Repository context supplies technical and procedural information, while the governance resource states the evidence, confirmation, and review conditions that apply to a change. The same rules consequently shape what contributors prepare and what maintainers verify, giving the workflow continuity across submission and review. Contributors, coding agents, maintainer-side review-support agents, and maintainers therefore work from the same project-defined expectations, while the repository's technical, submission, validation, and traceability interfaces retain their distinct roles.

On the contributor side, evidence preparation moves into the contribution phase so that review-relevant information is structured before submission. The contributor and coding agent read the repository context and AGM, prepare the change, and identify the risk zones affected by the modified files or components. Those classifications determine the applicable evidence obligations. A contributor-side agent can help draft the evidence package by listing affected files, summarizing the change, recording tests or checks, describing available provenance, flagging missing evidence, and preparing contributor-confirmation fields. The human contributor reviews and corrects this material, adds context where needed, and confirms the package when required. The evidence-bearing contribution then carries the change scope, risk classification, test results, missing-evidence status, observable AI involvement, and contributor-confirmation state into review.

On the maintainer side, AGM supports targeted verification under human authority. A maintainer, possibly assisted by a review-support agent, compares the submitted change and evidence package with the project rules carried by AGM\@. This comparison recovers the contribution's review state: the applicable risk class, the sufficiency and gaps in the evidence, the contributor-confirmation state, and the status of governance gates. Review-support artifacts can summarize these states and identify areas requiring attention. They organize review-relevant information for human maintainer judgment, which remains the basis for merge, revision, rejection, or escalation.

Agent roles remain differentiated. Contributor-side agents help prepare evidence for downstream review; maintainer-side agents organize completeness checks and review states under human maintainer authority. The same agentic infrastructure that accelerates contribution production~\citep{he2025cursorSpeedCostQuality,peng2023githubCopilotProductivity,song2026generativeAICollaborativeOSS} can therefore also support the preparation and organization of review-relevant information. The shared governance resource coordinates contributor preparation, human confirmation, maintainer-side verification, and final maintainer judgment while keeping these responsibilities distinct.

The workflow preserves a deliberate asymmetry in what crosses the contribution--review boundary: structured, verifiable evidence is transferred, while private reasoning and exploratory processes remain on the contributor side. This evidence boundary reduces uncertainty for maintainers~\citep{daft1986mediaRichness}. It avoids making prompts or reasoning traces a condition of review and responds to the ambiguity of AI attribution and disclosure~\citep{kraishan2025aiAttributionParadox}. The resulting separation preserves contributor process privacy and maintainer review independence.

\subsection{Canonical Governance Resource and Repository Interfaces}
\label{subsec:why_separate_agm}

The reference prototype implements AGM as a dedicated repository-hosted governance resource that gives canonical form to project-defined risk classifications, evidence obligations, contributor-confirmation states, and maintainer-facing review gates. The underlying design requirement is a canonical, repository-visible governance configuration that contributors, agents, maintainers, and workflow tools can inspect consistently.

Canonicality is an organizational property of this configuration: the relevant actors and interfaces identify the same rule source and the same relationships across preparation and review. Projects may realize that property in a dedicated resource or through explicitly linked resources, provided that those relationships remain clear. Explicit links preserve canonicality even when the carrier is distributed across more than one repository artifact. \texttt{AGENTS.md}, \texttt{CONTRIBUTING.md}, PR templates, CI workflows, and provenance systems can point to or invoke the governance rules while retaining their respective guidance, submission, validation, and traceability functions.

\subsection{Core Governance Mechanisms}
\label{subsec:agm_core_mechanisms}

AGM instantiates the governance design through four core mechanisms: risk-zoned governance, contributor-side evidence packaging, maintainer-side review support, and human-responsibility gating. Together, they give operational form to the bidirectional contract: risk zoning and evidence packaging organize preparation, while review support and responsibility gates structure verification and authority.

\textit{Risk-zoned governance} maps repository areas, file patterns, or functional components to risk levels. The specific risk zones are repository-dependent. In the prototype, documentation-only changes are treated as low risk, tests and non-critical configuration changes as medium risk, core application logic as high risk, and authentication logic, dependency files, workflow definitions, or security-sensitive configuration as critical risk. Each repository defines its own risk zones according to its technical architecture, security sensitivity, review capacity, and downstream dependency structure. Risk classification must be explicit and machine-readable~\citep{datla2025executableGovernanceAI} so that contributor-side and maintainer-side agents can interpret it consistently~\citep{mofidi2026contextEngineeringAIagents}.

\textit{Contributor-side evidence packaging} carries risk classification into submitted review material. Depending on the risk level, it may contain a contribution report, test report, trace manifest, linked-issue declaration, missing-evidence statement, known limitation statement, and contributor-confirmation declaration. This package brings review-relevant context alongside the diff and requires substantive evidence content in place of empty templates or placeholders. It responds to wider concerns about low-quality AI-generated contribution streams~\citep{baltes2026endlessStreamAISlop,sen2026githubAISlop} and the need for validation in AI-agent PR workflows~\citep{iyer2026aiCodingAgentsPrPipeline}.

\textit{Maintainer-side review support} organizes submitted evidence into structured review artifacts. A risk summary identifies affected risk zones and high-risk areas. A missing-evidence report identifies absent files, missing fields, or incomplete placeholder content. A test-evidence summary organizes reported test commands, results, and relevance. A provenance summary records observable AI assistance and contribution-process metadata while leaving prompts and private reasoning within the contributor-side process. A review checklist highlights areas that require maintainer attention. These artifacts organize the information needed for human review.

\textit{Human-responsibility gating} allows projects to require explicit contributor confirmation for high-risk or critical changes before submission or merge. Coding agents can generate or modify artifacts, but responsibility for submitted changes remains with accountable human actors. These gates are especially important for security-sensitive agentic PRs~\citep{siddiq2026securityAgeAITeammates} and for changes affecting authentication logic, dependencies, workflows, security-sensitive configuration, or other areas where emerging OSS AI contribution policies emphasize human accountability~\citep{fedora2025aiAssistedContributionsPolicy,hora2026aiPolicyDisclosureHumanLoop}.

Together, these mechanisms carry project-level governance rules into operational contribution and review routines. AGM calibrates evidence requirements to verification risk, applying stronger requirements to higher-risk changes while keeping low-risk pathways comparatively lightweight. This proportionality is important for OSS communities because excessive governance burden could discourage participation, especially in small or volunteer-driven projects.

\subsection{Governance Boundaries}
\label{subsec:agm_governance_boundaries}

AGM has a defined scope, and these boundaries are part of the design logic. Its role is to lower the cost of compliant behavior, make project expectations explicit, and help maintainers detect missing, inconsistent, placeholder, or incomplete evidence before final review. Projects seeking stronger guarantees against ignored rules, omitted evidence, or fabricated declarations require additional identity, signing, audit, attestation, cryptographic, or platform-level mechanisms.

Attribution and provenance remain partial contextual inputs~\citep{kraishan2025aiAttributionParadox,cursor2026agentTrace}. Public repository records capture explicit AI-related provenance only when contributors or tools preserve such markers; private use, removed attribution, manual editing, and submission through ordinary accounts remain difficult to observe. AGM therefore centers evidence obligations that contributors and their tools can prepare, disclose, and make available for review. The design governs the contribution presented for review even when perfect detection of AI use is unavailable.

Disclosure is evidence-centered and preserves contributor process privacy. Required materials concern the submitted change, its validation, known limitations, provenance, and human responsibility; original prompts, detailed intermediate reasoning, and private exploratory attempts remain within the contributor-side process. This boundary protects contributor autonomy and maintainer-side review independence in a setting where developers use and disclose GenAI in varied ways~\citep{tufano2026developersGenerativeAI}. It keeps the evidentiary burden aligned with review needs and the contributor's broader work process private.

Automated systems remain in a review-support role. They may organize submitted evidence, identify missing material, and highlight unresolved risk or gate states, but they do not make final review decisions. Their outputs remain contestable inputs to human judgment. Final authority remains with human maintainers, consistent with human-accountability principles in emerging OSS AI contribution policies~\citep{hora2026aiPolicyDisclosureHumanLoop,fedora2025aiAssistedContributionsPolicy}.

Configuration remains project-specific. Projects define risk zones, evidence thresholds, and responsibility gates according to their technical architecture, security exposure, and review capacity. This configurability lets a documentation-focused project apply lighter requirements than security-sensitive infrastructure while preserving the same governance logic. The design is therefore standardized in function while remaining adaptable in local thresholds and controls.

These boundaries position AGM as an evidence-oriented repository-hosted governance resource. It supports reviewability, accountability, and risk awareness under AI-mediated contribution while preserving human judgment, contributor autonomy, and proportionality, and it operates alongside existing OSS contribution norms, maintainer authority, and community-based review practices. Projects requiring stronger guarantees may deploy signed commits, CI enforcement, artifact attestation, provenance records, branch-protection rules, or platform-level policy gates as complementary controls outside the core scope of the current community-draft specification and reference prototype.

\section{Research Design and Evaluation Strategy}
\label{sec:research_design}

\subsection{Research Design Overview}
\label{subsec:research_design_overview}

This study adopts a design-science-informed mixed-method strategy organized around a diagnostic--design--evaluation sequence \citep{romme2023fromTheoriesToTools}. It diagnoses whether existing governance artifacts make AI-mediated contributions operationally reviewable, constructs a governance artifact around the observed gap, and evaluates whether that artifact improves reviewer-side governance judgment. Table~\ref{tab:research_design_overview} summarizes the three stages of the research design.

\begin{table}[!htbp]
\centering
\caption{Overview of the diagnostic--design--evaluation research design}
\label{tab:research_design_overview}
\footnotesize
\setlength{\tabcolsep}{2.5pt}
\renewcommand{\arraystretch}{1.02}
\begin{tabularx}{\textwidth}{P{0.18\linewidth} P{0.27\linewidth} Y Y}
\hline
\textbf{Stage} & \textbf{Purpose} & \textbf{Data or artifact} & \textbf{Main output} \\
\hline
Stage 1: Diagnostic governance audit
& Diagnose whether existing OSS governance artifacts make AI-mediated contributions operationally reviewable.
& Public GitHub governance documents, PRs, issues, workflows, repository metadata, and observable AI-related PR and issue signals.
& General-governance coverage measures, agent-readability and AI-governance cue measures, PR/issue signal measures, and governability-gap findings. \\

Stage 2: Artifact construction
& Construct AGM as a repository-hosted governance resource.
& AGM prototype repository, manifest, risk-zone rules, validation scripts, evidence templates, and review templates.
& Governance artifact specifying risk zones, evidence obligations, contributor-confirmation states, and maintainer-facing review gates. \\

Stage 3: Artifact evaluation
& Evaluate whether AGM improves reviewer-side governance-state recovery and whether contributor-side evidence preparation can be completed in practice.
& Controlled review tasks, ordinary and AGM-supported materials, contributor-side evidence-package tasks, participant questionnaires, and coded outputs.
& Objective review-output coding, contributor-side validation results, subjective usefulness and feasibility ratings, and participant feedback. \\
\hline
\end{tabularx}
\end{table}

The diagnostic--design--evaluation sequence gives each empirical component a distinct inferential role. Stage 1 provides the empirical problem grounding for AGM by examining whether public OSS governance artifacts make agent-mediated PRs operationally reviewable. It complements prior qualitative studies of OSS GenAI governance strategy and large-scale Agentic-PR datasets such as AIDev~\citep{li2026aiddev} by shifting attention from AI-related signals to contribution governability at review time.

The next two stages carry this diagnosis into artifact construction and evaluation. Stage 2 constructs AGM as the repository-hosted resource for project-side governability infrastructure, while Stage 3 evaluates two linked mechanisms: whether AGM improves reviewer-side recovery of governance-relevant states during review, and whether contributor-side evidence packages can be prepared and human-confirmed under pre-specified validation criteria. This staged design links empirical diagnosis, artifact construction, and controlled evaluation while keeping the audit, prototype, and evaluation at their appropriate levels of inference.

\subsection{Repository Sample and Data Collection}
\label{subsec:repository_sample_data_collection}

The repository-mining analysis uses a purposive sample of 50 public GitHub repositories selected to capture theoretically relevant variation in risk level, AI relevance, governance actor, and ecosystem role. The sample includes AI-native and AI-adjacent projects, agent-tool projects, infrastructure projects, security-sensitive repositories, enterprise-led repositories, foundation-led repositories, community-led repositories, and comparison projects. The sampling logic is diagnostic and design-oriented: it examines whether repositories likely to face AI-mediated contribution pressure provide infrastructure capable of supporting contribution governability at review time. Supplementary Material S1 reports the repository list, classification criteria, and sample distribution.

All repository artifacts and contribution records were bounded by a snapshot cutoff of June 14, 2026. The final API collection run was completed on June 15, 2026, and the resulting dataset was fixed to the June 14 observation boundary before analysis. The data pipeline collected repository metadata, governance documents, pull-request summaries, issue summaries, PR file-level change records, and repository-month panel observations. The processed dataset contains 50 repositories, 23,237 pull requests, 19,884 issues, 157,175 PR file-level change records, 1,445 governance documents, and 1,703 repository-month observations.

Repository-level governance analysis retained all 50 repositories. PR-level analysis covered 48 repositories, excluding \texttt{postgres/postgres} and \texttt{torvalds/linux} because their GitHub mirrors do not represent the projects' primary contribution workflows. Issue-level analysis covered 46 repositories, additionally excluding \texttt{django/django} and \texttt{encode/httpx} because GitHub Issues were unavailable at the observation cutoff. Supplementary Material S1 documents these analytical coverage differences.

The operational diagnosis draws on three complementary forms of evidence. Governance documents support coding of general OSS governance coverage, agent-readability, fragmented AI-governance cues, and project-wide governability arrangements. PR and issue data reveal AI-domain language, automation signals, and AI-contribution provenance signals in enacted contribution streams. PR-level process measures such as changed files, additions, review comments, and review counts provide descriptive context for review interaction and contribution complexity; they are not used to estimate causal effects of AI use on maintainer workload.

\subsection{Governance Coding and Measures}
\label{subsec:governance_coding_measures}

The project-level coding first distinguishes general OSS governance coverage from AI-specific governance cues and agent-readable artifacts. General governance coverage captures the publicly visible presence of ordinary OSS mechanisms such as testing requirements, CI workflows, security evidence, issue templates, PR templates, reproduction requirements, code ownership, and license or intellectual-property checks.

The initial AI-governance coding used two legacy binary variables: broad AI governance readiness and stricter AI evidence/execution mechanisms. A targeted second-coder check tested their construct boundaries. A second coder independently recoded a stratified subset of 15 repositories using the same raw evidence corpus. The checked subset showed full agreement on objective general-governance and agent-readable artifact indicators, including CI workflows, testing requirements, security policies, PR templates, \texttt{CODEOWNERS}, \texttt{AGENTS.md}, \texttt{CLAUDE.md}, and Copilot instructions.

Disagreements concentrated on the boundary between fragmented AI-governance cues and project-wide governability arrangements. The second coder often identified AI-use disclosure, accountability, or local workflow cues, whereas the original coding emphasized project-level allocation and coordination of review-facing governance functions. All 68 repository-level disagreements were adjudicated against the frozen evidence corpus, leaving no unresolved cases. The adjudication led us to replace the legacy binary AI-specific governance variable with the layered distinction among agent-readability, fragmented AI-governance cues, and project-wide governability arrangements reported in the findings.

The disagreement pattern clarified the construct boundary and prompted targeted recoding of all 50 repositories into three diagnostic strata. \textit{Agent-readability} artifacts tell agents how to work. \textit{Fragmented AI-governance cues} capture AI-use disclosure, accountability, evidence, risk, review-support, or local workflow-enforcement cues. A \textit{project-wide governability arrangement} is a canonical, repository-visible arrangement that specifies and coordinates risk classification, evidence obligations, human-accountability states, and maintainer-facing review gates across contribution and review workflows.

For reporting, agreement and kappa statistics retain the two coders' independent pre-adjudication labels, while the resolved codes support the final repository-level analytical dataset. Supplementary Material S2 reports the legacy-variable agreement and kappa values, the full codebook, the adjudication logic, and the targeted recoding outputs.

Pull-request-level AI-related signals are coded separately. AI-domain keywords capture topical references to AI, which may appear in AI-native projects or ordinary discussions about models and tools. Automation-bot signals capture ordinary automation that should not be conflated with AI-mediated contribution. Broad and conservative AI provenance signals capture explicit public markers that a PR was AI-assisted, agent-generated, or associated with coding agents. Public repository data do not reveal hidden or unreported AI use, so these provenance signals provide conservative lower-bound indicators.

\subsection{Artifact Construction and Evaluation Design}
\label{subsec:artifact_construction_evaluation_design}

The artifact was developed through staged prototyping and evaluation-driven refinement. End-to-end trials, concise-invocation tests, validator checks, controlled reviews of task materials, and internal contributor--maintainer trials progressively clarified the bidirectional workflow, repository-hosted process knowledge, evidence completeness, evidence-to-file mappings, canonical rule sources, and human-review states. Supplementary Material S4 records this design provenance in a concise problem--refinement--check table.

The resulting controlled prototype repository contains the AGM manifest, risk-zone definitions, evidence templates, validation scripts, sample tasks and evidence packages, missing-evidence reports, contributor-confirmation declarations, and maintainer-side review packets. Together, these components provide a bounded environment for testing governance feasibility under controlled conditions.

The evaluation examined AGM's bidirectional governance workflow through two complementary components: reviewer-side recovery of project-defined governance states and contributor-side preparation of those states. The reviewer-side study served as the primary controlled mechanism evaluation, testing whether governance states externalized before review could be recovered in task-level reviewer-side outputs. The contributor-side feasibility check assessed whether contributors and their agents could prepare the same states while preserving maintainer-side authority boundaries.

The reviewer-side evaluation used five task types that vary in governance risk: a low-risk documentation change, a medium-risk test/configuration change, a high-risk core logic change, a critical authentication-related change, and a critical workflow/evidence-placeholder scenario. Each task had two material conditions: ordinary materials and AGM-supported materials. Ordinary materials included a task description, diff, PR description, test output, and boundary notes. AGM-supported materials additionally included risk summaries, evidence indexes, contribution reports, test evidence summaries, provenance summaries, contributor-confirmation declarations where applicable, missing-evidence reports, and AGM review packets. The comparison tested whether externalizing project-defined risk, evidence, accountability, and gate states before review made those states more recoverable in task-level reviewer-side outputs.

The reviewer-side evaluation used a pre-specified within-participant design. Across participants, two condition patterns alternated: AGM-supported materials were assigned to T1, T3, and T5 and ordinary materials to T2 and T4, or vice versa. Task order was cyclically rotated so that each task occupied every sequence position.

Fifteen participants formed a heterogeneous technical pool spanning doctoral students, commercial software professionals, teachers or researchers, undergraduate and master's students, an independent developer, and one OSS project organization member. Their programming experience averaged 8.2 years (median = 8, range = 3--16). Nine reported high or very high Git/GitHub experience, eight reported at least moderate code-review experience, six had at least occasional OSS contribution experience, and all were frequent or intensive users of AI coding tools. This profile supports controlled mechanism-level evaluation of contribution governability at review time.

Together, the participants completed 75 task runs through agent-assisted reviewer configurations comprising the participant and their selected reviewer-side agent environment. The task-level reviewer-side output was the objective unit of analysis; participants separately provided human ratings and open-ended feedback. Outputs were coded for exact risk-label recovery, risk downgrading, contributor-confirmation status, governance-gate status, missing/placeholder evidence recognition, technical-review readiness, and final-acceptance boundary recognition. To assess output-coding reliability, a second coder independently coded a stratified subset of 20 outputs using the frozen objective rubric. Agreement was calculated from the two coders' pre-adjudication labels. Nine pre-adjudication disagreements are retained as part of the independent reliability record, with detailed results reported in Supplementary Material S6.

Because reviewer-side recovery depends on whether governance states can be prepared upstream, the contributor-side component examined the preparation side of the workflow. It used a non-overlapping cohort of 15 participants, eliminating participant-level carryover of reviewer-side experience into contributor-side evidence preparation. Each participant completed three AGM-supported contribution tasks---a low-risk documentation task, a high-risk parser-validation task, and a critical workflow-sensitive CI task---with the three task orders cyclically rotated across participants.

For each task, the participant's contributor-side agent was asked to produce a draft evidence package, pause for human contributor confirmation, and then produce a final evidence package. The validation protocol preserved role separation: contributors could confirm their own evidence package, while maintainer review could not be marked as completed on the contributor side. The resulting 45 draft and 45 final packages were checked with pre-specified validation criteria for structural completeness, governance-state consistency, and role-boundary compliance that distinguish strict structural validity from correctness of the core governance state. Participants also completed a compact post-experiment questionnaire covering perceived burden, requirement clarity, agent support, manageability, acceptability, and open-ended feedback. One participant's initial run was excluded because repeated agent-session interruptions affected task execution; the participant repeated the tasks under the same protocol, and the rerun outputs were used for final coding.

Across the two components, the evaluation operationalizes \textit{contribution governability} as a review-facing state formed across contribution preparation and review. Exact risk-label recovery captures risk classifiability; missing-evidence recognition captures evidence inspectability; contributor-confirmation and maintainer-review status capture accountability visibility and authority boundaries; gate-status judgment captures gate-state recoverability; and perceived review support captures whether participants experienced the materials as useful for review. Contributor-side structural completeness, governance-state consistency, and role-boundary compliance assess whether the required states can be prepared upstream without displacing maintainer authority. Together, these indicators test whether AGM externalizes contribution governance states in a form that supports contributor-side preparation, reviewer-side recovery, and human judgment under controlled conditions.

\subsection{Analysis Strategy and Reproducibility}
\label{subsec:analysis_strategy_reproducibility}

The analysis strategy is diagnostic and design-oriented. For the project-level diagnosis, we summarize general governance coverage, agent-readable artifacts, fragmented AI-governance cues, project-wide governability arrangements, and observable AI provenance traces. We compare repositories by risk level, AI relevance, governance actor, and ecosystem category to examine whether the diagnostic pattern is confined to a narrow stratum or appears across theoretically relevant project conditions.

For pull-request-level signals, we report descriptive contrasts between AI-associated and non-AI-associated PRs. These contrasts provide workflow context for observable AI-mediated contribution; they do not estimate the causal effect of AI assistance on maintainer workload. Because provenance is incomplete and PR measures vary with repository type, author role, PR size, and review culture, the signals are interpreted as contextual evidence for the governance design problem. AI use is not inferred from code style, contribution quality, or reviewer suspicion.

The artifact evaluation combines controlled comparisons, objective coding, contributor-side package validation, participant ratings, and open-ended feedback. The objective analyses test recovery and representation of project-defined governance states and role boundaries. Ratings capture reviewer-side support and contributor-side feasibility, while qualitative responses identify recurring benefits, friction points, and interface needs.

For reviewer-side objective outcomes, we report exact recovery rates, absolute risk differences, and risk ratios. Uncertainty was estimated with 10,000 percentile bootstrap resamples at the participant-cluster level, retaining all task outputs for each resampled participant. Reviewer-side ratings were averaged within participant and condition and summarized as paired differences; contributor-side ratings were averaged within participant across the three tasks, with task-level distributions retained descriptively.

Triangulation connects the study's evidence without collapsing its distinct inferential roles. The repository audit establishes the organizational gap, artifact construction specifies the proposed mechanism, and the reviewer-side and contributor-side evaluations examine its linked effects on verification and preparation. Ratings and open-ended feedback contextualize how participants experienced the arrangement. This separation preserves the distinction among diagnosis, artifact specification, objective mechanism evidence, and participant interpretation.

To support transparency and reproducibility, the study separates the public replication package from the versioned AGM artifact. The replication package preserves processed and derived repository data, governance coding and adjudication records, analysis code, de-identified and aggregated evaluation outputs, codebooks, and result-to-source mappings used to reproduce the reported analyses. The AGM prototype and specification are maintained separately as public research artifacts. Public-release boundaries and exclusions are stated in the Data Availability section.

\section{Empirical Findings}
\label{sec:empirical_findings}

This section reports the diagnostic governance audit of 50 public GitHub repositories. The sample was purposively stratified across repository risk, AI relevance, governance actor, and ecosystem category. Pull request-level analysis covers 48 repositories and 23,237 pull requests; issue-level analysis covers 46 repositories and 19,884 issues.

\begin{figure}[t]
\centering
\includegraphics[width=0.88\linewidth]{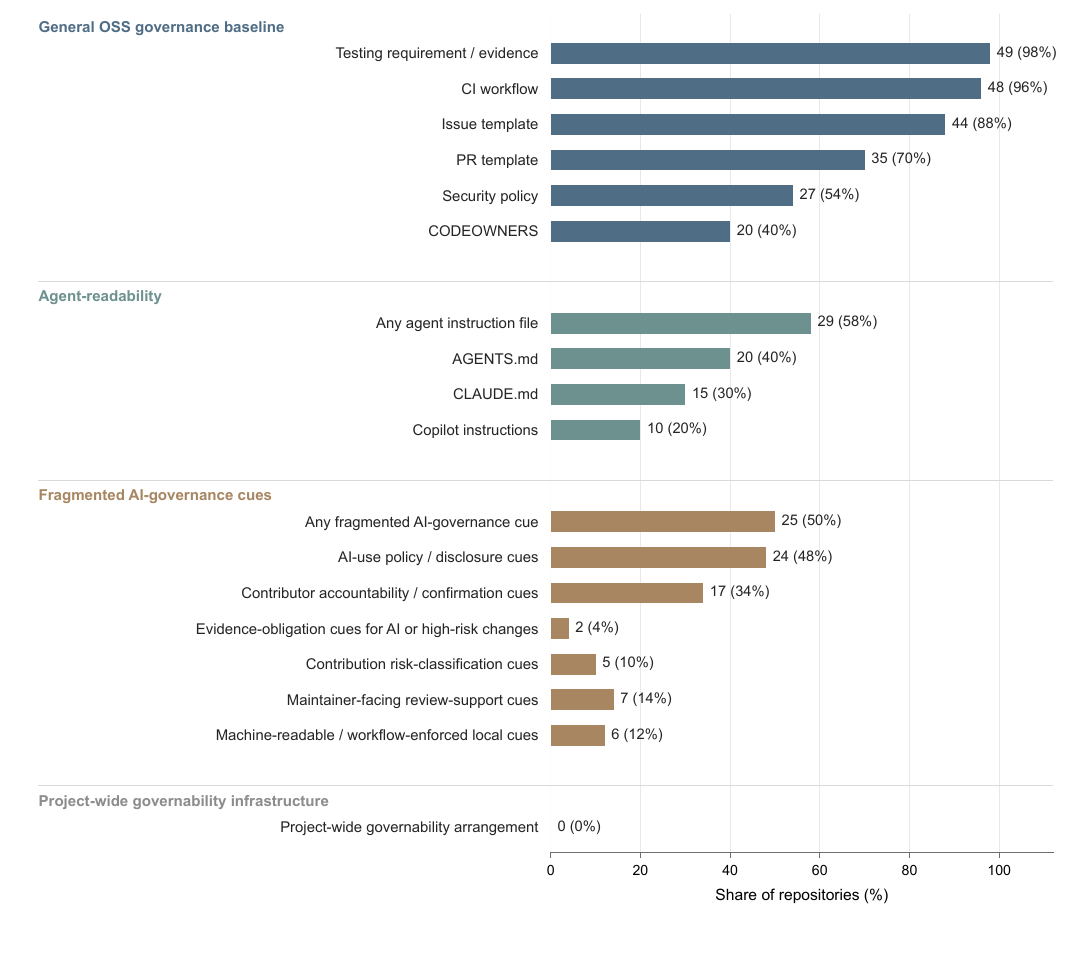}
\caption{Repository audit summary by governance-artifact category. Values indicate the number and percentage of audited repositories ($N=50$). The figure separates general OSS governance artifacts, agent-readable instruction artifacts, fragmented AI-governance cues, and project-wide governability arrangements. The project-wide category requires a canonical and repository-visible arrangement that coordinates risk, evidence, accountability, and review-gate functions across contribution and review workflows.}
\label{fig:governance_artifact_gap}
\end{figure}

The findings proceed from artifact-level governance coverage to cross-project persistence, the function of agent instruction files, and the interpretation of PR and issue signals. Figure~\ref{fig:governance_artifact_gap} provides the artifact-level overview used in Findings 1--3. It draws on the composite governance scores and targeted diagnostic recoding described in Section~\ref{subsec:governance_coding_measures}. Repository classification, coding details, composite-score distributions, group comparisons, PR/issue signal tables, targeted recoding outputs, and sensitivity checks are documented in Supplementary Material S1--S3.

\subsection{Finding 1: General Governance Coverage, Agent-Readability, and the Project-Wide Governability Gap}
\label{subsec:finding_general_ai_gap}

As shown in Figure~\ref{fig:governance_artifact_gap}, ordinary OSS governance artifacts are widespread: 98\% of repositories include testing requirements or evidence, 96\% include CI workflows, 88\% include issue templates, 70\% include PR templates, and 54\% include security policies. Agent-readability is also visible: 58\% of repositories contain an agent instruction file, including \texttt{AGENTS.md}, \texttt{CLAUDE.md}, or Copilot-related instructions.

The targeted recoding shows that AI-governance cues appear in fragmented and uneven forms. Fifty percent of repositories contain at least one fragmented AI-governance cue, most commonly an AI-use policy or disclosure cue. Thirty-four percent contain contributor-accountability or confirmation cues. More review-facing cues are much less common: 4\% contain evidence-obligation cues for AI-mediated or high-risk changes, 10\% contain contribution risk-classification cues, 14\% contain maintainer-facing review-support cues, and 12\% contain machine-readable or workflow-enforced local cues. These categories are not mutually exclusive; they capture local policies, templates, instruction-file clauses, or workflow markers.

The audit therefore identifies an architectural gap at the project level. Review-facing risk classification, evidence obligations, human-accountability states, and maintainer-facing review gates appear as sparse and unevenly distributed cues, and no repository in the audit satisfies the four-function criterion for a project-wide governability arrangement. The gap lies in the absence of a canonical arrangement that coordinates these functions across contribution preparation and review.

\subsection{Finding 2: The Project-Wide Governability Gap Appears Across Sampled Project Types}
\label{subsec:finding_group_comparison}

Across sampled risk levels, AI relevance categories, governance actors, and ecosystem roles, the same imbalance persists: general governance coverage is broad, agent-readability varies, and fragmented AI-governance cues appear locally, yet no repository provides a project-wide governability arrangement satisfying all four functional criteria.

The gap is not confined to repositories with limited ordinary governance or little exposure to AI\@. It also appears among enterprise- and foundation-led projects, AI-native and agent-tool repositories, and projects with extensive testing, CI, template, and security infrastructures. Formal sponsorship, mature general governance, and proximity to AI development therefore do not by themselves yield a project-wide arrangement connecting risk, evidence, accountability, and review gates.

Within the purposive sample, the need for project-side governability is not tied to a single risk profile, governance actor, AI category, or ecosystem role. The problem that motivates AGM is therefore cross-project rather than a niche deficit of one project type. Additional group-level descriptive statistics are reported in Supplementary Material S3.

\subsection{Finding 3: Agent Instruction Files Primarily Encode Development Guidance}
\label{subsec:finding_agent_instruction_files}

Agent-readable repository artifacts are visible across the sample and primarily encode development guidance. As shown in Figure~\ref{fig:governance_artifact_gap}, 29 repositories (58\%) contain an agent instruction file, 20 repositories (40\%) contain \texttt{AGENTS.md}, 15 repositories (30\%) contain \texttt{CLAUDE.md}, and 10 repositories (20\%) contain Copilot-related instructions. In addition, 30 repositories (60\%) contain some form of agent-related workflow or artifact.

Agent instruction files support contribution production by specifying repository-specific commands, coding conventions, test locations, structure, and workflow expectations. They reduce task-execution friction for coding agents. Project-wide governability serves a different function: it organizes review-facing obligations, accountability, and gate states across contribution preparation and verification.

The asymmetry lies in what repositories codify: operational knowledge for producing contributions is more widely encoded than the project-defined governance information needed to prepare and verify them.

\subsection{Finding 4: Observable AI-Associated PR Signals Provide Workflow Context for the Governance Diagnosis}
\label{subsec:finding_ai_provenance}

Observable AI provenance appears in public contribution records, although its distribution is highly uneven. At the PR level, 1,945 of 23,237 pull requests (8.4\%) contain broad AI contribution provenance signals, while 1,633 pull requests (7.0\%) satisfy the stricter conservative AI provenance criterion. Coding-agent traces appear in 1,808 pull requests (7.8\%), and suspected AI traces appear in 2,124 pull requests (9.1\%).

These observations complement large-scale Agentic-PR evidence by showing that public AI- and coding-agent-related contribution traces are also visible within the governance-audit sample. Explicit markers such as generated-with statements, coding-agent identifiers, tool accounts, AI-related co-authorship, or contributor disclosures appear in a measurable subset of PRs. Within this sample, their presence confirms that publicly observable AI-mediated contribution activity is relevant to the governance diagnosis.

AI provenance is highly uneven across repositories. One repository, \texttt{ClickHouse/ClickHouse}, accounts for a disproportionately high share of observable AI provenance. In that repository, 342 of 497 sampled pull requests (68.8\%) contain observable AI provenance, often associated with explicit Claude-related markers such as \texttt{Generated with [Claude Code]} or \texttt{Co-Authored-By: Claude}. When \texttt{ClickHouse/ClickHouse} is excluded, the broad AI provenance rate decreases from 8.4\% to 7.0\%, and the conservative AI provenance rate decreases from 7.0\% to 5.7\%.

The sensitivity analysis shows that observable AI-mediated contribution is concentrated in particular repositories and contribution cultures. Some projects preserve explicit AI markers, while others may use AI privately, remove attribution, or leave no visible AI provenance in public records.

Observable AI provenance is therefore a conservative lower-bound indicator of public AI-related contribution traces. Missing public markers cannot establish the absence of AI use.

\subsection{Finding 5: AI-Domain Keywords, Automation Traces, and AI Contribution Provenance Capture Distinct Phenomena}
\label{subsec:finding_signal_separation}

AI-domain keywords, ordinary automation, and AI contribution provenance capture distinct methodological and substantive phenomena. At the PR level, 9,334 of 23,237 pull requests (40.2\%) contain AI-domain keywords. This rate is much higher than the rate of AI contribution provenance because AI-native repositories naturally contain many references to models, agents, prompts, inference, embeddings, LLMs, AI APIs, and related technical terms. Such keywords indicate topical AI relevance, but they do not show that a particular contribution was generated or assisted by AI.

Ordinary automation is also common. Automation-bot traces appear in 1,982 of 23,237 pull requests (8.5\%). These traces include dependency-update bots, CI bots, release automation, formatting bots, stale issue bots, or other machine accounts. They capture routine machine-mediated workflow activity, a category distinct from generative AI or coding-agent contribution.

By contrast, AI contribution provenance refers to explicit public evidence that AI or coding agents participated in producing, modifying, reviewing, or submitting a contribution. This includes generated-with statements, AI co-authorship markers, coding-agent identifiers, AI tool accounts, assistant-generated comments, or contributor disclosures. Conservative AI provenance further restricts this measure to stronger AI or coding-agent markers.

This separation matters for empirical inference. AI-domain keywords conflate topical AI work with AI-mediated production when used as a contribution proxy; bot traces conflate ordinary automation with generative AI activity; and explicit provenance alone undercounts AI use when contributors do not preserve attribution markers.

These signals are informative but non-equivalent. In this study, they provide workflow context rather than direct measures of contribution governability or substitutes for contribution-specific review evidence.

\subsection{Summary of Empirical Findings}
\label{subsec:summary_empirical_findings}

General OSS governance is widespread and agent-readability is visible, yet AI-governance cues remain uneven and no audited repository satisfies all four functions of a project-wide governability arrangement. The gap persists across sampled risk, AI-relevance, governance-actor, and ecosystem strata, indicating a cross-project design problem rather than a deficit confined to one project type. Public contribution records also contain several non-equivalent signals: instruction files, AI-domain keywords, automation traces, and observable AI provenance capture different aspects of repository guidance, workflow automation, and contribution activity. These findings define the design problem for the artifact evaluation, which tests whether project-defined risk, evidence, accountability, and gate states become more recoverable during contribution review.

\section{Artifact Evaluation}
\label{sec:artifact_evaluation}

This section reports a controlled mechanism test of AGM, following the evaluation strategy in Section~\ref{sec:research_design}. It examines whether AGM addresses the project-wide governability gap identified in the audit by making risk zones, evidence status, contributor-confirmation declarations, human-accountability signals, and governance-gate states more recoverable than under ordinary contribution materials.

\subsection{Evaluation Design}
\label{subsec:evaluation_overview}

The reviewer-side evaluation produced 75 task-level outputs from 15 participants: 38 under AGM-supported materials and 37 under ordinary materials. Outputs were coded with the frozen objective rubric, while participant ratings and open-ended feedback were analyzed separately. A non-overlapping contributor-side cohort completed 45 AGM-supported tasks. The full task, condition-allocation, order-rotation, and measurement protocols are specified in Section~\ref{subsec:artifact_construction_evaluation_design} and Supplementary Material S5--S7.

\begin{figure}[t]
\centering
\includegraphics[width=0.88\linewidth]{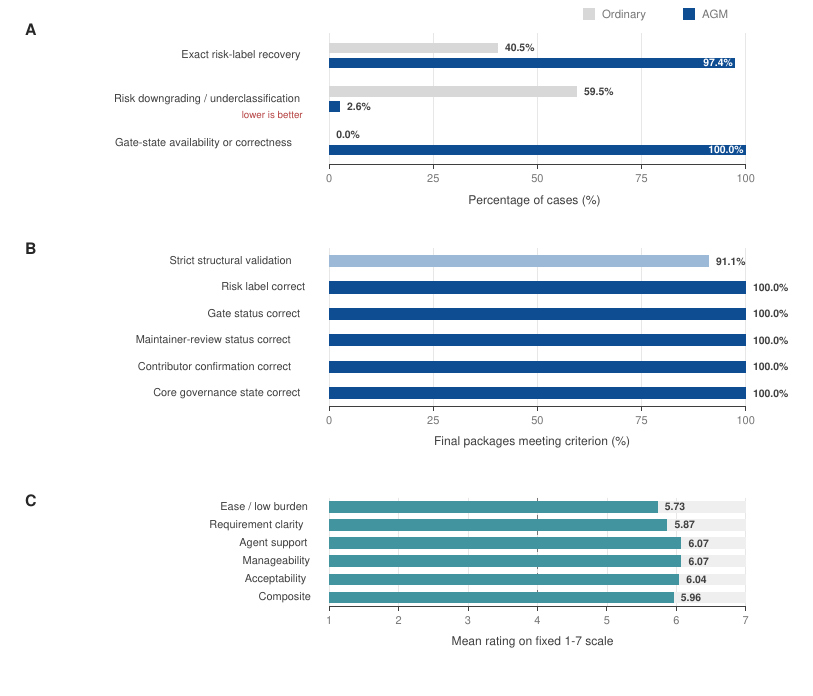}
\caption{Evaluation summary. Panel A compares ordinary and AGM-supported materials in the reviewer-side evaluation. Panel B summarizes contributor-side objective feasibility results. Panel C reports contributor-side questionnaire means on a fixed 1--7 scale, where higher values indicate stronger perceived support.}
\label{fig:evaluation_summary}
\end{figure}

Figure~\ref{fig:evaluation_summary} provides a visual roadmap for the evaluation results reported below. Panel A summarizes reviewer-side objective recovery under ordinary and AGM-supported materials. Panel B summarizes contributor-side evidence-package validation and governance-state correctness. Panel C reports contributor-side questionnaire means. Detailed item-level tables and task-level diagnostics are retained in Supplementary Material S5--S7.

\subsection{Objective Review-Output Results}
\label{subsec:objective_review_output_results}

Figure~\ref{fig:evaluation_summary}A reports the central objective results. Among AGM-supported task-level reviewer-side outputs, 37 of 38 (97.4\%) recovered the repository-defined risk level exactly, compared with 15 of 37 (40.5\%) ordinary-material outputs. The absolute difference was 56.8 percentage points (participant-clustered bootstrap 95\% CI: 42.1--71.8), corresponding to a risk ratio of 2.40 (95\% CI: 1.76--3.79).

The difference was especially visible in higher-risk tasks. Ordinary materials were sufficient for the low-risk documentation task, but they frequently led to governance-risk under-classification in medium-, high-, and critical-risk tasks. The single AGM-supported mismatch was a one-level under-classification in T5: the output recognized that the changed workflow file belonged to a critical risk zone but recommended a high rather than critical final risk label. The same output nevertheless recovered the invalid evidence status, required contributor-confirmation gate, blocked governance state, and non-eligibility for final acceptance. Under a close-or-correct risk-recognition criterion, AGM-supported outputs recovered the relevant risk zone in 38 of 38 cases.

The output-coding reliability check showed complete agreement for exact risk labels, contributor-confirmation states, and governance-gate states, with strong agreement for the two evidence-related coding fields ($\kappa=0.848$ and $\kappa=0.825$).

The task-level pattern locates the condition difference in the structured externalization of repository-defined governance states. For T1, ordinary materials were sufficient because the task was low risk and technically simple. For T2--T5, ordinary-material outputs often contained plausible technical comments while still under-classifying governance risk. Core logic and critical workflow tasks were frequently treated as ordinary technical modifications when project-defined risk-zone information was unavailable. AGM-supported materials made risk classification, evidence requirements, contributor-confirmation status, and gate status directly observable, whereas ordinary materials often left this information implicit even when outputs offered plausible technical comments. The remaining AGM-supported exact-label error also shows that agent-assisted review can depart from a manifest-defined risk level, reinforcing the role of human maintainer authority.

\subsection{Questionnaire Results}
\label{subsec:questionnaire_results}

To complement the objective coding of governance-state recovery, participants completed post-task ratings comparing ordinary and AGM-supported review materials. After averaging within participant and condition, the mean rating was 6.14 for AGM-supported materials and 3.27 for ordinary materials on the 1--7 scale. The paired mean difference was 2.87 points (participant-clustered bootstrap 95\% CI: 2.67--3.08), and the AGM-supported mean was higher for all 15 participants. The largest item-level differences concerned overall usefulness, contributor-confirmation visibility, provenance or source-trace usefulness, missing/placeholder evidence recognition, evidence sufficiency, and risk visibility. Detailed participant-level results are reported in Supplementary Material S6.

Participant ratings align with the objective coding: condition differences were concentrated in the visibility and interpretability of risk, evidence, accountability, provenance, and gate states. Human decision authority remained high in both conditions (6.76 vs.\ 6.41; paired difference 0.34, 95\% CI: $-0.04$ to 0.80). AGM's perceived support therefore centered on making contribution-governance states more inspectable while maintaining human decision authority.

The task-level ratings followed the same risk-sensitive pattern. The smallest condition difference appeared in the low-risk documentation task, while the largest appeared in the critical workflow and placeholder-evidence task. The contrast is consistent with proportional governance: condition differences were limited for low-risk work and larger when risk or evidence uncertainty was high.

\subsection{Participant Feedback and Design Implications}
\label{subsec:qualitative_feedback_design_implications}

Participant feedback helps explain the rating differences. Participants repeatedly identified contributor-confirmation declarations, missing-evidence reports, risk summaries, evidence indexes, test evidence, and gate status as the most useful parts of AGM\@. These comments locate AGM's value in making accountability and review readiness visible.

Participants also reported that the current presentation should become more intuitive. Several responses asked for structured summaries, risk cards, evidence checklists, gate-status panels, and automatic next-action suggestions. One participant compared the desired interface to a medical test report: the system should show required values, actual submitted evidence, missing or invalid items, and recommended next steps. In the controlled setting, AGM-supported review tasks were generally completed from a single initial prompt, indicating limited reviewer-side interaction friction. The feedback identifies human-facing presentation, rather than the underlying governance structure, as the main area for refinement.

The design implication is to keep the repository-hosted governance resource stable and machine-readable while presenting the same information through accessible review displays. The resource preserves the canonical schema and governance states; the interface can adapt their presentation to reviewer needs.

\subsection{Contributor-Side Feasibility Check}
\label{subsec:contributor_side_feasibility_check}

The reviewer-side evaluation shows whether AGM makes governance states recoverable during review. The contributor-side feasibility check examines whether agents can prepare the required evidence while preserving the boundary between contributor confirmation and maintainer review. A separate cohort of 15 participants completed 45 AGM-supported tasks, yielding 90 draft and final evidence packages.

Figure~\ref{fig:evaluation_summary}B summarizes the objective contributor-side results. Under the pre-specified validation criteria, 41 of 45 final packages passed strict structural validation. All 45 final packages correctly represented every component of the core governance state---risk label, review-gate status, maintainer-review requirement/status, and contributor confirmation. The four remaining strict-validation failures were limited to schema precision, such as missing explicit command/result fields for validation evidence or a missing boolean gate-required field; their core governance-state interpretations remained correct.

Across all 45 tasks, draft-to-final changes were limited to the package-stage and contributor-confirmation fields. No task changed the risk label, evidence content, review-gate status, or maintainer-review status during final confirmation. The contributor-side agent therefore prepared the governance evidence, the human contributor confirmed the package, and maintainer review remained outside the contributor-side workflow.

The post-task questionnaire reports the perceived feasibility of this workflow (Figure~\ref{fig:evaluation_summary}C). After averaging each participant across the three tasks, participant-level means were 5.87 for requirement clarity, 6.07 for agent support, 6.07 for manageability, 6.04 for acceptability, and 5.73 for reverse-coded ease. The composite mean was 5.96 (participant-clustered bootstrap 95\% CI: 5.79--6.12). Descriptive task-level differences were small; Supplementary Material S7 reports the task-level patterns and participant-level uncertainty estimates. Composite scores were 5.92 for Task A, 5.95 for Task B, and 6.00 for Task C. Thirteen of fifteen participants reported no clear difference in task difficulty, while two selected Task B as the most burdensome task. These questionnaire results support the feasibility and usability of the contributor-side workflow under the controlled protocol.

Open-ended responses and observations during the experiment suggest an additional design implication: agent-side workflow scaffolding matters. Some agents proactively reminded participants to complete the contributor-confirmation step and generated a follow-up prompt, whereas others treated evidence generation as a one-shot output task. The comments identify support for multi-step governance workflows as the relevant interface capability. Future AGM implementations should therefore treat workflow guidance---for example, pausing after draft preparation, prompting contributor confirmation, and preventing maintainer-review overclaiming---as part of the governance interface.

\subsection{Evaluation Summary}
\label{subsec:evaluation_summary}

Across the task-level indicators that map directly to contribution governability, AGM-supported materials improved risk-label recovery, missing-evidence recognition, gate-state judgment, and accountability visibility. Ordinary materials often supported plausible technical inspection while leaving governance risk under-classified. Reviewer ratings were strongest in the areas predicted by the framework: risk visibility, evidence sufficiency, missing or placeholder evidence detection, source-trace usefulness, and human-accountability visibility. Final approval, rejection, and merge authority remained with human maintainers. The contributor-side feasibility check further showed that the same governance contract can structure preparation before review: agents prepared draft and final evidence packages, contributors confirmed them, and maintainer-review authority remained separate.

These results provide controlled, mechanism-level evidence. Objective coding shows greater recovery of project-defined governance states, contributor-side validation shows how those states can be prepared before review, and participant ratings indicate stronger review support; open-ended feedback identifies the design features and friction points underlying that difference. Across both evaluations, the preparation--verification linkage shows how project-side governability infrastructure takes operational form in contribution preparation and review.

\section{Discussion}
\label{sec:discussion}

Prior research shows that agent-authored contributions are visible in GitHub workflows~\citep{li2026aiddev}, OSS communities are developing GenAI governance strategies~\citep{yang2026beyondBanningAI}, and provenance standards are making AI participation more traceable~\citep{cursor2026agentTrace}. This study theorizes their common technology-management implication: how to make agent-mediated contributions governable at review time. The repository diagnosis shows extensive general governance coverage, observable agent-readability, uneven AI-governance cues, and an architectural gap in the project-side coordination of risk, evidence, accountability, and review gates. The controlled evaluation shows stronger reviewer-side recovery of these states under AGM-supported materials, while the contributor-side feasibility check shows how agents can prepare structured evidence for human confirmation without absorbing maintainer-review authority.

\subsection{Theoretical Implications}
\label{subsec:theoretical_implications}

The controlled evaluation makes contribution governability empirically observable by showing how contributor-side workflows externalize project-defined governance states before review and reviewer-side outputs recover them during review. The central theoretical contribution explains how a production-side technological shift creates an organizational-design problem in distributed innovation. Coding agents lower the cost of generating contribution artifacts, while verification continues to depend on limited maintainer review capacity and project-specific judgment. Project-side governability infrastructure redistributes this information-processing work across the contribution workflow: projects state review-relevant requirements before submission, contributor-side actors prepare the corresponding evidence and accountability states, and maintainers move from open-ended reconstruction toward targeted verification while retaining final decision rights. Building on organizational information-processing research~\citep{tushmanNadler1978informationProcessing,daftWeick1984interpretationSystems} and open-collaboration research~\citep{brunswicker2025openCollaboration}, this mechanism shows how digital governance infrastructure can address uncertainty and equivocality through a temporal and relational redistribution of preparation, interpretation, and decision work across distributed actors.

The three-layer framework prevents distinct organizational problems from being collapsed into a single category of AI governance. Agent-readable guidance~\citep{gloaguen2026agentsMd} supports participation, and provenance~\citep{cursor2026agentTrace,vispute2026reasoningProvenanceAgents} preserves observable production histories, but these functions leave unanswered who must prepare review evidence, who may verify it, and how a contribution moves through project-defined gates. Governability supplies this normative and relational layer by allocating obligations, verification authority, and decision states across the workflow. The framework therefore positions project-side governability as a capability grounded in rules, relationships, and review authority.

Boundary-resource theory gains a further distinction between the governance resource and the object whose governance state it structures. Shared artifacts mediate external contribution and project control~\citep{ghazawnehHenfridsson2013boundaryResources}; AGM stabilizes project-defined governance meanings across contributors, agents, maintainers, validators, and workflow tools~\citep{handler2017ossBoundaryObject,mayer2025genAIBoundaryResource}. When those rules are enacted for a specific change, the evidence-bearing contribution becomes a governable boundary object interpretable across actor groups. Actor-level reputation remains relevant, while contribution-level inspectability becomes an additional basis for trust when contributor identity no longer reveals how a specific change was produced. This resource--object relationship explains how project-side infrastructure enters situated contribution work while preserving the distinction between the project arrangement, its repository-hosted carrier, and its contribution-level enactment.

The bidirectional governance contract extends this relational account by treating AI governance as an allocation of preparation work and decision authority across a distributed workflow. OSS governance research shows that authority is produced through community-legitimated roles and configurations of coordination processes~\citep{omahonyFerraro2007emergenceGovernance,shaikhHenfridsson2017coordinationProcesses}. In agent-mediated work, the contract couples contributor-side preparation obligations with maintainer-side verification and final decision rights, while assigning tools a supporting role. The common governance specification makes evidence inspectable, turns unresolved states into repair or escalation points, and keeps authority anchored in contribution confirmation and maintainer judgment. This account advances research on AI accountability~\citep{yang2026beyondBanningAI}, machine-contributor governance~\citep{manita2026machineContributorGovernance}, and executable governance~\citep{datla2025executableGovernanceAI} by explaining how obligations and decision rights are coordinated across the preparation--verification relationship.

Compliance enablement recasts formalization as a service to responsible participation. As an enabling form of governance~\citep{adler1996twoTypesBureaucracy,woutersWilderom2008enablingFormalization}, project rules help participants understand what responsible contribution requires, make incomplete governance states visible, and identify repair points before limited maintainer review capacity is committed to final judgment. This service-oriented formalization gives participants seeking to satisfy project requirements a clearer path through preparation and review. In a voluntary participation setting~\citep{shah2006motivationGovernance}, it can operate alongside higher-assurance controls for security-sensitive or high-consequence changes, allowing projects to configure governance portfolios according to verification risk while preserving participation and human decision authority.

Workflow scaffolding carries these implications into everyday human--AI work design. Repository-hosted governance resources can guide agents from shared project requirements to situated preparation, human confirmation, and handoff to maintainer review. Governance infrastructure thereby becomes a coordination script that sequences human and agent work without reallocating project authority. This interactional role complements participation-architecture research~\citep{westOmahony2008participationArchitecture} and speaks to private-ordering and agent-governance architecture accounts~\citep{su2026platformGovernancePrivateOrdering,alvarezTelena2026threeRingArchitecture} by showing how repository interfaces can make responsible participation more accessible while retaining project control over review and acceptance.

\subsection{Practical Implications}
\label{subsec:practical_implications}

For firms and OSS project stewards managing AI-exposed repositories, the findings identify project-side governability infrastructure as a distinct project capability alongside broad general governance coverage and localized AI-governance responses. Project-health practices and dashboards~\citep{linaker2022ossProjectHealth,chaoss2026terminology,linuxFoundation2025lfxInsights}, tests, CI, templates, security procedures, and AI-use disclosure provide important foundations. Project stewards still need a canonical arrangement that makes each contribution's governance state inspectable during review.

For contributors and tool builders, the findings point to a different organization of review-preparation work and accountability cues at the contributor--agent interface. Coding agents can prepare verification-relevant information before submission by identifying affected risk zones, organizing test and provenance evidence, flagging gaps, and preparing contributor-confirmation states for human review. This use of agentic infrastructure builds on productivity gains in AI-assisted development~\citep{he2025cursorSpeedCostQuality} while responding to concerns about low-quality AI-generated contribution streams~\citep{baltes2026endlessStreamAISlop,sen2026githubAISlop}. The contributor-side feasibility check shows that this workflow can be completed under controlled conditions and why interface support matters: agents that prompt users to review the draft and proceed to confirmation make the sequence easier to follow. The practical value lies in a clearer path for preparing what maintainers already need to review the change.

For platform operators, foundation stewards, and infrastructure providers, the results suggest that future OSS tools could support governance contracts directly through human-facing review interfaces that expose required, submitted, and unresolved governance information. Provenance standards such as Agent Trace could provide file- and line-level AI participation data, while AGM-like contracts would situate those signals within risk-sensitive evidence and review requirements. This direction aligns with platform-governance and ecosystem-smartification research on infrastructure for distributed governance and coordination~\citep{santalo2025strategicBlockchainGovernance,kowalski2026ecosystemSmartification}. Such infrastructure can help maintainers direct limited review capacity without requiring each reviewer to reconstruct the project's governance state from dispersed repository materials.

The compliance-enabling logic is designed for contributors willing to meet project requirements but needing clearer guidance on evidence, confirmation, and review readiness. It complements a broader governance portfolio in which security review, branch protection, \texttt{CODEOWNERS}, mandatory CI, signed commits, artifact attestation, and policy enforcement provide higher-assurance controls where risk warrants them. AGM supports upstream preparation and diagnosis by making compliant contribution easier to complete and unresolved evidence more visible before maintainer judgment.

\subsection{Boundary Conditions and Limitations}
\label{subsec:boundary_conditions_limitations}

The purposive repository sample examines whether project-wide governability gaps appear across different risk levels, AI relevance categories, governance actors, and ecosystem roles. The design therefore supports cross-stratum diagnosis rather than population-level prevalence estimation, which is better served by large-scale Agentic-PR datasets such as AIDev~\citep{li2026aiddev} and related PR studies~\citep{watanabe2026agenticCodingPullRequests}. The reliability check and targeted recoding strengthen the consistency of this diagnosis.

Observable AI provenance is incomplete. Public PR records capture AI-related contribution signals only when traces are preserved in titles, bodies, comments, commits, branch names, bot markers, or other public metadata. Hidden, manually edited, or unreported AI assistance cannot be fully measured, so pull-request-level AI signals should be interpreted as conservative lower-bound indicators, especially given the limits of provenance records~\citep{cursor2026agentTrace} and the strategic ambiguity of AI attribution~\citep{kraishan2025aiAttributionParadox}.

Pull-request-level AI signals provide workflow-contextual evidence for the governance design problem. They do not identify causal effects of AI assistance on maintainer workload, which remain outside the study's scope.

The artifact evaluation uses controlled tasks rather than production pull requests. This design isolates the mechanism of AGM and observes contributor-side feasibility under comparable conditions, but it does not measure long-term adoption, contributor compliance, community acceptance, maintainer workload, or production maintainability. The contributor-side questionnaire should therefore be interpreted as a controlled feasibility and usability signal, not as evidence of production-community adoption. Field deployments and action research, as emphasized in OSS GenAI roadmap and governance-practice research~\citep{feng2026chartingUncertainWaters,oliveira2026governancePracticeRoles}, would be needed to evaluate how AGM-like mechanisms operate in live OSS communities.

Participants used heterogeneous agent environments, so the analysis concerns governance-workflow behavior across those environments. Model-level performance comparisons remain outside the study's inferential scope, and observed workflow differences may depend on model version, interface design, prompt handling, context-window behavior, tool integration, and deployment context. This heterogeneity limits attribution of the observed workflow behavior to any particular model or interface.

AGM can introduce burden if evidence requirements are too broad, vague, or heavy. Effective use requires careful calibration of risk zones, evidence thresholds, and responsibility gates, consistent with information-processing views of uncertainty management~\citep{daft1986mediaRichness} and agent-governance architecture concerns~\citep{alvarezTelena2026threeRingArchitecture}. Stronger requirements should be reserved for changes that impose greater verification risk.

\subsection{Future Research}
\label{subsec:future_research}

Longitudinal research should examine how agent-readability~\citep{gloaguen2026agentsMd}, traceability standards~\citep{cursor2026agentTrace}, localized AI-governance responses~\citep{fedora2025aiAssistedContributionsPolicy,feng2026chartingUncertainWaters}, and project-wide governability arrangements develop, interact, or diverge over time. Such work could assess how these governance functions diffuse unevenly across ecosystems and become embedded in different project-level architectures.

Field evaluations should test AGM-like mechanisms under production conditions. Controlled evaluation establishes mechanism feasibility, while production adoption raises questions about maintainer perception, contributor compliance, workload effects, review quality, and community legitimacy~\citep{hoffman2026aiBreaksSystems}. Live OSS deployments, action research, and foundation- or platform-supported pilots would help evaluate how AGM-like mechanisms operate under real community norms, contributor incentives, and review constraints. Component-level evaluations could examine the relative contribution of risk summaries, evidence indexes, confirmation states, and review packets to governance-state recovery.

Measurement research should connect AIDev-like Agentic-PR datasets~\citep{li2026aiddev,watanabe2026agenticCodingPullRequests}, public repository traces, contributor surveys, platform metadata, tool-level telemetry, and provenance standards~\citep{cursor2026agentTrace} where ethically and legally appropriate. Better measurement would help distinguish AI-assisted production, ordinary automation, topical AI work, and governance responses.

Standardization and interface design form another research priority as agentic software development expands into scientific agent ecosystems~\citep{xu2026claw4science} and broader human--AI programming workflows~\citep{cruz2025redefiningProgrammer}. Standards for repository-hosted governability resources should preserve stable machine-readable schemas while allowing human-facing interfaces to render the same governance state in accessible forms.

\section{Conclusion}
\label{sec:conclusion}

Coding agents are changing the conditions under which OSS communities generate and review contributions. They can help contributors produce code, tests, documentation, configuration changes, and contribution materials more quickly, but responsibility for verifying whether those contributions are correct, safe, maintainable, sufficiently tested, and accountable remains concentrated on maintainers.

This paper distinguishes three functions within that governance challenge. Agent instruction files make repositories readable to agents, provenance logs make AI participation traceable, and governability organizes project rules so that specific contributions become risk-classifiable, evidence-supported, human-accountable, and ready for review.

The argument is supported by a 50-repository diagnostic governance audit, a controlled reviewer-side assessment with 15 participants and 75 task-level outputs, and a contributor-side feasibility check with 15 participants and 45 tasks. The audit finds broad general governance coverage, observable agent-readability, and fragmented AI-governance cues, but no project-wide arrangement satisfying all four governability criteria. Under controlled conditions, AGM-supported materials improve recovery of risk, evidence, accountability, and gate states during review, while contributor-side agents prepare evidence packages whose core governance states are correctly represented for human confirmation.

AGM instantiates project-side governability as a repository-hosted governance resource that defines risk-sensitive evidence obligations, human confirmation, and review gates across contribution and review. By moving review-relevant preparation upstream and keeping final judgment with maintainers, this arrangement connects machine-scaled production with human-accountable review. AI-mediated OSS governance therefore depends on project-side infrastructure that can give shared rules contribution-level force while preserving project-governed decision rights.

\section*{Ethics and Informed Consent}

Participation was voluntary, and all participants provided informed consent before beginning the study. Participants were informed about the study procedures and the intended use of the collected data. Participant records were de-identified before analysis and reporting, and no identifiable personal information is reported.

\section*{Data Availability}

A public replication package containing the processed and derived repository data, governance coding and adjudication records, analysis code, de-identified and aggregated evaluation results, codebooks, and result-to-source mappings supporting the reported analyses will be made available at \url{https://github.com/AnonymousResearchersAGM/AGM_Data_Availability}. The package excludes raw API data, participant-specific task materials and outputs, free-text questionnaire responses, prompts, and manuscript-production assets.

\appendix
\section{Artifact Availability}
\label{app:artifact_availability}

The public AGM resources are available as versioned research artifacts. These include the reference prototype repository (\url{https://github.com/agent-governance-manifest/agent-governance-manifest}) and the community-draft specification (\url{https://agent-governance-manifest.github.io/agent-governance-manifest/AGM_SPEC_v0.1.html}). The prototype provides an inspectable implementation of the manifest structure, evidence-package workflow, validation logic, and review-support process. The specification documents the intended governance semantics of AGM, including risk zones, evidence obligations, contributor-confirmation states, maintainer-review gates, and artifact boundaries. These resources represent the AGM v0.1.0 research prototype and community-draft specification; field adoption, certification, and production-service evaluation remain outside the evidentiary scope of this study.

\section{Supplementary Material Index}
\label{app:supplementary_material_index}

The Supplementary Material provides methodological details, coding evidence, participant information, and additional evaluation outputs that support the main text. Its structure is summarized below.

\begin{center}
\begin{minipage}{\textwidth}
\centering
\footnotesize
\setlength{\tabcolsep}{2.5pt}
\renewcommand{\arraystretch}{0.96}
\captionof{table}{Guide to the Supplementary Material.}
\label{tab:supplementary_index}
\begin{tabularx}{\textwidth}{@{}p{0.12\linewidth} p{0.35\linewidth} X@{}}
\hline
\textbf{Section} & \textbf{Contents} & \textbf{Main-text support} \\
\hline
S1 & Repository sample and classification & Sample design and audit scope \\
S2 & Coding scheme, second-coder check, and adjudication & Construct validity and coding reliability \\
S3 & Repository-level findings and sensitivity checks & Empirical findings \\
S4 & AGM artifact, specification, and public repository & Artifact design and public research artifact access \\
S5 & Reviewer-side evaluation protocol, materials, and participant profile & Evaluation design, task materials, and participant background \\
S6 & Reviewer-side results, output-coding reliability, questionnaire ratings, and robustness checks & Objective recovery, coding reliability, perceived usefulness, and authority compatibility \\
S7 & Contributor-side feasibility protocol and questionnaire & Contributor-side validation and feasibility \\
S8 & Artifact boundary, adjacent mechanisms, and versioning notes & AGM positioning and reproducibility boundary \\
\hline
\end{tabularx}
\end{minipage}
\end{center}

\FloatBarrier
\bibliographystyle{elsarticle-num}
\bibliography{references}

% ================================================================
% Supplementary Material (merged for the arXiv version)
% ================================================================
\FloatBarrier
\clearpage

% The main manuscript has already entered appendix mode.  Reset the
% displayed counters and the hyperref anchor counters so that the merged
% supplementary sections, tables, and figures use unique S-prefixed labels.
\renewcommand{\thesection}{S\arabic{section}}
\renewcommand{\thesubsection}{S\arabic{section}.\arabic{subsection}}
\renewcommand{\thetable}{S\arabic{table}}
\renewcommand{\thefigure}{S\arabic{figure}}
\renewcommand{\theequation}{S\arabic{equation}}
\setcounter{section}{0}
\setcounter{subsection}{0}
\setcounter{table}{0}
\setcounter{figure}{0}
\setcounter{equation}{0}
\makeatletter
\renewcommand{\theHsection}{supp.\arabic{section}}
\renewcommand{\theHsubsection}{supp.\arabic{section}.\arabic{subsection}}
\renewcommand{\theHtable}{supp.\arabic{table}}
\renewcommand{\theHfigure}{supp.\arabic{figure}}
\renewcommand{\theHequation}{supp.\arabic{equation}}
\makeatother
\section*{Supplementary Material for ``Making Agent-Mediated Contributions Governable: A Project-Level Governance Manifest for Open-Source AI Collaboration''}

This supplementary material provides the audit trails, item-level details, and reproducibility documentation supporting the main manuscript. The main text presents the conceptual framing, design logic, core empirical patterns, and controlled evaluation results; this file documents the sample, coding decisions, second-coder checks, artifact implementation, reviewer-side evaluation, contributor-side feasibility check, and scope boundaries.

Table~\ref{tab:app_reader_guide} directs readers to the evidence supporting each main-text claim.

\begin{table}[!htbp]
\centering
\caption{Reader guide to the supplementary material}
\label{tab:app_reader_guide}
\scriptsize
\setlength{\tabcolsep}{3pt}
\renewcommand{\arraystretch}{1.06}
\begin{adjustbox}{max width=\textwidth}
\begin{tabular}{p{0.16\linewidth} p{0.38\linewidth} p{0.36\linewidth}}
\hline
\textbf{Section} & \textbf{What to find} & \textbf{Main manuscript claim supported} \\
\hline
S1 & Repository sample, public data units, sampling logic, and classification dimensions & Diagnostic sample construction and audit scope \\
S2 & Governance coding protocol, composite indexes, second-coder diagnostic check, and adjudication logic & Reliability of objective artifact detection and construct-boundary clarification \\
S3 & Sample composition, item-level prevalence, adjudicated layered governance variables, and AI-trace statistics & Empirical finding of uneven fragmented cues and the absence of a project-wide governability arrangement \\
S4 & AGM artifact map, public repository, and relationship to adjacent artifacts & Artifact design and boundary between AGM and existing mechanisms \\
S5 & Reviewer-side evaluation protocol, materials, rubric, and participant profile & Controlled mechanism test and participant-background transparency \\
S6 & Reviewer-side results, output-coding reliability, error patterns, questionnaire ratings, feedback categories, and robustness checks & Contribution governability recovery and coding reliability under AGM-supported materials \\
S7 & Contributor-side feasibility protocol, frozen validation rule set, draft-to-final audit, and questionnaire results & Contributor-side feasibility and perceived manageability \\
S8 & Scope boundaries, adjacent-mechanism distinctions, and versioning notes & Interpretation and reproducibility boundary for the submitted research snapshot \\
\hline
\end{tabular}
\end{adjustbox}
\end{table}

\section{Repository Sample and Classification Criteria}
\label{app:sample_design}

This section documents the public data sources, sampling logic, and classification rules used to build the 50-repository diagnostic sample. The purposive sample covers theoretically relevant variation in downstream risk, AI relevance, governance actor, and ecosystem role and supports diagnostic, theory-building analysis. Section~\ref{app:additional_statistics} reports the resulting sample composition and descriptive statistics.

The repository audit uses public GitHub data and publicly visible repository artifacts. Repository-level governance artifacts are used to diagnose formal governance infrastructure, while pull-request- and issue-level records are used to observe contribution streams and public AI-related traces. Table~\ref{tab:app_data_sources_units} summarizes the main data units and their analytical roles.

\begin{table}[!htbp]
\centering
\caption{Data sources and analytical units used in the empirical audit}
\label{tab:app_data_sources_units}
\scriptsize
\setlength{\tabcolsep}{3pt}
\renewcommand{\arraystretch}{1.05}
\begin{adjustbox}{max width=\textwidth}
\begin{tabular}{p{0.23\linewidth} p{0.31\linewidth} p{0.36\linewidth}}
\hline
\textbf{Analytical unit} & \textbf{Main public sources} & \textbf{Use in the study} \\
\hline
Repository & Repository metadata and project-level files & Classifies ecosystem role, governance actor, AI relevance, and repository-level downstream risk. \\
Governance document & README, contribution files, security policy, templates, workflows, code ownership, agent-instruction files, and AI-policy artifacts & Codes general governance coverage, agent-readability, fragmented AI-governance cues, and indicators of project-wide governability arrangements. \\
Pull request & PR metadata, titles, bodies, comments, review activity, merged/closed status, and file-level change records & Observes AI-domain language, automation traces, AI provenance signals, changed-file surfaces, and review-interaction proxies. \\
Issue & Issue metadata, titles, bodies, comments, labels, and state & Observes public issue-level AI language, automation traces, AI provenance signals, and security-related discussion. \\
Repository-month panel & Time-indexed repository activity aggregates where available & Supports descriptive context for project activity and longitudinal contribution volume. \\
\hline
\end{tabular}
\end{adjustbox}
\end{table}

All repository artifacts and contribution records were bounded by a snapshot cutoff of June 14, 2026. The final API collection run was completed on June 15, 2026, and the resulting dataset was fixed to the June 14 observation boundary before analysis. Repository-level governance analysis retained all 50 repositories; the contribution-level coverage differences are documented below.

The sample was purposively stratified to examine whether repositories facing plausible AI-mediated contribution pressure have public governance mechanisms that support contribution governability at review time. This design concentrates the diagnosis on consequential OSS settings while preserving variation in project type and governance context. Table~\ref{tab:app_sampling_logic} summarizes the sampling logic.

\begin{table}[!htbp]
\centering
\caption{Sampling logic and inclusion principles}
\label{tab:app_sampling_logic}
\scriptsize
\setlength{\tabcolsep}{3pt}
\renewcommand{\arraystretch}{1.05}
\begin{adjustbox}{max width=\textwidth}
\begin{tabular}{p{0.22\linewidth} p{0.34\linewidth} p{0.34\linewidth}}
\hline
\textbf{Sampling principle} & \textbf{Operational meaning} & \textbf{Reason for inclusion} \\
\hline
AI exposure & Include AI-native, agent-tool, agent-sensitive, and AI-affected projects as well as comparison repositories & Captures contexts where AI-mediated contribution is plausible or institutionally relevant. \\
Downstream consequence & Include infrastructure, package/supply-chain, security-sensitive, cloud-native, database, and developer-tool projects & Tests whether governability gaps appear in projects where insufficiently verified changes may have broader consequences. \\
Governance actor variation & Include enterprise-led, foundation-led, community-led, and individual-led projects & Examines whether formal organizational capacity maps onto AI-related governance responses. \\
Ecosystem heterogeneity & Include different technical domains and contribution surfaces & Avoids treating AI governance as a problem limited to AI-native repositories. \\
Comparison logic & Retain some general OSS or case-oriented repositories & Provides contrast without claiming population-level representativeness. \\
\hline
\end{tabular}
\end{adjustbox}
\end{table}

Repositories were classified along five dimensions before the main descriptive analysis. These dimensions stratify and contextualize the governance diagnosis. Table~\ref{tab:app_classification_dimensions} defines the classification dimensions. The detailed repository-level risk rubric and composition counts are reported in Supplementary Section~\ref{app:additional_statistics}.

\begin{table}[!htbp]
\centering
\caption{Repository classification dimensions used for stratified diagnosis}
\label{tab:app_classification_dimensions}
\scriptsize
\setlength{\tabcolsep}{3pt}
\renewcommand{\arraystretch}{1.05}
\begin{adjustbox}{max width=\textwidth}
\begin{tabular}{p{0.20\linewidth} p{0.36\linewidth} p{0.34\linewidth}}
\hline
\textbf{Dimension} & \textbf{Coding question} & \textbf{Analytical role} \\
\hline
Repository-level risk & What would be the likely downstream consequence of accepting insufficiently verified contributions in this project context? & Distinguishes high-risk infrastructure from medium- and low-risk technical contexts; separate from AGM path-level risk zones. \\
AI relevance & How directly is the repository connected to AI systems, coding agents, AI-affected development, or general OSS comparison? & Tests whether AI-related governability cues appear only in AI-native contexts or also in broader OSS infrastructure. \\
Governance actor & What dominant organizational form appears to govern the repository? & Compares enterprise-led, foundation-led, community-led, and individual-led governance capacity. \\
Ecosystem category & What technical or ecological role does the repository play? & Interprets governance patterns across infrastructure, AI/ML, agent tools, package/supply-chain, databases, and applications. \\
Sample role & Is the repository used for core project-level diagnosis, comparison, or case-oriented interpretation? & Separates project-level governance evidence from PR/issue-level comparability constraints. \\
\hline
\end{tabular}
\end{adjustbox}
\end{table}

Repository-level risk is coded as a project-level attribute capturing expected downstream consequence. The empirical audit and AGM apply risk at different analytical levels. The empirical audit classifies repositories by project-level downstream consequence; AGM classifies changed paths, evidence obligations, and review gates inside a repository. The two measures are related and serve different analytical purposes.

Repository-level governance analysis retained all 50 repositories. PR-level analysis covered 48 repositories, excluding \texttt{postgres/postgres} and \texttt{torvalds/linux} because their GitHub mirrors do not represent the projects' primary contribution workflows. Issue-level analysis covered 46 repositories: the same two case-oriented mirrors were excluded, and GitHub Issues were unavailable for \texttt{django/django} and \texttt{encode/httpx} at the observation cutoff. These coverage differences preserve heterogeneous projects for repository-level governance diagnosis without treating unavailable or non-primary GitHub contribution channels as comparable workflow observations. The findings should therefore be interpreted as diagnostic evidence across theoretically relevant contexts rather than as prevalence estimates for all GitHub repositories.

\section{Governance Coding Protocol, Composite Measures, and Second-Coder Check}
\label{app:governance_coding}

This section documents the governance coding and second-coder checks summarized in the main manuscript. It explains how public repository artifacts and contribution-level records are coded as diagnostic indicators, how legacy composite indexes are constructed, and how adjudication produced the final layered variables. Section~\ref{app:additional_statistics} reports item-level prevalence, score distributions, and the three-tier diagnostic stratification.

The coding protocol has two levels. Repository-level coding captures publicly visible governance mechanisms. Contribution-level coding captures public PR and issue signals related to AI-domain topics, ordinary automation, and observable AI provenance. Table~\ref{tab:app_repo_coding_constructs} summarizes the repository-level constructs.

\begin{table}[!htbp]
\centering
\caption{Repository-level governance coding constructs}
\label{tab:app_repo_coding_constructs}
\scriptsize
\setlength{\tabcolsep}{3pt}
\renewcommand{\arraystretch}{1.05}
\begin{adjustbox}{max width=\textwidth}
\begin{tabular}{p{0.25\linewidth} p{0.32\linewidth} p{0.33\linewidth}}
\hline
\textbf{Construct} & \textbf{Coding purpose} & \textbf{Indicator families} \\
\hline
General governance coverage & Captures the publicly visible coverage of ordinary OSS contribution coordination, quality-control, and review-routing mechanisms & CI workflows, testing expectations, security policy, PR/issue templates, code ownership, reproduction guidance, benchmark expectations, license/IP checks. \\
Agent-readability & Captures repository artifacts that make the repository easier for coding agents to interpret & Agent instruction files, \texttt{AGENTS.md}, \texttt{CLAUDE.md}, Copilot instructions, tool-specific coding guidance, build/test/navigation instructions. \\
Fragmented AI-governance cues & Captures local AI-related governance signals that recognize AI-mediated contribution but remain scattered across files or workflows & AI-use policy/disclosure cues, contributor accountability or confirmation cues, evidence or risk cues, maintainer-facing review-support cues, machine-readable or workflow-enforced local cues. \\
Project-wide governability arrangement & Captures whether a canonical and repository-visible project arrangement coordinates the four functions required for contribution governability & Risk classification, evidence obligations, human-accountability states, and maintainer-facing review gates, expressed through one repository artifact or through explicitly linked repository resources. \\
\hline
\end{tabular}
\end{adjustbox}
\end{table}

The coding scheme distinguishes agent-readability, fragmented AI-governance cues, and project-wide governability arrangements. Agent instructions capture production-side guidance; disclosure, accountability, evidence, risk, review-support, and workflow markers capture localized governance cues; and the project-wide stratum requires canonical allocation and coordination of all four contribution-governability functions. This layered structure provides the construct boundary used in the targeted recoding.

Contribution-level signals are coded conservatively because public records do not reveal all AI use. Table~\ref{tab:app_contribution_signal_rules} defines the signal families used for PR and issue analysis.

\begin{table}[!htbp]
\centering
\caption{Contribution-level AI and automation signal coding rules}
\label{tab:app_contribution_signal_rules}
\scriptsize
\setlength{\tabcolsep}{3pt}
\renewcommand{\arraystretch}{1.05}
\begin{adjustbox}{max width=\textwidth}
\begin{tabular}{p{0.24\linewidth} p{0.34\linewidth} p{0.32\linewidth}}
\hline
\textbf{Signal family} & \textbf{Coding meaning} & \textbf{Interpretation rule} \\
\hline
AI-domain keyword & Topical AI-related language in a PR or issue & Indicates that the contribution concerns AI-related subject matter; not evidence of AI-assisted production by itself. \\
Automation-bot trace & Ordinary bot, dependency-update, CI, stale-issue, formatting, or release-automation activity & Indicates workflow automation; not treated as coding-agent contribution unless explicit AI or coding-agent markers appear. \\
Broad AI provenance & Public evidence suggesting AI-assisted, generated, or coding-agent-mediated contribution & Coded when explicit AI/coding-agent identifiers, generated-with statements, tool accounts, co-authorship, or contributor disclosures appear. \\
Conservative AI provenance & Stricter subset of observable AI provenance & Requires stronger AI or coding-agent markers; reduces false positives but likely underestimates hidden AI assistance. \\
High-risk path touch & Changed files or contribution surfaces associated with higher review sensitivity & Used as a review-surface indicator; not treated as evidence of AI use. \\
\hline
\end{tabular}
\end{adjustbox}
\end{table}

Composite scores are constructed from binary indicators within each construct. For repository $i$ and construct $c$, the score is calculated as
\[
S_{ic}=\frac{1}{K_c}\sum_{k=1}^{K_c} I_{ick},
\]
where $I_{ick}$ is a binary indicator and $K_c$ is the number of indicators included in construct $c$. Scores are scaled from 0 to 1. They are diagnostic indexes of publicly visible governance breadth, not causal measures or quality ratings. Table~\ref{tab:app_score_construction_protocol} summarizes the construction logic.

\begin{table}[!htbp]
\centering
\caption{Composite score construction protocol}
\label{tab:app_score_construction_protocol}
\scriptsize
\setlength{\tabcolsep}{3pt}
\renewcommand{\arraystretch}{1.05}
\begin{adjustbox}{max width=\textwidth}
\begin{tabular}{p{0.24\linewidth} p{0.34\linewidth} p{0.32\linewidth}}
\hline
\textbf{Score} & \textbf{Included indicator logic} & \textbf{Interpretation} \\
\hline
General-governance coverage score & Average of binary indicators for publicly visible ordinary OSS governance artifacts & Higher values indicate broader coverage of public contribution and review mechanisms. \\
Legacy AI-governance readiness index & Average of legacy binary indicators for AI-aware governance cues or agent-readable guidance & Higher values indicate more visible recognition of AI or coding agents in repository guidance or policy; this score is retained as a descriptive legacy index. \\
Legacy AI evidence-execution index & Average of legacy stricter binary indicators for evidence-oriented AI governance and review support & Higher values indicate more structured mechanisms for making AI-mediated contributions reviewable and accountable; targeted recoding further separates local cues from project-wide coordinated arrangements. \\
Group-level gaps & Difference between general-governance coverage and legacy AI-specific scores & Larger gaps indicate a wider separation between ordinary governance coverage and visible AI-specific governance mechanisms. \\
\hline
\end{tabular}
\end{adjustbox}
\vspace{2pt}
\begin{minipage}{0.95\textwidth}
\scriptsize
\emph{Note.} The score formula uses equal weights within each construct. The legacy AI indexes are descriptive continuity measures; the final diagnosis uses the adjudicated layered variables reported in Section~\ref{app:additional_statistics}.
\end{minipage}
\end{table}

The coding rules separate development guidance from governance mechanisms. A file such as \texttt{AGENTS.md} or \texttt{CLAUDE.md} is coded as an agent-readable artifact when it provides instructions for coding agents. It is not coded as a legacy AI evidence-execution mechanism unless it also specifies review-relevant evidence obligations, risk classification, human responsibility, test reporting, provenance reporting, or maintainer-side review diagnostics. Similarly, topical AI language is not coded as AI provenance, and ordinary automation-bot activity is not coded as coding-agent contribution.

Table~\ref{tab:app_coding_quality_controls} summarizes the conservative interpretation and reproducibility controls used in coding.

\begin{table}[!htbp]
\centering
\caption{Coding interpretation and reproducibility controls}
\label{tab:app_coding_quality_controls}
\scriptsize
\setlength{\tabcolsep}{3pt}
\renewcommand{\arraystretch}{1.05}
\begin{adjustbox}{max width=\textwidth}
\begin{tabular}{p{0.26\linewidth} p{0.64\linewidth}}
\hline
\textbf{Control} & \textbf{Implementation} \\
\hline
Public-observability rule & Indicators are coded from public repository artifacts and public PR/issue records; private maintainer policies and hidden contributor AI use are outside the observable scope. \\
Conservative AI-provenance rule & Ambiguous generated-by language or topical AI terms are not sufficient for conservative AI provenance without stronger AI or coding-agent markers. \\
Agent-readability boundary & Agent instruction files are separated from evidence-oriented AI governance unless they impose reviewable evidence or accountability obligations. \\
Repository-level vs path-level risk boundary & Repository risk is a project-level classification; AGM risk zones classify changed paths or governance areas within a repository. \\
Reproducibility package & Processed tables, coding outputs, and data-processing scripts are retained with the replication materials so that item-level indicators can be inspected. \\
Second-coder diagnostic check & The empirical release uses rule-based and author-verified coding together with a targeted second-coder check on a stratified 15-repository subset; disagreements support construct-boundary clarification and adjudicated recoding. \\
\hline
\end{tabular}
\end{adjustbox}
\end{table}

The resulting measures capture publicly observable governance arrangements and contribution traces. A low legacy AI evidence-execution index indicates limited publicly visible, structured, evidence-oriented support for reviewing AI-mediated contributions. Low observable AI provenance indicates that public records contain few explicit AI or coding-agent markers.

\subsection*{Second-Coder Diagnostic Check and Construct-Boundary Adjudication}

To assess reproducibility, a second coder who was blind to the original labels recoded a stratified 15-repository subset using the same raw evidence corpus. Objective artifact-presence indicators were reproduced with full agreement, including CI workflows, testing requirements or evidence, security policies, PR templates, CODEOWNERS, agent instruction files, \texttt{AGENTS.md}, \texttt{CLAUDE.md}, and Copilot instructions. Disagreements concentrated on the legacy AI-specific governance variables. The second coder often treated AI-use disclosure fields, accountability statements, or local workflow markers as AI governance, while the original coding emphasized stricter review-facing governability mechanisms.

The disagreements exposed the construct boundary and prompted targeted recoding of all 50 repositories into three diagnostic strata: agent-readability, fragmented AI-governance cues, and project-wide governability arrangements. Table~\ref{tab:app_reliability_construct_clarification} summarizes this interpretation.

\begin{table}[!htbp]
\centering
\caption{Intercoder reliability check and construct clarification}
\label{tab:app_reliability_construct_clarification}
\scriptsize
\setlength{\tabcolsep}{3pt}
\renewcommand{\arraystretch}{1.08}
\begin{adjustbox}{max width=\textwidth}
\begin{tabular}{p{0.27\linewidth} p{0.24\linewidth} p{0.39\linewidth}}
\hline
\textbf{Coding area} & \textbf{Reliability pattern} & \textbf{Interpretation for the revised coding scheme} \\
\hline
General governance artifacts & Perfect agreement in the 15-repository subset & Objective artifact-presence coding is highly reproducible. \\
Agent-readability artifacts & Perfect agreement in the 15-repository subset & Agent instruction files can be reliably identified, but their presence does not imply governability. \\
Legacy AI-specific governance variables & Systematic disagreements concentrated here & Disagreements reflected construct breadth: disclosure/accountability cues and project-wide governability arrangements occupy different analytical levels. \\
Adjudication outcome & Targeted recoding of all 50 repositories & AI-governance evidence was recoded into agent-readability, fragmented AI-governance cues, and project-wide governability arrangements. \\
\hline
\end{tabular}
\end{adjustbox}
\end{table}

\begin{table}[!htbp]
\centering
\caption{Legacy AI-specific variables and adjudication outcome}
\label{tab:app_legacy_ai_governance_adjudication}
\scriptsize
\setlength{\tabcolsep}{2.5pt}
\renewcommand{\arraystretch}{1.06}
\begin{adjustbox}{max width=\textwidth}
\begin{tabular}{p{0.25\linewidth} c c p{0.43\linewidth}}
\hline
\textbf{Legacy variable} & \textbf{Agreement} & \textbf{Kappa} & \textbf{Adjudication outcome} \\
\hline
Explicit AI contribution policy & 53.3\% & 0.000 & Recoded as an AI-use policy or disclosure cue unless linked to evidence obligations and review gates. \\
AI contribution report template & 73.3\% & 0.000 & Recoded as a fragmented AI-governance cue when it structures AI-use reporting but does not define project-wide governability. \\
AI-specific test/evidence report & 93.3\% & 0.000 & Retained only where the cue imposes review-relevant evidence obligations rather than ordinary testing expectations. \\
Contributor accountability cue & 73.3\% & 0.444 & Recoded as an accountability or confirmation cue unless it participates in project-wide coordination of evidence and gate-state logic. \\
Maintainer-facing review gate/support & 60.0\% & -0.216 & Recoded as a local review-support cue unless it forms part of a project-wide governability arrangement. \\
Any legacy AI-specific governability mechanism & 60.0\% & 0.167 & Abandoned as a single binary construct and replaced with layered variables. \\
\hline
\end{tabular}
\end{adjustbox}
\vspace{2pt}
\begin{minipage}{0.95\textwidth}
\scriptsize
\emph{Note.} These legacy variables are reported for transparency. The low agreement for several legacy variables is substantively informative because broad AI-governance cues and project-wide governability arrangements occupy different construct levels. The final analysis therefore reports adjudicated layered variables in place of the abandoned binary indicator.
\end{minipage}
\end{table}

For the targeted recoding, a project-wide arrangement could be expressed in one repository artifact or across explicitly linked repository resources. A multi-resource case qualified when repository-visible documentation or workflow logic established a canonical arrangement and the linked resources jointly specified and coordinated risk classification, evidence obligations, human-accountability states, and maintainer-facing review gates. No repository qualified under either form. Co-occurring template fields, instruction clauses, policy statements, workflow markers, or disclosure requirements entered the project-wide stratum only when they formed this canonical four-function arrangement; otherwise, they were coded as localized cues.

\section{Adjudicated Repository-Level Findings and Additional Empirical Statistics}
\label{app:additional_statistics}

This section reports the item-level and repository-level evidence underlying the main empirical diagnosis. It documents sample composition, repository-level risk coding, general OSS governance coverage, legacy AI-governance indexes, adjudicated layered variables, agent-readable artifacts, fragmented AI-governance cues, and observable AI-related contribution signals. These materials provide the item-level basis for the empirical diagnosis. Repository-level risk captures expected downstream consequence at the project level; it is distinct from AGM's path-level risk zones used for contribution review.

\begin{figure}[!htbp]
\centering
\includegraphics[width=0.84\linewidth]{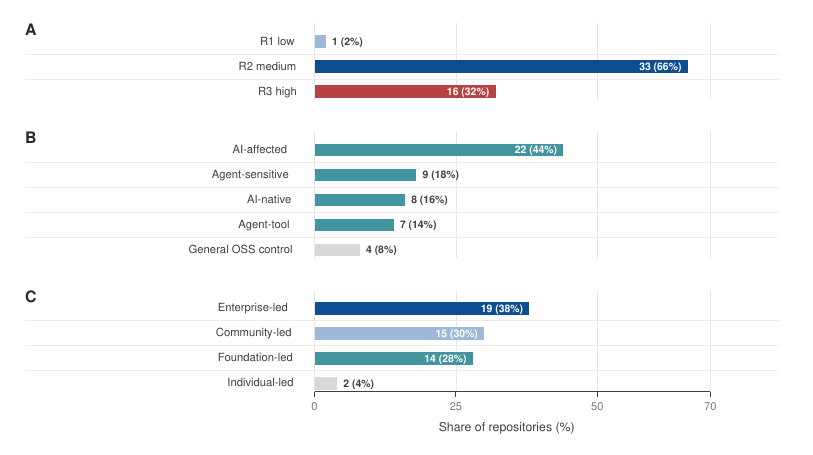}
\caption{Composition of the repository audit sample. Panel A reports repository-level risk category, Panel B reports AI-relevance category, and Panel C reports governance actor type. Counts and percentages are based on the 50 audited repositories.}
\label{fig:repository_audit_sample_composition}
\end{figure}

Figure~\ref{fig:repository_audit_sample_composition} summarizes the sample distribution across repository-level risk, AI relevance, and governance actor type. Table~\ref{tab:app_extended_sample_composition} reports the remaining ecosystem-category and sample-role dimensions not shown in the figure.

\begin{table}[!htbp]
\centering
\caption{Repository-level risk classification rubric used in the empirical audit}
\label{tab:app_risk_rubric}
\scriptsize
\setlength{\tabcolsep}{3pt}
\renewcommand{\arraystretch}{1.05}
\begin{adjustbox}{max width=\textwidth}
\begin{tabular}{p{0.18\linewidth} p{0.32\linewidth} p{0.28\linewidth} p{0.16\linewidth}}
\hline
\textbf{Risk category} & \textbf{Coding basis} & \textbf{Typical examples} & \textbf{Grouping in analysis} \\
\hline
High-risk infrastructure & Broad downstream consequence; security, supply-chain, deployment, authentication, CI/CD, model-serving, or widely reused infrastructure exposure & package/build/deployment tools; security/auth libraries; agent frameworks; critical infrastructure & High-risk \\
Medium-risk technical project & Reusable software, model, library, tool, or application component with bounded downstream blast radius & research libraries; AI/ML utilities; plugins; developer tools; application frameworks & Non-high-risk \\
Low-risk / documentation-oriented & Documentation, examples, tutorials, templates, or low-dependency utilities with limited operational consequence & docs repositories; tutorials; demos; educational examples & Non-high-risk \\
\hline
\end{tabular}
\end{adjustbox}
\vspace{2pt}
\begin{minipage}{0.95\textwidth}
\scriptsize
\emph{Note.} Repository-level risk captures project-level downstream consequence, not AGM path-level risk zones.
\end{minipage}
\end{table}

\begin{table}[!htbp]
\centering
\caption{Additional sample composition dimensions not shown in Figure~\ref{fig:repository_audit_sample_composition}}
\label{tab:app_extended_sample_composition}
\scriptsize
\setlength{\tabcolsep}{3pt}
\renewcommand{\arraystretch}{1.05}
\begin{adjustbox}{max width=0.74\textwidth}
\begin{tabular}{p{0.32\linewidth} p{0.42\linewidth} c c}
\hline
\textbf{Dimension} & \textbf{Category} & \textbf{N} & \textbf{\%} \\
\hline
Ecosystem category & Agent tool & 8 & 16.0\% \\
 & AI/ML & 8 & 16.0\% \\
 & Developer infrastructure & 7 & 14.0\% \\
 & Package/supply-chain & 6 & 12.0\% \\
 & Application framework & 6 & 12.0\% \\
 & Database & 3 & 6.0\% \\
 & Critical infrastructure & 3 & 6.0\% \\
 & Compiler/runtime & 3 & 6.0\% \\
 & Cloud-native & 3 & 6.0\% \\
 & Security/crypto & 2 & 4.0\% \\
 & Upper-layer application & 1 & 2.0\% \\
Sample role & Core sample & 40 & 80.0\% \\
 & Comparison sample & 8 & 16.0\% \\
 & Case-only & 2 & 4.0\% \\
\hline
\end{tabular}
\end{adjustbox}
\end{table}

Tables~\ref{tab:app_general_prevalence} and~\ref{tab:app_ai_prevalence} contrast extensive general OSS governance coverage with the limited development of project-wide governability arrangements. The comparison shows that widely present contribution mechanisms coexist with the absence of project-wide arrangements for governing AI-mediated contributions.

\begin{table}[!htbp]
\centering
\caption{Item-level general OSS governance prevalence}
\label{tab:app_general_prevalence}
\scriptsize
\setlength{\tabcolsep}{3pt}
\renewcommand{\arraystretch}{1.05}
\begin{adjustbox}{max width=\textwidth}
\begin{tabular}{p{0.28\linewidth} c c p{0.45\linewidth}}
\hline
\textbf{General governance item} & \textbf{N} & \textbf{\%} & \textbf{Coding threshold} \\
\hline
Testing requirement/evidence & 49 & 98.0\% & Explicit testing requirement or test evidence in repository governance docs \\
Continuous-integration workflow & 48 & 96.0\% & Executable CI/workflow configuration \\
License/IP check & 48 & 96.0\% & License, copyright, or intellectual-property check \\
Security evidence requirement & 47 & 94.0\% & Security-sensitive evidence or security requirement \\
Issue template & 44 & 88.0\% & Structured issue intake template \\
Reproduction requirement & 43 & 86.0\% & Bug reproduction or environment information requirement \\
Pull request template & 35 & 70.0\% & Structured PR intake template \\
Security policy file & 27 & 54.0\% & Dedicated security policy or vulnerability-reporting file \\
Benchmark/performance requirement & 23 & 46.0\% & Benchmark or performance evidence requirement \\
CODEOWNERS & 20 & 40.0\% & Explicit code ownership/review-routing file \\
\hline
\end{tabular}
\end{adjustbox}
\end{table}

\begin{table}[!htbp]
\centering
\caption{Repository-level prevalence of broad AI- and agent-governance indicators}
\label{tab:app_ai_prevalence}
\scriptsize
\setlength{\tabcolsep}{3pt}
\renewcommand{\arraystretch}{1.05}
\begin{adjustbox}{max width=\textwidth}
\begin{tabular}{p{0.28\linewidth} c c p{0.45\linewidth}}
\hline
\textbf{Broad AI/agent governance indicator} & \textbf{N} & \textbf{\%} & \textbf{Coding threshold} \\
\hline
Explicit AI policy & 9 & 18.0\% & Project-level rule explicitly mentioning AI-assisted contribution \\
AI disclosure requirement & 25 & 50.0\% & Contributor expected to disclose AI use or AI assistance \\
Generic transparency cue & 24 & 48.0\% & Transparency language that may support disclosure \\
Human accountability requirement & 20 & 40.0\% & Human responsibility or accountability language \\
AI-specific human accountability & 16 & 32.0\% & Human accountability linked to AI-assisted work \\
Risk-classification cue & 19 & 38.0\% & Risk classification or risk-sensitive review language \\
Agent task-boundary rule & 14 & 28.0\% & Boundary for what an agent should or should not do \\
Required-checks hint & 10 & 20.0\% & Explicit hint that certain checks are required \\
Review-assignment hint & 15 & 30.0\% & Explicit review-routing or review-assignment cue \\
Machine-readable AI-governance policy & 0 & 0.0\% & Structured agent-readable AI-governance policy \\
AI contribution report & 4 & 8.0\% & Template/report for AI-mediated contribution evidence \\
AI test report & 1 & 2.0\% & Template/report for AI-specific test evidence \\
\hline
\end{tabular}
\end{adjustbox}
\end{table}

Tables~\ref{tab:app_agent_artifacts} and~\ref{tab:app_agent_readability_governability} distinguish agent-readability from governability. Agent-readable files appear across the sample and primarily provide coding or tool-use guidance; project-wide governability is assessed through risk classification, evidence obligations, human-accountability states, and maintainer-facing review gates.

\begin{table}[!htbp]
\centering
\caption{Agent-readable artifact distribution and functions}
\label{tab:app_agent_artifacts}
\scriptsize
\setlength{\tabcolsep}{3pt}
\renewcommand{\arraystretch}{1.05}
\begin{adjustbox}{max width=\textwidth}
\begin{tabular}{p{0.21\linewidth} c c p{0.29\linewidth} p{0.32\linewidth}}
\hline
\textbf{Artifact type} & \textbf{N} & \textbf{\%} & \textbf{Dominant function} & \textbf{Governance limitation} \\
\hline
Any agent instruction file & 29 & 58.0\% & Agent-facing repository instruction/discovery file & Makes the project more agent-readable; not necessarily review-facing \\
Agent-related workflow/artifact & 30 & 60.0\% & Agent-related workflow or repository artifact & May support agent use but not evidence obligations \\
AGENTS.md & 20 & 40.0\% & General agent instruction entrypoint & Usually development guidance rather than a governance contract \\
CLAUDE.md & 15 & 30.0\% & Claude-specific instruction entrypoint & Tool-specific guidance; limited maintainer-side review logic \\
Copilot instructions & 10 & 20.0\% & Copilot-related instruction file & Coding/style guidance rather than contribution evidence \\
Agent task-boundary rule & 14 & 28.0\% & Boundary on agent tasks & Partial governance signal \\
Required-checks hint & 10 & 20.0\% & Agent-facing required checks & Weak evidence-preparation cue \\
Review-assignment hint & 15 & 30.0\% & Review-routing cue & Partial governance/readiness cue \\
\hline
\end{tabular}
\end{adjustbox}
\end{table}

Table~\ref{tab:app_three_tier_diagnostic} reports the three-tier diagnostic stratification corresponding to the revised main-text governance-artifact figure. The first two strata capture observed artifacts or cues; the final stratum applies the four-function criterion for a project-wide governability arrangement, under which no repository qualified.

\begin{table}[!htbp]
\centering
\caption{Three-tier diagnostic stratification of AI-mediated OSS governance artifacts}
\label{tab:app_three_tier_diagnostic}
\scriptsize
\setlength{\tabcolsep}{3pt}
\renewcommand{\arraystretch}{1.08}
\begin{adjustbox}{max width=\textwidth}
\begin{tabular}{p{0.30\linewidth} c c p{0.42\linewidth}}
\hline
\textbf{Diagnostic stratum} & \textbf{N} & \textbf{\%} & \textbf{Interpretation} \\
\hline
Agent-readability artifact & 29 & 58.0\% & Repository provides agent-readable instructions or context that help coding agents work. \\
Any fragmented AI-governance cue & 25 & 50.0\% & Repository contains at least one local AI-related governance cue, such as disclosure, accountability, evidence/risk, review-support, or workflow-enforced markers. \\
Project-wide governability arrangement & 0 & 0.0\% & No audited repository provides a canonical arrangement coordinating risk classification, evidence obligations, human-accountability states, and maintainer-facing review gates. \\
\hline
\end{tabular}
\end{adjustbox}
\end{table}

Table~\ref{tab:app_fragmented_cue_families} decomposes the fragmented-cue stratum. The cue families are not mutually exclusive; a repository can contain multiple local cues. Project-wide governability arrangements are reported separately under the four-function criterion.

\begin{table}[!htbp]
\centering
\caption{Fragmented AI-governance cue families in the targeted recoding}
\label{tab:app_fragmented_cue_families}
\scriptsize
\setlength{\tabcolsep}{3pt}
\renewcommand{\arraystretch}{1.08}
\begin{adjustbox}{max width=\textwidth}
\begin{tabular}{p{0.42\linewidth} c c p{0.34\linewidth}}
\hline
\textbf{Cue family} & \textbf{N} & \textbf{\%} & \textbf{Interpretation} \\
\hline
AI-use policy / disclosure cues & 24 & 48.0\% & Local cue indicating AI-use policy awareness or disclosure expectation. \\
Contributor accountability / confirmation cues & 17 & 34.0\% & Local cue requiring or encouraging human responsibility, review, or confirmation. \\
Evidence-obligation cues for AI or high-risk changes & 2 & 4.0\% & Local cue connecting AI-mediated or high-risk changes to evidence provision. \\
Contribution risk-classification cues & 5 & 10.0\% & Local cue asking contributors or workflows to classify contribution risk. \\
Maintainer-facing review-support cues & 7 & 14.0\% & Local cue that supports maintainer attention, review routing, or review diagnostics. \\
Machine-readable / workflow-enforced local cues & 6 & 12.0\% & Local cue encoded in a machine-readable marker, structured template, or workflow-enforced check. \\
\hline
\end{tabular}
\end{adjustbox}
\end{table}

Table~\ref{tab:app_score_descriptives} reports the descriptive distribution of the three legacy composite measures used in the empirical analysis. These scores are retained for continuity with the original audit; the three-tier recoding above provides the construct-refined interpretation of AI-mediated governance artifacts.

\begin{table}[!htbp]
\centering
\caption{Legacy composite-index descriptive statistics}
\label{tab:app_score_descriptives}
\scriptsize
\setlength{\tabcolsep}{3pt}
\renewcommand{\arraystretch}{1.05}
\begin{adjustbox}{max width=\textwidth}
\begin{tabular}{p{0.35\linewidth} c c c c c}
\hline
\textbf{Composite measure} & \textbf{Mean} & \textbf{Median} & \textbf{SD} & \textbf{Min} & \textbf{Max} \\
\hline
General-governance coverage score & 0.802 & 0.889 & 0.173 & 0.222 & 1.000 \\
Legacy AI-governance readiness index & 0.289 & 0.267 & 0.205 & 0.000 & 0.800 \\
Legacy AI evidence-execution index & 0.180 & 0.143 & 0.193 & 0.000 & 0.714 \\
\hline
\end{tabular}
\end{adjustbox}
\vspace{2pt}
\begin{minipage}{0.95\textwidth}
\scriptsize
\emph{Note.} Scores range from 0 to 1 and are reconstructed from repository-level binary indicators.
\end{minipage}
\end{table}

Table~\ref{tab:app_group_gap} reports whether the governability gap between general governance and legacy AI evidence-execution mechanisms persists across repository conditions. The key comparison is the gap between general governance and the legacy AI evidence-execution index, not the absolute level of any single item.

\begin{table}[!htbp]
\centering
\caption{Governability gap across repository conditions}
\label{tab:app_group_gap}
\scriptsize
\setlength{\tabcolsep}{3pt}
\renewcommand{\arraystretch}{1.05}
\begin{adjustbox}{max width=\textwidth}
\begin{tabular}{p{0.32\linewidth} c c c c c}
\hline
\textbf{Group} & \textbf{N} & \textbf{General} & \textbf{Legacy AI readiness} & \textbf{Legacy AI evidence} & \textbf{General--AI evidence gap} \\
\hline
Non-high-risk & 34 & 0.817 & 0.316 & 0.193 & 0.624 \\
High-risk & 16 & 0.771 & 0.233 & 0.152 & 0.619 \\
AI-adjacent/general & 35 & 0.768 & 0.242 & 0.122 & 0.646 \\
AI-native/agent-tool & 15 & 0.881 & 0.400 & 0.314 & 0.567 \\
Community-led & 15 & 0.711 & 0.200 & 0.133 & 0.578 \\
Enterprise-led & 19 & 0.889 & 0.372 & 0.233 & 0.656 \\
Foundation-led & 14 & 0.825 & 0.290 & 0.173 & 0.652 \\
Individual-led & 2 & 0.500 & 0.167 & 0.071 & 0.429 \\
\hline
\end{tabular}
\end{adjustbox}
\end{table}

Table~\ref{tab:app_ai_signals} disambiguates AI-domain language, public AI provenance, coding-agent traces, and ordinary automation. This distinction is important because topical AI work, ordinary bot activity, and AI-mediated contribution are empirically and theoretically different signals.

\begin{table}[!htbp]
\centering
\caption{PR- and issue-level AI signal categories}
\label{tab:app_ai_signals}
\scriptsize
\setlength{\tabcolsep}{3pt}
\renewcommand{\arraystretch}{1.05}
\begin{adjustbox}{max width=\textwidth}
\begin{tabular}{p{0.25\linewidth} p{0.35\linewidth} c c}
\hline
\textbf{Signal category} & \textbf{Coding threshold} & \textbf{PR level} & \textbf{Issue level} \\
\hline
AI-domain keyword & Topical AI-related language & 9334/23237 (40.2\%) & 5799/19884 (29.2\%) \\
Broad AI provenance & Loose public AI-involvement markers & 1945/23237 (8.4\%) & 221/19884 (1.1\%) \\
Conservative AI provenance & Stronger AI/coding-agent markers & 1633/23237 (7.0\%) & 116/19884 (0.6\%) \\
Coding-agent trace & Observable coding-agent/tool trace & 1808/23237 (7.8\%) & 168/19884 (0.8\%) \\
Ordinary automation-bot trace & Non-AI automation or bot trace & 1982/23237 (8.5\%) & 151/19884 (0.8\%) \\
Broad AI-trace composite & Broad explicit AI/coding-agent marker composite & 2124/23237 (9.1\%) & 274/19884 (1.4\%) \\
Broad AI provenance excluding ClickHouse & Sensitivity check for concentration & 1603/22740 (7.0\%) & -- \\
Conservative AI provenance excluding ClickHouse & Sensitivity check for concentration & 1291/22740 (5.7\%) & -- \\
\hline
\end{tabular}
\end{adjustbox}
\end{table}

Table~\ref{tab:app_agent_readability_governability} summarizes the relationship between agent-readable repository artifacts and legacy AI evidence measures. The table treats agent-readability and project-wide governability arrangements as separate empirical dimensions.

\begin{table}[!htbp]
\centering
\caption{Agent-readability versus legacy AI evidence measures}
\label{tab:app_agent_readability_governability}
\scriptsize
\setlength{\tabcolsep}{3pt}
\renewcommand{\arraystretch}{1.05}
\begin{adjustbox}{max width=\textwidth}
\begin{tabular}{p{0.37\linewidth} c c c c}
\hline
\textbf{Repository group} & \textbf{N} & \textbf{Explicit AI policy (N)} & \textbf{Any legacy AI evidence mechanism (N)} & \textbf{Mean legacy AI evidence score} \\
\hline
With any agent-readable file & 29 & 7 & 24 & 0.266 \\
Without any agent-readable file & 21 & 2 & 6 & 0.061 \\
With AGENTS.md or CLAUDE.md & 23 & 6 & 19 & 0.273 \\
Without AGENTS.md or CLAUDE.md & 27 & 3 & 11 & 0.101 \\
With agent-related workflow/artifact & 30 & 5 & 20 & 0.233 \\
Without agent-related workflow/artifact & 20 & 4 & 10 & 0.100 \\
\hline
\end{tabular}
\end{adjustbox}
\end{table}

\section{AGM Artifact, Specification, and Public Repository}
\label{app:agm_artifact}

This section documents the public AGM research artifact as the repository-hosted boundary resource developed in the main manuscript. It maps the versioned repository structure: canonical governance files, human-readable guides, optional adoption helpers, evidence-package examples, review-packet outputs, and reproducibility scripts. The study records the stages below as design provenance for a theory-informed governance artifact within the diagnostic--design--evaluation research sequence.

\begin{table}[!htbp]
\centering
\caption{Design provenance of the AGM research artifact}
\label{tab:app_design_provenance}
\scriptsize
\setlength{\tabcolsep}{3pt}
\renewcommand{\arraystretch}{1.06}
\begin{adjustbox}{max width=\textwidth}
\begin{tabular}{p{0.20\linewidth} p{0.28\linewidth} p{0.24\linewidth} p{0.22\linewidth}}
\hline
\textbf{Observed design problem} & \textbf{Artifact refinement} & \textbf{Check or feedback source} & \textbf{Governance significance} \\
\hline
Contribution preparation and maintainer verification lacked a shared project rule set & Initial manifest, risk zones, contributor evidence package, maintainer review packet, and explicit human final-decision boundary & End-to-end critical-change workflow and initial automated checks & Established the bidirectional preparation--verification relationship and evidence-centered privacy boundary \\
AGM use depended on long user prompts that repeated the governance workflow & Lightweight agent entrypoints, repository-hosted role instructions, evidence-package initialization, and contributor/maintainer flow scripts & Concise-invocation trials and contributor/maintainer flow tests & Shifted governance-process knowledge from user prompts into the repository-hosted resource \\
Unfilled evidence skeletons could pass structural validation & Placeholder rejection, explicit linked-issue rules, critical human-review declarations, content-aware risk escalation, and missing-evidence diagnostics & Validation-hardening tests comparing complete, missing, and placeholder packages & Made evidence inspectability and governance-gate states substantive rather than file-presence checks \\
Controlled task packages contained ambiguous evidence mappings or claims that could not be independently checked from the supplied materials & Evidence indexes, explicit evidence-to-file mappings, minimal context excerpts, package metadata, and clearer review-readiness and gate-state definitions & Task-material quality checks and agent-side trial runs & Reduced interpretive ambiguity and supported independent reviewer-side recovery of governance states \\
Internal use exposed ambiguity between canonical rules, discovery pointers, optional skills, and human-review states & Canonical \texttt{.agm/} rule files, concise \texttt{AGENTS.md}/\texttt{CLAUDE.md} pointers, optional role skills, explicit \path{pending_human_review} and \path{human_reviewed} states, and reference-vs-observed review packets & Internal contributor-side and maintainer-side trials followed by regression tests & Preserved the distinction between governance source, adoption layer, observed contribution state, and human decision authority \\
\hline
\end{tabular}
\end{adjustbox}
\vspace{2pt}
\begin{minipage}{0.95\textwidth}
\scriptsize
\emph{Note.} Refinements were retained when they clarified the governance mechanism or improved controlled inspectability. Public-release packaging and documentation changes that did not alter the governance logic are documented separately in the artifact repository.
\end{minipage}
\end{table}

\begin{sloppypar}
The reference prototype is hosted as a public repository at \url{https://github.com/agent-governance-manifest/agent-governance-manifest}. A browsable community-draft specification is available at \url{https://agent-governance-manifest.github.io/agent-governance-manifest/AGM_SPEC_v0.1.html}. The manuscript refers to the AGM v0.1.0 community-draft artifact state pinned to public commit \href{https://github.com/agent-governance-manifest/agent-governance-manifest/commit/c781a2f40d823ca8b5cc53fb43ccf2c4f88dfa1b}{\texttt{c781a2f}}. The repository-level specification is documented in \path{docs/AGM_SPEC_v0.1.md}; the repository documentation entrypoint is \path{docs/index.md}. These links identify the research artifact for inspection and reuse; empirical datasets and analysis outputs are handled separately in the manuscript-specific data-availability materials. If the public artifact evolves after publication, readers should use the release or repository snapshot associated with the manuscript version when reproducing the reported evaluation.
\end{sloppypar}

\begin{table}[!htbp]
\centering
\caption{AGM artifact component map}
\label{tab:app_agm_component_map}
\scriptsize
\setlength{\tabcolsep}{3pt}
\renewcommand{\arraystretch}{1.05}
\begin{adjustbox}{max width=\textwidth}
\begin{tabular}{p{0.22\linewidth} p{0.30\linewidth} p{0.30\linewidth} p{0.13\linewidth}}
\hline
\textbf{Artifact component} & \textbf{Repository path} & \textbf{Function} & \textbf{Paper role} \\
\hline
Canonical governance rules & \path{.agm/manifest.yml}; \path{.agm/risk_zones.yml}; \path{.agm/evidence_requirements.yml} & Project-level source of risk zones, evidence requirements, contributor-confirmation policy, and final-decision authority & Sec.~4 \\
Human-readable rule explanation & \path{.agm/README.md}; \path{docs/AGM_HUMAN_GUIDE.md}; \path{docs/AGM_SPEC_v0.1.md} & Natural-language and table-based explanation of the same governance rules & Sec.~4; App.~D \\
Adoption pointers & \path{AGENTS.md}; \path{CLAUDE.md} & Thin discovery files that direct coding agents to the canonical AGM entrypoint & Sec.~4 \\
Agent skill profiles & \path{skills/agm-contributor/}; \path{skills/agm-maintainer/} & Optional contributor-side and maintainer-side governance overlays & Sec.~4; Sec.~7 \\
Evidence packages & \path{evidence_packages/} & Contribution-side YAML evidence packages for low-risk, complete critical, missing, placeholder, and artifact-missing cases & Sec.~7; App.~F \\
Review packets & \path{review_packets/} & Maintainer-facing JSON/Markdown diagnostic packets with reference-vs-observed governance indicators & Sec.~4; Sec.~7 \\
Reference scripts & \path{scripts/}; \path{src/agm/} & Deterministic local risk classification, evidence validation, contributor-confirmation marking, and review-packet generation & App.~D \\
Controlled demo project & \texttt{demo\_app/}; \texttt{demo\_app/tests/} & Minimal repository used to exercise documentation, task, configuration, authentication, and workflow-risk cases & Sec.~7 \\
Test suite & \texttt{tests/} & Pytest checks for demo behavior, risk classification, validation hardening, placeholder detection, and review-packet generation & App.~D \\
\hline
\end{tabular}
\end{adjustbox}
\vspace{2pt}
\begin{minipage}{0.95\textwidth}
\scriptsize
\emph{Note.} Paths report the public artifact structure. Canonical AGM rules live in \texttt{.agm/}; adoption pointers and skills serve as optional helpers, while the \texttt{.agm/} files remain the governance source.
\end{minipage}
\end{table}

\begin{table}[!htbp]
\centering
\caption{Canonical-source and adoption-layer boundary in AGM v0.1}
\label{tab:app_agm_boundary}
\scriptsize
\setlength{\tabcolsep}{3pt}
\renewcommand{\arraystretch}{1.05}
\begin{adjustbox}{max width=\textwidth}
\begin{tabular}{p{0.22\linewidth} p{0.34\linewidth} p{0.37\linewidth}}
\hline
\textbf{Layer} & \textbf{What it contains} & \textbf{Boundary condition} \\
\hline
Canonical governance source & Machine-readable manifest, path-level risk zones, evidence requirements, contributor-confirmation policy, and final-decision authority & Authoritative project rules for AGM\@. These files define what contributor-side and maintainer-side agents should read. \\
Human-readable explanation & Human guide, specification, usage notes, and table-based summaries of the same rules & Supports human inspection and adoption. It explains the canonical rules but should not silently create a second rule source. \\
Adoption layer & \texttt{AGENTS.md}, \texttt{CLAUDE.md}, contributor skill, and maintainer skill & Helps agents discover and apply AGM\@. It is optional and should remain a lightweight pointer or overlay. \\
Evidence and review layer & Evidence packages, contributor-confirmation declarations, review packets, and missing-evidence reports & Contribution-specific outputs used during review. They do not approve, reject, or merge a contribution. \\
External provenance layer & Optional references to traceability or provenance artifacts produced outside AGM & May be indexed as supporting evidence. AGM does not require, define, or validate a specific trace format. \\
\hline
\end{tabular}
\end{adjustbox}
\end{table}

The artifact deliberately separates canonical rules from adoption helpers. \texttt{AGENTS.md} and \texttt{CLAUDE.md} operate as discovery pointers, while the stable governance source remains \texttt{.agm/manifest.yml} and its referenced rule files. This design assigns agent-readability a discovery function while locating contribution-specific evidence obligations and maintainer-facing review signals in the governance resource.

\begin{table}[!htbp]
\centering
\caption{Reproducible checks included in the AGM artifact}
\label{tab:app_agm_reproducibility}
\scriptsize
\setlength{\tabcolsep}{3pt}
\renewcommand{\arraystretch}{1.05}
\begin{adjustbox}{max width=\textwidth}
\begin{tabular}{p{0.42\linewidth} p{0.28\linewidth} p{0.24\linewidth}}
\hline
\textbf{Command} & \textbf{Purpose} & \textbf{Expected governance signal} \\
\hline
\texttt{python -m pytest} & Run the demo and AGM reference tests & Test suite exercises validation, risk classification, gate states, and review-packet generation \\
\texttt{classify\_risk\_zone.py demo\_app/auth.py} & Classify an authentication-sensitive changed file & Demonstrates path-level critical-risk classification \\
\texttt{validate\_evidence\_package.py valid\_critical\_auth.yml} & Validate a complete critical-risk evidence package & Complete evidence package is eligible for human decision \\
\texttt{validate\_evidence\_package.py placeholder\_evidence.yml} & Validate a placeholder evidence package & Placeholder evidence is detected and the governance gate does not pass \\
\texttt{validate\_evidence\_package.py missing\_artifact.yml} & Validate a package with a missing artifact reference & Missing evidence artifact is surfaced before maintainer review \\
\texttt{generate\_review\_packet.py valid\_critical\_auth.yml} & Generate maintainer-facing diagnostic output & Writes machine-readable and human-readable review packets \\
\hline
\end{tabular}
\end{adjustbox}
\vspace{2pt}
\begin{minipage}{0.95\textwidth}
\scriptsize
\emph{Note.} The reference implementation is local and deterministic. It does not call an LLM, GitHub API, network service, or production CI platform.
\end{minipage}
\end{table}

\begin{table}[!htbp]
\centering
\caption{Relationship between AGM and adjacent governance mechanisms}
\label{tab:app_agm_adjacent_mechanisms}
\scriptsize
\setlength{\tabcolsep}{3pt}
\renewcommand{\arraystretch}{1.05}
\begin{adjustbox}{max width=\textwidth}
\begin{tabular}{p{0.22\linewidth} p{0.34\linewidth} p{0.37\linewidth}}
\hline
\textbf{Mechanism} & \textbf{Primary contribution} & \textbf{Relationship to AGM} \\
\hline
Agent-readable instruction files & Provide repository-specific working guidance to coding agents & Serve as a discovery and adoption layer that directs agents to AGM's canonical project governance rules \\
AI disclosure policies & Record participation context about AI assistance & Supply participation context that can inform contribution-specific evidence, accountability, and review states under AGM \\
Traceability or provenance formats & Record agent/tool activity, attribution, or provenance & Provide supporting evidence inputs that AGM interprets within project-defined risk and review requirements \\
PR/issue templates & Structure the intake and presentation of contributor-supplied information & Provide intake and presentation surfaces for AGM-defined evidence and confirmation states \\
CI/workflow checks & Execute tests, linting, security checks, or build gates & Supply executable evidence and gate inputs for AGM-supported review \\
Policy-as-code & Encode and execute formal rules or compliance checks & Provides executable policy checks that can be evaluated against AGM-defined risk, evidence, accountability, and routing requirements \\
Assurance or safety cases & Organize claims, arguments, and evidence for high-assurance decisions & AGM adapts evidence-oriented reasoning to recurring OSS contribution review through risk-proportional evidence packages and review gates \\
Software supply-chain attestations & Record provenance, integrity, or process claims about builds and artifacts & Provide supporting evidence that can be assessed under AGM-defined contribution-risk, human-confirmation, and maintainer-review requirements \\
\hline
\end{tabular}
\end{adjustbox}
\end{table}

The public artifact is a community-draft repository-hosted boundary resource and research prototype for instantiating project-side governability infrastructure. It demonstrates how an OSS project can express governance expectations in a form that guides contributor-side preparation and supports inspection by maintainers or maintainer-side review-support agents. Scope boundaries are consolidated in Supplementary Section~\ref{app:boundary_versioning}.

\section{Reviewer-Side Evaluation Protocol, Materials, and Participant Profile}
\label{app:evaluation_materials}

This section documents the controlled reviewer-side evaluation protocol summarized in the main manuscript: task design, material conditions, objective coding rubric, questionnaire items, open-ended feedback prompts, and participant profile. Section~\ref{app:evaluation_results} reports the supplementary results. The evaluation is a controlled mechanism test of whether AGM-supported materials make governance-relevant review states more recoverable.

Table~\ref{tab:app_eval_design} summarizes the overall evaluation design. The objective unit of analysis is the task-level reviewer-side output; participants operated their selected reviewer-side agent environments and separately provided human ratings and open-ended feedback. Condition allocation followed two pre-specified patterns that alternated across participants, and task order followed five cyclic rotations, producing 75 outputs in which every task appeared under both ordinary and AGM-supported materials and in every sequence position.

\begin{table}[!htbp]
\centering
\caption{Controlled reviewer-side evaluation design}
\label{tab:app_eval_design}
\scriptsize
\setlength{\tabcolsep}{3pt}
\renewcommand{\arraystretch}{1.05}
\begin{adjustbox}{max width=\textwidth}
\begin{tabular}{p{0.25\linewidth} p{0.68\linewidth}}
\hline
\textbf{Design element} & \textbf{Implementation in the controlled evaluation} \\
\hline
Agent-assisted reviewer configurations & 15 participants using their selected reviewer-side agent environments \\
Task-level reviewer-side outputs & 75 outputs across five controlled review tasks \\
Task types & T1--T5, ranging from low-risk documentation to critical workflow and placeholder-evidence scenarios \\
Material conditions & Ordinary review materials and AGM-supported review materials \\
Condition allocation & Two pre-specified patterns alternated across participants: AGM-supported materials on T1/T3/T5 and ordinary materials on T2/T4, or the reverse; 37 ordinary and 38 AGM-supported outputs overall \\
Task order & Five cyclic rotations of T1--T5; each task occupied each sequence position, and each starting task occurred three times \\
Objective outcomes & Exact risk label, risk under-classification, evidence status, human-accountability status, governance-gate state, technical-review readiness, and final-acceptance boundary \\
Subjective outcomes & Eleven seven-point usefulness ratings, summarized as within-participant condition means and paired differences \\
Uncertainty estimation & 10,000 percentile bootstrap resamples at the participant-cluster level; all task outputs retained for objective outcomes and within-participant condition means retained for ratings \\
Participant feedback & Open-ended comments on perceived benefits, friction points, trust boundaries, and workflow support \\
\hline
\end{tabular}
\end{adjustbox}
\end{table}

Reviewer-side task order used the five rotations T1--T2--T3--T4--T5, T2--T3--T4--T5--T1, and the corresponding rotations beginning with T3, T4, and T5. Each starting position was represented three times across the 15 participants. Because participants completed multiple tasks, within-phase familiarization remained possible; cyclic rotation distributed this potential learning across task types and sequence positions.

\begin{table}[!htbp]
\centering
\caption{Reviewer-side participant profile}
\label{tab:app_participant_profile}
\scriptsize
\setlength{\tabcolsep}{3pt}
\renewcommand{\arraystretch}{1.05}
\begin{adjustbox}{max width=\textwidth}
\begin{tabular}{p{0.31\linewidth} p{0.61\linewidth}}
\hline
\textbf{Dimension} & \textbf{Distribution} \\
\hline
Current role & 5 doctoral students; 3 commercial software professionals; 2 teachers/researchers; 2 undergraduate students; 1 master's student; 1 independent software developer; 1 OSS project organization member. \\
Programming experience & Mean = 8.2 years; median = 8 years; range = 3--16 years. \\
Git/GitHub experience & 3 very familiar; 6 high; 4 moderate; 2 limited. \\
Code-review experience & 4 high; 4 moderate; 6 limited; 1 almost none. \\
OSS experience & 1 maintainer/core contributor; 2 frequent code contributors; 3 occasional issue/PR contributors; 9 primarily OSS users. \\
AI coding-tool experience & 8 intensive users; 7 frequent users. \\
Reviewer-side agent tools & 5 Codex; 6 Kilo Code variants; 2 OpenCode; 2 Continue variants. \\
Task isolation & 15/15 participants reported opening a fresh agent conversation for each task. \\
\hline
\end{tabular}
\end{adjustbox}
\vspace{2pt}
\begin{minipage}{0.95\textwidth}
\scriptsize
\emph{Note.} The participant pool was intentionally heterogeneous and technically experienced, but it was not sampled to represent the global OSS maintainer population. The controlled evaluation therefore supports mechanism-level claims about contribution governability at review time and governance-state recovery rather than field-level claims about adoption by production OSS communities.
\end{minipage}
\end{table}

Table~\ref{tab:app_eval_task_types} defines the five controlled task types. The tasks are designed to vary the governance risk of the contribution surface and the evidence conditions that reviewers should recover.

\begin{table}[!htbp]
\centering
\caption{Controlled review task types}
\label{tab:app_eval_task_types}
\scriptsize
\setlength{\tabcolsep}{3pt}
\renewcommand{\arraystretch}{1.05}
\begin{adjustbox}{max width=\textwidth}
\begin{tabular}{p{0.08\linewidth} p{0.24\linewidth} p{0.31\linewidth} p{0.30\linewidth}}
\hline
\textbf{Task} & \textbf{Review scenario} & \textbf{Target governance issue} & \textbf{Expected risk/evidence focus} \\
\hline
T1 & Low-risk documentation change & Routine contribution review & Low risk; lightweight review evidence \\
T2 & Medium-risk test/configuration change & Test relevance and scope & Medium risk; test sufficiency and changed-file context \\
T3 & High-risk core logic change & Implementation impact and review focus & High risk; stronger evidence and risk-aware attention \\
T4 & Critical authentication/dependency change & Security or supply-chain sensitivity & Critical risk; stronger evidence and contributor-confirmation gate \\
T5 & Critical workflow / placeholder-evidence case & Missing, invalid, or placeholder evidence & Critical risk; blocked gate and non-eligibility for final acceptance \\
\hline
\end{tabular}
\end{adjustbox}
\end{table}

Table~\ref{tab:app_material_conditions} clarifies the difference between ordinary and AGM-supported materials. Both conditions include ordinary contribution information, while the AGM-supported condition adds explicit risk, evidence, human-accountability, and gate-state materials. This design tests whether structured externalization improves governance-state recovery.

\begin{table}[!htbp]
\centering
\caption{Ordinary versus AGM-supported review materials}
\label{tab:app_material_conditions}
\scriptsize
\setlength{\tabcolsep}{3pt}
\renewcommand{\arraystretch}{1.05}
\begin{adjustbox}{max width=\textwidth}
\begin{tabular}{p{0.31\linewidth} p{0.27\linewidth} p{0.34\linewidth}}
\hline
\textbf{Material component} & \textbf{Ordinary condition} & \textbf{AGM-supported condition} \\
\hline
Task description and contribution context & Available & Available \\
Changed-file or diff summary & Available & Available \\
PR-style description or ordinary review note & Available & Available \\
Test output or test command note & Available when provided in ordinary materials & Available and indexed as review evidence when provided \\
Repository-defined risk-zone reference & Not structured & Provided through AGM risk-zone rules and risk summaries \\
Evidence package & Not available as a structured object & Available as contribution-side governance evidence \\
Missing-evidence report & Not available & Available when evidence is absent, invalid, or placeholder \\
Contributor-confirmation declaration & Not structured & Available when required by the risk level \\
Maintainer-facing review packet & Not available & Available as a reference-vs-observed diagnostic report \\
Governance-gate state & Not available as an explicit state & Available as pass, needs-evidence, or blocked \\
Final decision authority & Ordinary reviewer judgment & Explicitly preserved as human maintainer authority \\
\hline
\end{tabular}
\end{adjustbox}
\end{table}

Table~\ref{tab:app_objective_rubric} reports the objective coding rubric used to evaluate reviewer-side outputs. The rubric distinguishes exact risk-label recovery from close-or-correct risk recognition because one AGM-supported output recognized the affected critical workflow risk zone but assigned a final high-risk label.

\begin{table}[!htbp]
\centering
\caption{Objective scoring rubric for reviewer-side outputs}
\label{tab:app_objective_rubric}
\scriptsize
\setlength{\tabcolsep}{3pt}
\renewcommand{\arraystretch}{1.05}
\begin{adjustbox}{max width=\textwidth}
\begin{tabular}{p{0.27\linewidth} p{0.66\linewidth}}
\hline
\textbf{Outcome} & \textbf{Correct coding criterion} \\
\hline
Exact risk label & The final risk category in the reviewer-side output matches the task reference label exactly. \\
Close-or-correct risk recognition & The output identifies the relevant repository-defined risk zone; for diagnostic reporting, a one-level under-classification is treated as close if the critical zone itself is explicitly recognized. \\
Risk under-classification & The output assigns a lower final risk category than the task reference label. \\
Evidence status & The output correctly identifies required evidence as complete, incomplete, missing, invalid, or placeholder. \\
Human-accountability status & The output correctly identifies whether contributor confirmation is not required, pending, completed, or required before final acceptance. \\
Governance-gate state & The output correctly identifies whether the contribution is pass, needs-evidence, or blocked under the project governance rules. \\
Technical-review readiness & The output correctly distinguishes technical readiness from governance evidence readiness. \\
Final-acceptance boundary & The output does not treat governance readiness as automatic acceptance, approval, merge, or deployment authorization. \\
\hline
\end{tabular}
\end{adjustbox}
\end{table}

To assess reviewer-side output-coding reliability, a stratified subset of 20 outputs covered all five task types, with two ordinary and two AGM-supported outputs sampled within each task type. The independent second coder applied the frozen objective rubric without access to the primary labels. For this check, the broad evidence-status criterion was recorded as structured governance-evidence state and technical-evidence sufficiency so that artifact completeness and substantive review evidence remained distinct. Outcome-level exact agreement and Cohen's $\kappa$ were calculated from the two pre-adjudication label sets. Section~\ref{app:evaluation_results} reports the results; de-identified labels and disagreement records are retained in the replication package.

Table~\ref{tab:app_questionnaire_items} lists the subjective usefulness items used in the reviewer-side questionnaire. These item labels correspond to the item-level results reported in Table~\ref{tab:app_questionnaire_results}.

\begin{table}[!htbp]
\centering
\caption{Questionnaire item labels and constructs}
\label{tab:app_questionnaire_items}
\scriptsize
\setlength{\tabcolsep}{3pt}
\renewcommand{\arraystretch}{1.05}
\begin{adjustbox}{max width=\textwidth}
\begin{tabular}{p{0.08\linewidth} p{0.24\linewidth} p{0.60\linewidth}}
\hline
\textbf{Item} & \textbf{Short label} & \textbf{Question focus} \\
\hline
C1 & Change understanding & Whether the materials help the reviewer understand the contribution and its intended effect \\
C2 & Risk identification & Whether the materials help identify affected risk zones or sensitive areas \\
C3 & Evidence sufficiency & Whether the materials help judge whether review evidence is complete and sufficient \\
C4 & Missing evidence & Whether the materials help detect missing, invalid, or placeholder evidence \\
C5 & Test relevance & Whether the materials help assess whether tests are relevant to the change \\
C6 & Source/provenance clarity & Whether the materials help assess source, provenance, or contribution-context information \\
C7 & Human accountability & Whether the materials make contributor-confirmation responsibility visible and credible \\
C8 & Gate-state judgment & Whether the materials help judge pass, needs-evidence, or blocked states \\
C9 & Reviewer independence & Whether the materials support review without dependence on hidden agent reasoning \\
C10 & Human decision authority & Whether the materials preserve human authority over acceptance, rejection, or merge decisions \\
C11 & Overall usefulness & Overall usefulness of the materials for reviewer-side governance diagnosis \\
\hline
\end{tabular}
\end{adjustbox}
\end{table}

Open-ended feedback prompts asked participants to describe perceived benefits, missing information, concerns about trust or fabrication, workflow friction, and desired interface or workflow support. Responses were coded at the participant level into recurring design-feedback categories and are summarized in Table~\ref{tab:app_feedback_themes}.

\section{Reviewer-Side Evaluation Results and Robustness Checks}
\label{app:evaluation_results}

This section reports the supplementary results for the controlled reviewer-side evaluation, including task allocation, objective accuracy, output-coding reliability, error patterns, evidence and accountability visibility, questionnaire ratings, participant-feedback categories, and sensitivity checks supporting the main-text findings.

\begin{table}[!htbp]
\centering
\caption{Controlled task and condition allocation}
\label{tab:app_task_allocation}
\scriptsize
\setlength{\tabcolsep}{3pt}
\renewcommand{\arraystretch}{1.05}
\begin{adjustbox}{max width=\textwidth}
\begin{tabular}{p{0.09\linewidth} p{0.42\linewidth} c c c}
\hline
\textbf{Task} & \textbf{Risk/evidence condition} & \textbf{Ordinary materials} & \textbf{AGM-supported materials} & \textbf{Total} \\
\hline
T1 & Low-risk documentation & 7 & 8 & 15 \\
T2 & Medium-risk test/configuration & 8 & 7 & 15 \\
T3 & High-risk core logic & 7 & 8 & 15 \\
T4 & Critical authentication/dependency & 8 & 7 & 15 \\
T5 & Critical workflow/placeholder evidence & 7 & 8 & 15 \\
\hline
\end{tabular}
\end{adjustbox}
\end{table}

The main objective contrast was summarized with participant-clustered uncertainty because the 75 task-level outputs were nested within 15 participants. Each bootstrap resample drew participants with replacement and retained all outputs belonging to each selected participant.

\begin{table}[!htbp]
\centering
\caption{Participant-clustered uncertainty for exact risk-label recovery}
\label{tab:app_clustered_objective_uncertainty}
\scriptsize
\setlength{\tabcolsep}{4pt}
\renewcommand{\arraystretch}{1.05}
\begin{adjustbox}{max width=0.82\textwidth}
\begin{tabular}{p{0.38\linewidth} c c}
\hline
\textbf{Metric} & \textbf{Point estimate} & \textbf{Bootstrap 95\% CI} \\
\hline
AGM exact recovery rate & 97.4\% (37/38) & 91.9--100.0\% \\
Ordinary exact recovery rate & 40.5\% (15/37) & 25.0--55.6\% \\
Absolute difference & 56.8 percentage points & 42.1--71.8 percentage points \\
Risk ratio & 2.40 & 1.76--3.79 \\
\hline
\end{tabular}
\end{adjustbox}
\vspace{2pt}
\begin{minipage}{0.92\textwidth}
\scriptsize
\emph{Note.} Intervals are percentile estimates from 10,000 participant-clustered bootstrap resamples. Leave-one-participant-out estimates ranged from 54.3 to 60.0 percentage points for the absolute difference and from 2.27 to 2.62 for the risk ratio.
\end{minipage}
\end{table}

Table~\ref{tab:app_reviewer_output_reliability} reports the independent reliability check for reviewer-side output coding. Agreement was calculated before adjudication so that the statistics reflect the two coders' independent labels.

\begin{table}[!htbp]
\centering
\caption{Independent reliability check for reviewer-side output coding}
\label{tab:app_reviewer_output_reliability}
\scriptsize
\setlength{\tabcolsep}{4pt}
\renewcommand{\arraystretch}{1.05}
\begin{adjustbox}{max width=0.90\textwidth}
\begin{tabular}{p{0.43\linewidth} c c c}
\hline
\textbf{Outcome} & \textbf{Exact agreement} & \textbf{Cohen's $\kappa$} & \textbf{Disagreements} \\
\hline
Final risk label & 100\% & 1.000 & 0 \\
Relevant risk-zone recognition & 95\% & 0.487 & 1 \\
Governance-evidence state & 90\% & 0.848 & 2 \\
Technical-evidence sufficiency & 90\% & 0.825 & 2 \\
Contributor-confirmation state & 100\% & 1.000 & 0 \\
Governance-gate state & 100\% & 1.000 & 0 \\
Technical-review readiness & 100\% & Not estimable & 0 \\
Final-acceptance boundary & 100\% & Not estimable & 0 \\
\hline
\end{tabular}
\end{adjustbox}
\vspace{2pt}
\begin{minipage}{0.94\textwidth}
\scriptsize
\emph{Note.} Agreement statistics use the two coders' pre-adjudication labels. Cohen's $\kappa$ was not estimable for technical-review readiness and final-acceptance boundary because neither field showed between-case variation. Relevant risk-zone recognition had 95\% exact agreement; its lower $\kappa$ reflects a highly skewed category distribution. An auxiliary final-acceptance eligibility field reached 80\% agreement ($\kappa=0.474$), while the distinct final-acceptance boundary variable reached 100\% agreement.
\end{minipage}
\end{table}

\begin{table}[!htbp]
\centering
\caption{Objective governance judgment accuracy by task and condition}
\label{tab:app_objective_accuracy}
\scriptsize
\setlength{\tabcolsep}{3pt}
\renewcommand{\arraystretch}{1.05}
\begin{adjustbox}{max width=\textwidth}
\begin{tabular}{p{0.08\linewidth} p{0.16\linewidth} c c c c}
\hline
\textbf{Task} & \textbf{Condition} & \textbf{N} & \textbf{Exact risk label} & \textbf{Risk under-classified} & \textbf{Governance state correct} \\
\hline
T1 & ordinary & 7 & 100.0\% & 0.0\% & 100.0\% \\
T1 & AGM & 8 & 100.0\% & 0.0\% & 100.0\% \\
T2 & ordinary & 8 & 37.5\% & 62.5\% & 100.0\% \\
T2 & AGM & 7 & 100.0\% & 0.0\% & 100.0\% \\
T3 & ordinary & 7 & 42.9\% & 57.1\% & 100.0\% \\
T3 & AGM & 8 & 100.0\% & 0.0\% & 100.0\% \\
T4 & ordinary & 8 & 25.0\% & 75.0\% & 100.0\% \\
T4 & AGM & 7 & 100.0\% & 0.0\% & 100.0\% \\
T5 & ordinary & 7 & 0.0\% & 100.0\% & 85.7\% \\
T5 & AGM & 8 & 87.5\% & 12.5\% & 100.0\% \\
\hline
\end{tabular}
\end{adjustbox}
\vspace{2pt}
\begin{minipage}{0.95\textwidth}
\scriptsize
\emph{Note.} Exact risk label requires the final risk category to match the task reference label. The single AGM-supported mismatch occurs in T5, where the output recognized the changed workflow path as critical but assigned a final high-risk label. It still recovered the blocked governance state and invalid evidence condition; under a close-or-correct risk-recognition criterion, AGM-supported materials recover the relevant risk zone in 38/38 observations.
\end{minipage}
\end{table}

Tables~\ref{tab:app_ordinary_errors} and~\ref{tab:app_observability_detection} decompose the main failure modes under ordinary review materials. These tables show that ordinary materials often made high-risk changes appear routine and made missing, invalid, or placeholder evidence difficult to detect.

\begin{table}[!htbp]
\centering
\caption{Error patterns under ordinary review materials}
\label{tab:app_ordinary_errors}
\scriptsize
\setlength{\tabcolsep}{3pt}
\renewcommand{\arraystretch}{1.05}
\begin{adjustbox}{max width=\textwidth}
\begin{tabular}{p{0.28\linewidth} c c p{0.45\linewidth}}
\hline
\textbf{Error pattern} & \textbf{N} & \textbf{\% of ordinary tasks} & \textbf{Review implication} \\
\hline
Risk exact mismatch & 22 & 59.5\% & Repository-defined risk not recovered \\
Risk under-classified & 22 & 59.5\% & Governance-critical change treated as lower risk \\
Risk not close/correct & 11 & 29.7\% & Risk label not adjacent to expected category \\
Governance state not fully correct & 1 & 2.7\% & Governance state partially misread \\
Final strict state not matched & 5 & 13.5\% & Final-acceptance state too permissive/strict \\
Final safe state not matched & 1 & 2.7\% & Unsafe final judgment \\
Risk over-classified & 0 & 0.0\% & Risk overstated \\
\hline
\end{tabular}
\end{adjustbox}
\end{table}

\begin{table}[!htbp]
\centering
\caption{Review-state observability and evidence detection by condition}
\label{tab:app_observability_detection}
\scriptsize
\setlength{\tabcolsep}{3pt}
\renewcommand{\arraystretch}{1.05}
\begin{adjustbox}{max width=\textwidth}
\begin{tabular}{p{0.32\linewidth} p{0.12\linewidth} c p{0.45\linewidth}}
\hline
\textbf{Indicator} & \textbf{Condition} & \textbf{N} & \textbf{Observed response distribution} \\
\hline
Missing evidence detected & ordinary & 37 & Not applicable / not observable: 37 \\
Missing evidence detected & AGM & 38 & No: 28; Yes: 9; Not applicable / not observable: 1 \\
Placeholder/empty-template evidence detected & ordinary & 37 & Not applicable / not observable: 37 \\
Placeholder/empty-template evidence detected & AGM & 38 & No: 27; Yes: 9; Not applicable / not observable: 2 \\
Contributor-confirmation declaration required & ordinary & 37 & Not applicable / not observable: 37 \\
Contributor-confirmation declaration required & AGM & 38 & No: 7; Yes: 28; Not applicable / not observable: 3 \\
AGM gate state visible & ordinary & 37 & Not observable: 37 \\
AGM gate state visible & AGM & 38 & Pass: 15; Needs additional evidence: 15; Blocked by missing or invalid evidence: 8 \\
\hline
\end{tabular}
\end{adjustbox}
\end{table}

Table~\ref{tab:app_human_accountability_visibility} reports human-accountability visibility, which anchors AGM-supported review in human contributor confirmation and maintainer authority.

\begin{table}[!htbp]
\centering
\caption{Human accountability visibility by material condition}
\label{tab:app_human_accountability_visibility}
\scriptsize
\setlength{\tabcolsep}{3pt}
\renewcommand{\arraystretch}{1.05}
\begin{adjustbox}{max width=\textwidth}
\begin{tabular}{p{0.16\linewidth} p{0.50\linewidth} c c}
\hline
\textbf{Condition} & \textbf{Contributor-confirmation status response} & \textbf{N} & \textbf{\%} \\
\hline
ordinary & Not observable & 36 & 97.3\% \\
ordinary & Not required & 1 & 2.7\% \\
AGM & Not observable & 0 & 0.0\% \\
AGM & Not required & 14 & 36.8\% \\
AGM & Missing & 11 & 28.9\% \\
AGM & Invalid / placeholder & 7 & 18.4\% \\
AGM & Pending & 1 & 2.6\% \\
AGM & Completed and declared & 5 & 13.2\% \\
\hline
\end{tabular}
\end{adjustbox}
\end{table}

Tables~\ref{tab:app_questionnaire_results}--\ref{tab:app_feedback_themes} report subjective usefulness ratings, participant-level variation, and participant-feedback categories. After averaging within participant and condition, the overall mean was 6.14 under AGM-supported materials and 3.27 under ordinary materials. The paired difference was 2.87 points (participant-clustered bootstrap 95\% CI: 2.67--3.08), and all 15 participants had a higher AGM-supported mean.

\begin{table}[!htbp]
\centering
\caption{Reviewer questionnaire participant-level results with clustered uncertainty}
\label{tab:app_questionnaire_results}
\scriptsize
\setlength{\tabcolsep}{3pt}
\renewcommand{\arraystretch}{1.05}
\begin{adjustbox}{max width=\textwidth}
\begin{tabular}{p{0.07\linewidth} p{0.31\linewidth} c c c p{0.18\linewidth}}
\hline
\textbf{Item} & \textbf{Short label} & \textbf{Ordinary mean} & \textbf{AGM mean} & \textbf{Paired difference} & \textbf{Bootstrap 95\% CI} \\
\hline
C1 & Understand/review change & 3.06 & 6.04 & +2.99 & 2.67--3.31 \\
C2 & Identify risk zones & 3.02 & 6.09 & +3.07 & 2.80--3.36 \\
C3 & Judge evidence sufficiency & 2.90 & 5.96 & +3.06 & 2.83--3.29 \\
C4 & Detect missing/placeholder evidence & 2.77 & 6.07 & +3.30 & 3.08--3.53 \\
C5 & Understand test relevance & 2.94 & 5.87 & +2.92 & 2.59--3.23 \\
C6 & Use source/provenance information & 2.71 & 6.04 & +3.33 & 2.96--3.70 \\
C7 & See human accountability & 2.53 & 6.04 & +3.51 & 3.13--3.91 \\
C8 & Judge AGM gate state & 3.16 & 6.03 & +2.88 & 2.62--3.14 \\
C9 & Support review without reasoning dependence & 3.76 & 6.31 & +2.56 & 2.12--2.94 \\
C10 & Preserve human decision authority & 6.41 & 6.76 & +0.34 & [$-0.04$, 0.80] \\
C11 & Overall usefulness & 2.67 & 6.31 & +3.64 & 3.30--3.96 \\
\hline
\end{tabular}
\end{adjustbox}
\vspace{2pt}
\begin{minipage}{0.95\textwidth}
\scriptsize
\emph{Note.} Means are calculated after averaging task ratings within participant and condition. Differences are paired participant-level contrasts. Intervals use 10,000 percentile bootstrap resamples of participants.
\end{minipage}
\end{table}

\begin{table}[!htbp]
\centering
\caption{Participant-level exact risk-label recovery}
\label{tab:app_participant_variation}
\scriptsize
\setlength{\tabcolsep}{3pt}
\renewcommand{\arraystretch}{1.05}
\begin{adjustbox}{max width=\textwidth}
\begin{tabular}{c c c c c c c c}
\hline
\textbf{Participant} & \textbf{Ord.\ N} & \textbf{Ord.\ exact} & \textbf{AGM N} & \textbf{AGM exact} & \textbf{Difference (pp)} & \textbf{Ord.\ underclass.} & \textbf{AGM underclass.} \\
\hline
P01 & 2 & 2/2 (100.0\%) & 3 & 3/3 (100.0\%) & +0.0 & 0 & 0 \\
P02 & 3 & 1/3 (33.3\%) & 2 & 2/2 (100.0\%) & +66.7 & 2 & 0 \\
P03 & 2 & 0/2 (0.0\%) & 3 & 2/3 (66.7\%) & +66.7 & 2 & 1 \\
P04 & 3 & 1/3 (33.3\%) & 2 & 2/2 (100.0\%) & +66.7 & 2 & 0 \\
P05 & 2 & 1/2 (50.0\%) & 3 & 3/3 (100.0\%) & +50.0 & 1 & 0 \\
P06 & 3 & 2/3 (66.7\%) & 2 & 2/2 (100.0\%) & +33.3 & 1 & 0 \\
P07 & 2 & 0/2 (0.0\%) & 3 & 3/3 (100.0\%) & +100.0 & 2 & 0 \\
P08 & 3 & 1/3 (33.3\%) & 2 & 2/2 (100.0\%) & +66.7 & 2 & 0 \\
P09 & 2 & 0/2 (0.0\%) & 3 & 3/3 (100.0\%) & +100.0 & 2 & 0 \\
P10 & 3 & 2/3 (66.7\%) & 2 & 2/2 (100.0\%) & +33.3 & 1 & 0 \\
P11 & 2 & 0/2 (0.0\%) & 3 & 3/3 (100.0\%) & +100.0 & 2 & 0 \\
P12 & 3 & 2/3 (66.7\%) & 2 & 2/2 (100.0\%) & +33.3 & 1 & 0 \\
P13 & 2 & 0/2 (0.0\%) & 3 & 3/3 (100.0\%) & +100.0 & 2 & 0 \\
P14 & 3 & 1/3 (33.3\%) & 2 & 2/2 (100.0\%) & +66.7 & 2 & 0 \\
P15 & 2 & 2/2 (100.0\%) & 3 & 3/3 (100.0\%) & +0.0 & 0 & 0 \\
\hline
\end{tabular}
\end{adjustbox}
\vspace{2pt}
\begin{minipage}{0.95\textwidth}
\scriptsize
\emph{Note.} Exact risk-label recovery was higher under AGM-supported materials for 13 of 15 participants and equal for the remaining two.
\end{minipage}
\end{table}

\begin{table}[!htbp]
\centering
\caption{Participant feedback categories and design implications}
\label{tab:app_feedback_themes}
\scriptsize
\setlength{\tabcolsep}{3pt}
\renewcommand{\arraystretch}{1.05}
\begin{adjustbox}{max width=\textwidth}
\begin{tabular}{p{0.27\linewidth} c p{0.25\linewidth} p{0.35\linewidth}}
\hline
\textbf{Feedback category} & \textbf{Participants mentioning} & \textbf{Example participant IDs} & \textbf{Design implication} \\
\hline
Risk visibility / risk-zone attention & 11 & P01, P03, P04, P05, P06, P08, P09, P10\ldots & Keep concise risk summaries and risk-zone labels. \\
Evidence index / missing-evidence report / tests & 10 & P01, P02, P03, P05, P06, P07, P09, P10\ldots & Preserve evidence index, missing-evidence report, and test-evidence summary. \\
Contributor confirmation / accountability visibility & 13 & P01, P02, P03, P04, P05, P06, P07, P08\ldots & Keep contributor-confirmation declaration explicit but separate from final acceptance. \\
Gate-state visibility & 8 & P01, P02, P03, P04, P05, P06, P11, P12 & Maintain reference-vs-observed gate-state presentation. \\
Usability simplification needed & 10 & P01, P02, P03, P04, P05, P08, P09, P10\ldots & Separate short summary from full diagnostic report. \\
Structured UI or next-step prompts desired & 11 & P01, P02, P03, P04, P05, P06, P07, P08\ldots & Consider risk cards, checklist panels, and next-action prompts. \\
Suitable for large/high-risk/low-trust projects & 13 & P01, P03, P04, P05, P06, P07, P08, P09\ldots & Position AGM for high-risk, multi-party, unfamiliar-contributor settings. \\
Changed reviewer-agent interaction & 15 & P01, P02, P03, P04, P05, P06, P07, P08\ldots & AGM can structure reviewer-side prompting and triage behavior. \\
\hline
\end{tabular}
\end{adjustbox}
\vspace{2pt}
\begin{minipage}{0.95\textwidth}
\scriptsize
\emph{Note.} Feedback-category counts are participant-level counts from D1--D10 open responses using conservative, text-supported coding; each participant was counted at most once within a category.
\end{minipage}
\end{table}

Table~\ref{tab:app_sensitivity_checks} reports supplementary diagnostic checks. These checks assess whether the main pattern depends on a single task type or task subset.

\begin{table}[!htbp]
\centering
\caption{Supplementary sensitivity checks}
\label{tab:app_sensitivity_checks}
\scriptsize
\setlength{\tabcolsep}{3pt}
\renewcommand{\arraystretch}{1.05}
\begin{adjustbox}{max width=\textwidth}
\begin{tabular}{p{0.34\linewidth} c c c c c c}
\hline
\textbf{Check} & \textbf{Ord.\ N} & \textbf{Ord.\ exact risk} & \textbf{Ord.\ underclass.} & \textbf{AGM N} & \textbf{AGM exact risk} & \textbf{AGM underclass.} \\
\hline
All tasks & 37 & 40.5\% & 59.5\% & 38 & 97.4\% & 2.6\% \\
High/critical tasks only (T3--T5) & 22 & 22.7\% & 77.3\% & 23 & 95.7\% & 4.3\% \\
Excluding T5 blocked-evidence task & 30 & 50.0\% & 50.0\% & 30 & 100.0\% & 0.0\% \\
Excluding T1 low-risk task & 30 & 26.7\% & 73.3\% & 30 & 96.7\% & 3.3\% \\
\hline
\end{tabular}
\end{adjustbox}
\end{table}

\section{Contributor-Side Feasibility Protocol and Questionnaire Results}
\label{app:contributor_feasibility}

This section documents the contributor-side feasibility check summarized in the main manuscript: the task procedure, frozen validation rule set, draft-to-final audit, and questionnaire results used to assess AGM-based evidence preparation by contributor-side agents and confirmation by human contributors under controlled conditions.

\subsection*{Task Design and Procedure}

The feasibility check used a separate cohort of 15 participants with no overlap with the reviewer-side evaluation. Each participant completed three tasks: Task A represented a low-risk documentation contribution, Task B a high-risk parser-validation contribution requiring test evidence, and Task C a critical workflow-sensitive CI contribution requiring maintainer review to remain pending. Task order followed the cyclic rotations A--B--C, B--C--A, and C--A--B, with each starting position represented five times. Cohort separation eliminated participant-level transfer of reviewer-side risk, evidence, and gate-state experience into contributor-side evidence-package preparation. For each task, the participant asked a contributor-side agent to read the project AGM and the task file, generate a draft evidence package, wait for contributor confirmation, and then generate a final evidence package. The expected file-level workflow proceeded from draft preparation to contributor confirmation and finalization.

One participant's initial run was excluded because repeated agent-session interruptions affected task execution. The participant repeated the three tasks under the same protocol, and the rerun outputs were used for final coding. The original interrupted run was retained only as an audit record and was not used in the main analysis.

\subsection*{Validation Rule}

The frozen validation rule set distinguishes strict structural validity from correctness of the core governance state. Strict structural validation checks package stage, required fields, evidence items, gate fields, contributor-confirmation fields, and maintainer-review fields. For test and validation evidence, separate command and result fields are required, while narrative content is optional. The core governance state is coded as correct when the risk label, review-gate status, maintainer-review requirement/status, and contributor-confirmation handling are all correct. This distinction is important because a package can contain minor schema-format errors while still correctly representing the governance state relevant to review.

\begin{figure}[!htbp]
\centering
\includegraphics[width=0.84\linewidth]{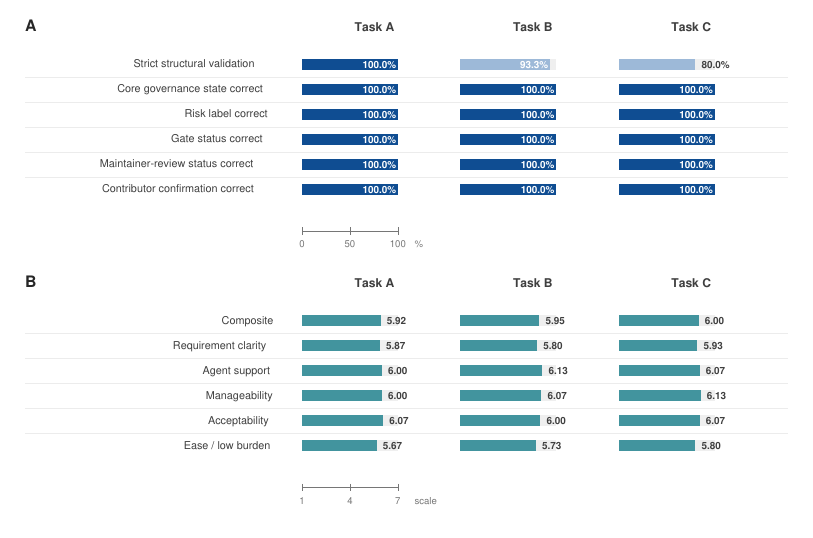}
\caption{Task-level contributor-side robustness. Panel A reports objective contributor-side package-validation and governance-state metrics by task. Panel B reports descriptive task-level questionnaire means on a fixed 1--7 scale.}
\label{fig:task_level_robustness}
\end{figure}

Figure~\ref{fig:task_level_robustness} summarizes task-level robustness across Tasks A, B, and C and supports the main-text feasibility result.

As summarized in Figure~\ref{fig:task_level_robustness}A, all 45 final packages represented the core governance state correctly, and 41 of 45 passed strict structural validation. The four remaining strict-validation failures were schema-precision issues. Two Task C packages contained validation evidence but did not provide separate command and result fields. Two packages had a correct gate status but omitted the explicit boolean \texttt{review\_gate.required} field. None of these failures changed the coded risk label, review-gate status, maintainer-review requirement/status, or contributor-confirmation state.

\subsection*{Draft-to-Final Audit}

All 45 observations produced both draft and final evidence packages. Across all observations, draft-to-final changes were limited to the package-stage and contributor-confirmation fields. No task changed the risk label, evidence content, review-gate status, or maintainer-review status during the final confirmation step. This audit supports the role separation intended by AGM: contributor-side agents prepare evidence, human contributors confirm the package, and maintainer review remains a separate authority state.

\subsection*{Questionnaire Items and Summary}

After completing all three tasks, participants completed a compact questionnaire. For each task, they rated five items on a seven-point scale: evidence-preparation burden, requirement clarity, agent support, manageability, and acceptability. The burden item was reverse-coded into an ease score using \texttt{ease = 8 - burden}. Participant-level estimates first averaged each participant across Tasks A--C; 95\% confidence intervals used 10,000 percentile bootstrap resamples of participants. The 45 task-level ratings were retained for descriptive task patterns. Participants also reported whether they experienced technical interruptions, which task felt most difficult, and open-ended comments about burden, clarity, usefulness, and agent behavior.

Figure~\ref{fig:task_level_robustness}B reports the corresponding task-level questionnaire patterns. All 15 participants reported no major interruption in the final dataset. Thirteen participants reported no clear difference in task difficulty, while two selected Task B as the most burdensome task. Open comments locate the main burden in the initial effort required to understand AGM, with less emphasis on evidence preparation itself. Several participants suggested standard AGM documents, dedicated AGM agents, or skill-like workflow automation. These responses reinforce the interpretation that workflow scaffolding is part of the practical usability of AGM-supported governance workflows.

\begin{table}[!htbp]
\centering
\caption{Contributor-side questionnaire participant-level summary}
\label{tab:app_contributor_questionnaire_summary}
\scriptsize
\setlength{\tabcolsep}{3pt}
\renewcommand{\arraystretch}{1.05}
\begin{adjustbox}{max width=\textwidth}
\begin{tabular}{p{0.33\linewidth} c c c c c}
\hline
\textbf{Metric} & \textbf{Participants} & \textbf{Mean} & \textbf{Median} & \textbf{Bootstrap 95\% CI} & \textbf{Task-level ratings} \\
\hline
Ease / low burden & 15 & 5.73 & 6.00 & 5.33--6.11 & 45 \\
Requirement clarity & 15 & 5.87 & 6.00 & 5.64--6.07 & 45 \\
Agent support & 15 & 6.07 & 6.00 & 5.84--6.31 & 45 \\
Manageability & 15 & 6.07 & 6.00 & 5.80--6.33 & 45 \\
Acceptability & 15 & 6.04 & 6.00 & 6.00--6.13 & 45 \\
Composite & 15 & 5.96 & 6.00 & 5.79--6.12 & 45 \\
\hline
\end{tabular}
\end{adjustbox}
\vspace{2pt}
\begin{minipage}{0.95\textwidth}
\scriptsize
\emph{Note.} Each participant was first averaged across the three tasks. Intervals use 10,000 percentile bootstrap resamples of participants. The task-level count is retained only to document the descriptive distribution. The burden item is reverse-coded as ease/low burden using \texttt{ease = 8 - burden}.
\end{minipage}
\end{table}

\section{Artifact Boundary, Adjacent Mechanisms, and Versioning Notes}
\label{app:boundary_versioning}

This section consolidates the scope boundaries, adjacent-mechanism distinctions, and versioning boundary for the submitted research snapshot. AGM is treated as a repository-hosted governance resource for making AI-mediated contributions risk-classifiable, evidence-inspectable, accountability-visible, and review-gate-ready. Detection, surveillance, provenance standardization, and automated decision authority remain outside this role.

\subsection*{Scope Boundaries of AGM}

Table~\ref{tab:app_boundary_conditions} summarizes the main scope boundaries. They position AGM as a compliance-enabling governance resource that clarifies the responsible contribution path while preserving low-barrier participation and human maintainer authority.

\begin{table}[!htbp]
\centering
\caption{Boundary conditions for interpreting AGM}
\label{tab:app_boundary_conditions}
\scriptsize
\setlength{\tabcolsep}{3pt}
\renewcommand{\arraystretch}{1.06}
\begin{adjustbox}{max width=\textwidth}
\begin{tabular}{p{0.22\linewidth} p{0.34\linewidth} p{0.34\linewidth}}
\hline
\textbf{Boundary} & \textbf{What AGM does} & \textbf{What AGM does not claim} \\
\hline
AI-use governance
& Defines project-side expectations for AI-mediated contributions, including risk zones, evidence obligations, accountability boundaries, and review gates.
& Does not prohibit all AI use, certify that a contribution is or is not AI-generated, or estimate population-level AI adoption. \\

Evidence inspectability
& Turns contribution evidence into structured artifacts that reviewers can inspect, validate, and challenge.
& Does not treat templates, empty fields, placeholder reports, or unverifiable claims as sufficient evidence. \\

Human accountability
& Makes contributor-confirmation requirements and responsibility declarations visible for high-risk or critical changes.
& Does not replace maintainer judgment, assign legal responsibility, or automatically approve or reject a contribution. \\

Agent-readability
& Provides machine-readable governance rules that agents and review-support tools can consume.
& Does not reduce governance to \texttt{AGENTS.md}, \texttt{CLAUDE.md}, prompt files, or coding instructions. \\

Traceability
& Allows external provenance or trace artifacts to be attached as supporting evidence when available.
& Does not require disclosure of private reasoning or mandate a single provenance format. \\

Enforcement
& Supports maintainers by making missing evidence, blocked gates, and risk-sensitive review requirements visible.
& Does not function as punitive surveillance, an anti-fraud guarantee, or a complete defense against deliberate misrepresentation. \\

Model and deployment scope
& Externalizes governance rules, evidence requirements, validation states, and review packets into structured project-side artifacts that can be used with deterministic checks and supporting tools.
& Does not establish model-size independence, comparative deployment costs, or effectiveness across models and organizational settings. \\

\hline
\end{tabular}
\end{adjustbox}
\end{table}

\subsection*{Versioning and Reproducibility Boundary}

The empirical diagnosis, artifact design, and controlled evaluations should be interpreted as a fixed research snapshot. The public AGM artifact may continue to evolve, whereas the claims reported in this study refer to the artifact state, processed data, task materials, and validation rules used in the reported analyses. Table~\ref{tab:app_versioning_notes} records these reproducibility boundaries.

\begin{table}[!htbp]
\centering
\caption{Versioning and reproducibility notes}
\label{tab:app_versioning_notes}
\scriptsize
\setlength{\tabcolsep}{3pt}
\renewcommand{\arraystretch}{1.06}
\begin{adjustbox}{max width=\textwidth}
\begin{tabular}{p{0.25\linewidth} p{0.65\linewidth}}
\hline
\textbf{Element} & \textbf{Reproducibility boundary} \\
\hline

AGM artifact state
& The study refers to the AGM v0.1.0 community-draft artifact state pinned to public commit \href{https://github.com/agent-governance-manifest/agent-governance-manifest/commit/c781a2f40d823ca8b5cc53fb43ccf2c4f88dfa1b}{\texttt{c781a2f}}, including the specification, example manifests, validation scripts, review-packet generation scripts, and documentation described in Supplementary Section~\ref{app:agm_artifact}. \\

Artifact development status
& AGM v0.1.0 is a research prototype and community-draft specification intended to instantiate and evaluate the proposed governance mechanism. Subsequent releases may extend its schema, integrations, interfaces, and validation capabilities. \\

Repository audit data
& The empirical findings refer to the processed 50-repository diagnostic dataset bounded by the June 14, 2026 observation cutoff. The final API collection run was completed on June 15, 2026. The sampling and analytical boundaries are documented in Supplementary Sections~\ref{app:sample_design}--\ref{app:additional_statistics}. \\

Reviewer-side evaluation
& The reviewer-side results refer to 15 participants and 75 task-level reviewer-side outputs generated under ordinary-material and AGM-supported conditions. The task materials, objective coding rubric, condition assignments, and supplementary results are documented in Supplementary Sections~\ref{app:evaluation_materials}--\ref{app:evaluation_results}. \\

Contributor-side feasibility check
& The contributor-side results refer to 15 participants, 45 AGM-supported contribution tasks, and 90 draft and final evidence packages. All packages were evaluated using the frozen validation rule set archived with the replication materials and documented in Supplementary Section~\ref{app:contributor_feasibility}. \\

Interpretive boundary
& The artifact, data, and evaluation materials jointly support the design-theoretic and diagnostic claims reported in the study. Later changes to the artifact or its validation rules constitute subsequent design development and are outside the evidentiary boundary of the reported analyses. \\
\hline
\end{tabular}
\end{adjustbox}
\end{table}

Supplementary Sections~\ref{app:sample_design}--\ref{app:additional_statistics} document the repository diagnosis, governance coding, second-coder check, and adjudicated variables. Supplementary Section~\ref{app:agm_artifact} documents the AGM artifact and its relationship to adjacent governance mechanisms. Supplementary Sections~\ref{app:evaluation_materials}--\ref{app:evaluation_results} cover the reviewer-side evaluation, and Supplementary Section~\ref{app:contributor_feasibility} covers the contributor-side feasibility check. These materials define the fixed research snapshot underlying the reported findings.

\FloatBarrier
\end{document}